\documentclass[11pt,onecolumn,english,showpacs,preprintnumbers,amsmath,amssymb,floatfix,nofootinbib]{revtex4-1}

\usepackage[T1]{fontenc}
\usepackage[latin9]{inputenc}
\usepackage{color}
\usepackage{array}
\usepackage{amstext}
\usepackage{graphicx}
\usepackage{esint}
\usepackage{cleveref}
\usepackage{rotating}
\usepackage{slashed}
\usepackage[normalem]{ulem}
\usepackage{siunitx}
\usepackage[font=small,labelfont=bf]{caption}
\usepackage{xcolor}
\usepackage[colorlinks = true,
linkcolor = magenta,
urlcolor  = blue,
citecolor = red,
anchorcolor = blue]{hyperref}

\usepackage{ulem}

\newcommand{\pom} {I\!\!P}
\newcommand{\reg} {I\!\!R}
\newcommand{\xpom}{x_{\xpom}}

\makeatletter

\providecommand{\tabularnewline}{\\}

\@ifundefined{textcolor}{}
{%
 \definecolor{BLACK}{gray}{0}
 \definecolor{WHITE}{gray}{1}
 \definecolor{RED}{rgb}{1,0,0}
 \definecolor{GREEN}{rgb}{0,1,0}
 \definecolor{BLUE}{rgb}{0,0,1}
 \definecolor{CYAN}{cmyk}{1,0,0,0}
 \definecolor{MAGENTA}{cmyk}{0,1,0,0}
 \definecolor{YELLOW}{cmyk}{0,0,1,0}
 }

\@ifundefined{definecolor}
 {\usepackage{color}}{}
\@ifundefined{definecolor}
 {\usepackage{color}}{}
\makeatother
\usepackage{babel}

\def\Re{{\cal R \mskip-4mu \lower.1ex \hbox{\it e}\,}}
\def\Im{{\cal I \mskip-5mu \lower.1ex \hbox{\it m}\,}}

\def\tev{\,{\ifmmode\mathrm {TeV}\else TeV\fi}}
\def\gev{\,{\ifmmode\mathrm {GeV}\else GeV\fi}}
\def\mev{\,{\ifmmode\mathrm {MeV}\else MeV\fi}}
\def\to{\rightarrow}

\begin{document}


\title { First global next-to-leading order determination of diffractive parton distribution functions and their uncertainties within the {\tt xFitter} framework  }

\author{Muhammad Goharipour$^{1}$}
\email{Muhammad.Goharipour@gmail.com}

\author{Hamzeh Khanpour$^{2,1}$}
\email{Hamzeh.Khanpour@mail.ipm.ir}

\author{Vadim Guzey$^{3,4,5}$}
\email{Guzey\_va@pnpi.nrcki.ru}

\affiliation {
$^{(1)}$School of Particles and Accelerators, Institute for Research in Fundamental Sciences (IPM), P.O.Box 19395-5531, Tehran, Iran            \\
$^{(2)}$Department of Physics, University of Science and Technology of Mazandaran, P.O.Box 48518-78195, Behshahr, Iran    \\ 
$^{(3)}$Department  of Physics,  University  of  Jyv\"askyl\"a, P.O. Box 35, 40014  University  of  Jyv\"askyl\"a,  Finland \\  
$^{(4)}$Helsinki Institute of Physics, P.O.  Box  64,  00014  University  of  Helsinki,  Finland   \\  
$^{(5)}$National Research Center ``Kurchatov Institute'', Petersburg Nuclear Physics Institute (PNPI), Gatchina, 188300, Russia}

\date{\today}

%
\begin{abstract}\label{abstract}

We present {\tt GKG18-DPDFs}, a next-to-leading order (NLO) QCD analysis of diffractive parton distribution functions (diffractive PDFs) and their uncertainties.
This is the first global set of diffractive PDFs determined within the {\tt xFitter} framework. This analysis is motivated by all available and most up-to-date data on inclusive diffractive deep inelastic scattering (diffractive DIS). Heavy quark contributions are considered within the framework of the Thorne-Roberts (TR) general mass variable flavor number scheme (GM-VFNS).
We form a mutually consistent set of diffractive PDFs due to the inclusion of high-precision data from H1/ZEUS combined inclusive diffractive cross sections measurements. We study the impact of the H1/ZEUS combined data by producing a variety of determinations based on reduced data sets. We find that these data sets have a significant impact on the diffractive PDFs with some substantial reductions in uncertainties.
The predictions based on the extracted diffractive PDFs are compared to the analyzed diffractive DIS data and with other determinations of the diffractive PDFs.

\end{abstract}
%


\maketitle
\tableofcontents{}

%
\section{Introduction}\label{sec:intro}

High precision calculations of hard scattering cross sections in lepton-hadron deep inelastic scattering (DIS) and hadron-hadron collider experiments can be done within the framework of perturbative quantum chromodynamics (pQCD). The computations of cross sections can be performed using the so-called factorization theorem that allows for a systematic separation of perturbative and nonperturbative physics~\cite{Aktas:2007hn,Collins:1997sr}.
Some examples for describing the latter in various processes are the well-known parton distribution functions (PDFs)~\cite{Ball:2017nwa,Bourrely:2015kla,Harland-Lang:2014zoa,Hou:2017khm,Alekhin:2017kpj}, nuclear PDFs~\cite{Khanpour:2016pph,Eskola:2016oht,Kovarik:2015cma,Wang:2016mzo}, and polarized  PDFs~\cite{Khanpour:2017cha,Shahri:2016uzl,Jimenez-Delgado:2014xza,Sato:2016tuz,Nocera:2014gqa,Ethier:2017zbq,Khanpour:2017fey},
which are rather tightly constrained by global QCD fits to DIS and hadron collider data. In fact, they are crucial assets in all scattering processes involving hadrons (nucleons and nuclei) in the initial state. In this respect, phenomenological and experimental studies over the past three decades have provided important information on the structure of hadrons. A 
significant  amount of PDF sets has been determined considering the most precise data from LHC Run I and II~\cite{Harland-Lang:2014zoa,Alekhin:2017kpj,Ball:2017nwa,Ball:2014uwa,Dulat:2015mca,Abramowicz:2015mha,Jimenez-Delgado:2014twa,Accardi:2016qay,Butterworth:2015oua}. In the literature, the relative importance of LHC data has been subject to considerable discussion. These new and up-to-date sets of PDFs have played an
important role in the search for new physics, for example in the top quark and Higgs boson sectors~\cite{Rojo:2015acz,Ball:2017nwa}.

Diffractive processes, $e p \to e p X$, where $X$ represents hadronic final state separated from  the recoiled proton by a rapidity gap  and the proton in the final state carries most of the beam momentum (see Fig.~\ref{fig:Feynman}), have been studied extensively in the H1 and ZEUS experiments at the electron-proton ($ep$) collider  HERA~\cite{Aktas:2006hy,Aktas:2007bv,Aktas:2007hn,Aktas:2006up,Chekanov:2007aa,Chekanov:2008fh,Chekanov:2009aa}.
At HERA, a substantial fraction of up to 10\% of  all $ep$ DIS interactions proceeds via the diffractive scattering process initiated by a highly virtual photon.
In the framework of the collinear factorization theorem, the theoretical calculation of diffractive cross sections requires a special type of
nonperturbative functions as input, so that the universal diffractive PDFs may be defined.
To be more precise, the factorization theorem predicts that the cross section can be expressed 
as the convolution of nonperturbative diffractive PDFs and partonic cross sections of the hard subprocess calculable within the framework of pQCD. 
Consequently, the dynamics of the diffractive processes can be formulated in terms of quark and gluon densities.
The diffractive PDFs have properties similar to the PDFs of the free nucleon, but with the constraint of a leading proton or its low mass excitations being
present in the final state. Like PDFs, it is well established  that the diffractive PDFs are universal quantities, which can be extracted from diffractive DIS data through global QCD analyses.
The knowledge of diffractive PDFs for different hadron species as well as the estimation of their uncertainties
is therefore vital for precise theoretical and experimental calculations and, hence, has received quite some interests in the past (see, for
example, Ref.~\cite{Levonian:2017bqk} for a recent review).

The main sources to constrain the diffractive PDFs are the inclusive diffractive DIS data measured at HERA.
Given the diffractive PDFs, perturbative QCD calculations are expected to be
applicable to other processes such as the jet and heavy quark production in diffractive DIS at HERA~\cite{Chekanov:2002qm,Aktas:2006up,Aktas:2007bv,Chekanov:2007aa,Aaron:2011mp,Andreev:2014yra}.
A full discussion of diffractive dijet production at HERA will be the main subject of our future work.
Indeed, the next-to-leading order (NLO) QCD predictions using diffractive PDFs describe these measurements rather well.
There are several studies in which the diffractive PDFs have been determined from the QCD analyses of diffractive DIS data~\cite{Ceccopieri:2016rga,Hautmann:2000pw,Martin:2004xw,Martin:2005hd,Royon:2000wg,Monfared:2011xf,Chekanov:2009aa,Aktas:2006hy}. 
In this paper, we present a new set of diffractive PDFs, referred to as {\tt GKG18-DPDFs}, through a comprehensive NLO QCD analysis. The {\tt GKG18-DPDFs} diffractive PDFs are
determined using all available and up-to-date data from diffractive DIS cross section~\cite{Aaron:2012zz,Aaron:2012ad}, including, for the first time, the H1 and ZEUS combined inclusive diffractive cross section measurements~\cite{Aaron:2012hua}. 

The outline of this paper is as follows: In Section~\ref{sec:DDIS}, we briefly present the theoretical formalism
adopted for describing the diffractive DIS at HERA. After reviewing the QCD factorization theorem in
Section~\ref{sec:factorisation}, we explain the heavy flavor contributions to the diffractive DIS structure function in Section~\ref{sec:heavy-flavour}.
The phenomenological framework used in {\tt GKG18-DPDFs} global QCD analysis is presented in Sec.~\ref{sec:statistical}. This section includes our parametrizations of the diffractive PDFs (Sec.~\ref{sec:Parametrizations}), a detailed discussion of the description of different data sets included in {\tt GKG18-DPDFs} global fit (Sec.~\ref{sec:data}), and the method of minimization and diffractive PDF uncertainties (Sec.~\ref{sec:minimization}). 
In Section~\ref{sec:results}, we present {\tt GKG18-DPDFs} results for diffractive PDFs obtained from global fits to H1 diffractive DIS cross sections~\cite{Aaron:2012zz,Aaron:2012ad}, 
and H1 and ZEUS combined inclusive diffractive data~\cite{Aaron:2012hua}.
In Sec.~\ref{sec:comparison-to-other-DPDFs}, we compare the diffractive PDFs obtained in this work to the previously 
determined by other groups. Section~\ref{sec:comparison-to-DDISdata} is also devoted to comparing the theoretical 
predictions based on the extracted diffractive PDFs with the analyzed diffractive DIS data.
Finally, in Section~\ref{sec:Discussion}, we present our summary and conclusions.

%
\section{Theoretical framework and assumptions}\label{sec:framework}
%

In the following we describe the standard theoretical framework adopted for the diffractive DIS. Although, there are different theoretical approaches to describe the diffractive processes in literature~\cite{Royon:2006by}, it is well known now that the approach, where the diffractive DIS is mediated by the exchange of 
the hard Pomeron and a secondary Reggeon can be remarkably successful for the description of most of diffractive DIS data.

\subsection{Cross section for diffractive DIS}\label{sec:DDIS}

In order to discuss the cross section for diffractive DIS, one needs to introduce the kinematic variables first. 
The common variables in any DIS process are as follows:
the photon virtuality $Q^{2} = - q^2$, where $q=k - k^\prime$ is the difference of the four-momenta of the incoming ($k$) and outgoing ($k^\prime$) leptons;
the longitudinal momentum fraction $x=\frac{-q^2}{2P.q}$, where $P$ is the four-momentum of the incoming proton; and the inelasticity $y=\frac{P.q}{P.k}$.

\begin{figure*}[htb]
\vspace{0.50cm}
\includegraphics[clip,width=0.50\textwidth]{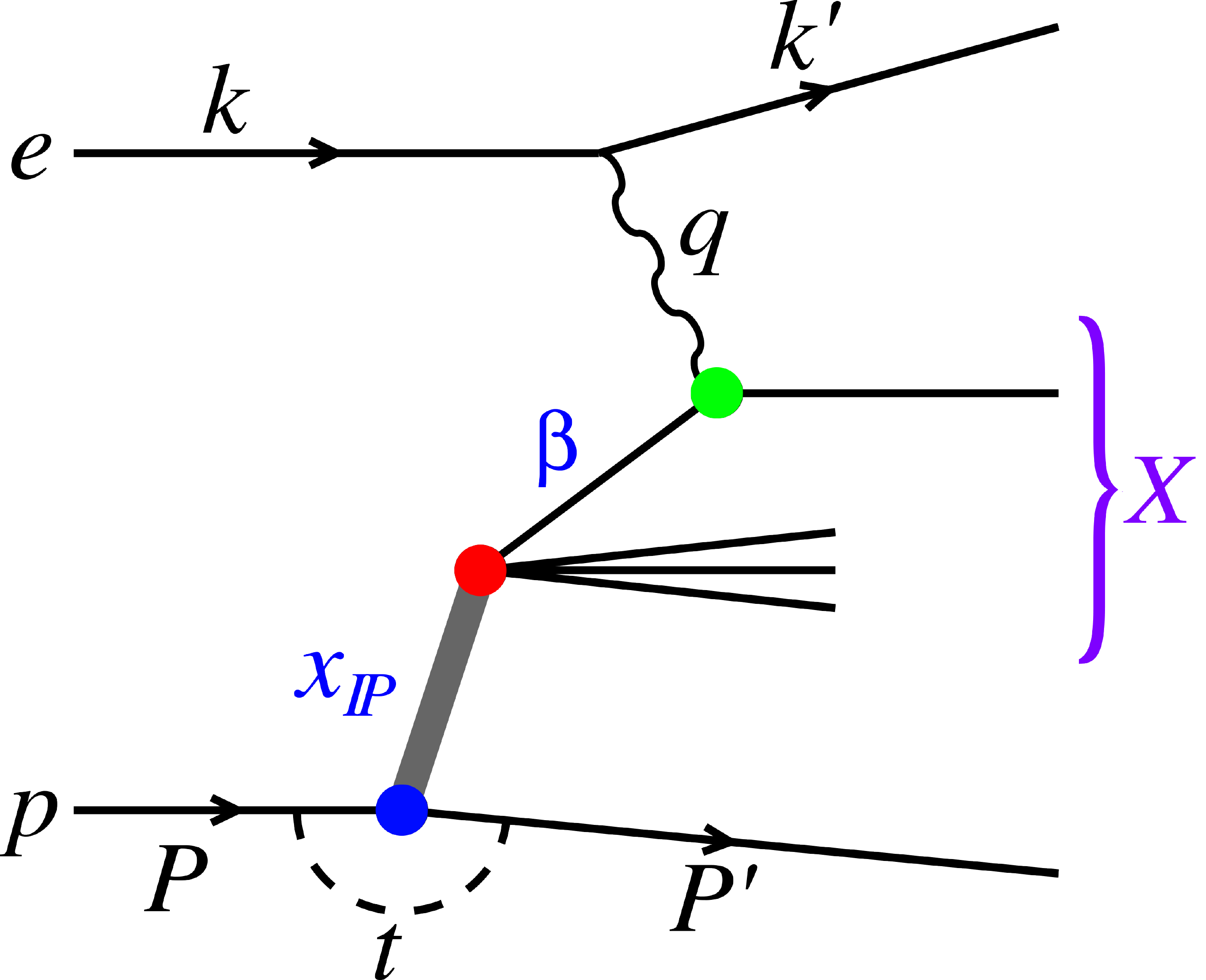}
\begin{center}
\caption{{\small Representative Feynman diagram for the neutral current diffractive DIS process $e p \to e p X$. } \label{fig:Feynman}}
\end{center}
\end{figure*}

The representative Feynman diagram for the neutral current diffractive DIS process $e p \to e p X$, proceeding via 
a virtual photon exchange, is depicted in Fig.~\ref{fig:Feynman}.
In the case of diffractive DIS, as illustrated in Fig.~\ref{fig:Feynman}, the additional variables are the squared four-momentum transferred 
$t=(P-P^\prime)^2$, where $P^\prime$ is the four-momentum of the outgoing proton, and the mass $M_{X}$ of the diffractive
final state, which is produced by diffractive dissociation of the exchanged virtual photon. This mass is much smaller than the invariant
photon-proton energy and should be considered as a further degree of freedom.
It is usually replaced by the light-cone momentum fraction of the diffractive exchange $\beta$,
\begin{eqnarray}\label{beta}
\beta = \frac{Q^2}{2 (P - P^\prime).q} = \frac{ Q^{2}}{M_{X}^{2} + Q^{2} - t }\,.
\end{eqnarray}

The $t$-integrated differential cross section for the diffractive process, $e p \to e p X$, is presented 
in the form of a diffractive reduced cross section $\sigma_{r}^{D(3)} (\beta, Q^2; x_{\pom})$ as
\begin{eqnarray}\label{dsigma}
\frac{d \sigma^{e p \to e p X}}{ d \beta dQ^{2} dx_{\pom}} = \frac{2 \pi \alpha^2}{\beta Q^4} \biggl[1 + (1-y)^2\biggr]\sigma_r^{D(3)}(\beta, Q^2; x_{\pom})\,,
\end{eqnarray}
where $x_{\pom} = \frac{(P-P^\prime).q}{P.q}$ refers to the longitudinal momentum fraction lost by the incoming proton, which 
is carried away by the diffractive exchange; and $t$ is the four-momentum transfer squared at the proton vertex.
Note that the longitudinal momentum fraction $\beta$ of the struck parton with respect to the colourless exchange can be also expressed as $\beta = x/x_{\pom}$.
The diffractive reduced cross section is given by
\begin{eqnarray}\label{sigma}
\sigma_r^{D(3)}(\beta, Q^2; x_{\pom}) = F_2^{D(3)}(\beta, Q^2; x_{\pom}) - {y^2 \over 1 + (1-y)^2}\ F_{L}^{D(3)}(\beta, Q^2; x_{\pom})\,,
\end{eqnarray}
where $F_2^{D(3)}$ and $F_{L}^{D(3)}$ are the diffractive structure functions.
It should be emphasized here that  for the $y$ not to close to unity, one can neglect the contribution from $F_{L}^{D(3)}$ and 
$\sigma_r^{D(3)}(\beta, Q^2; x_{\pom}) \approx F_2^{D(3)}(\beta, Q^2; x_{\pom})$ holds to very good approximation. Since our analysis is based on recent measurements of inclusive diffractive DIS at HERA for the reduced cross sections, we consider the contributions of both $F_2^{D(3)}$ and the longitudinal diffractive structure function $F_{L}^{D(3)}$.

\subsection{QCD factorization theorem}\label{sec:factorisation}

It has been shown that the diffractive DIS cross sections at HERA~\cite{Aktas:2006hy,Aktas:2007bv,Chekanov:2009aa} are 
well interpreted assuming the ``proton vertex factorization'' approach which
provides a good description of diffractive DIS data in terms of a resolved Pomeron ($\pom$)~\cite{Ingelman:1984ns,Donnachie:1987xh}.
Within the Regge phenomenology~\cite{Regge:1959mz}, the cross sections of diffractive processes at high energies are described by the exchange of so-called Regge trajectories.
The diffractive cross section is dominated by a trajectory usually called the Pomeron, 
while the subleading Reggeon ($\reg$) contribution is significant  only  for $x_{\pom} > 0.01$.
It has been shown that the QCD factorization theorem and the well-known DGLAP parton evolution equations can be applied to describe the
dependence of the cross section on $\beta$ and $Q^{2}$, while a Regge inspired approach is used to express the dependence on $x_{\pom}$
and $t$.

In the QCD factorization approach, the diffractive structure functions can be written as a convolution
of hard scattering coefficient functions with the diffractive PDFs,
\begin{eqnarray}\label{eq:factorisation}
F_{2/L}^{D(4)}(\beta, Q^2; x_{\pom}, t) = \sum_i \int_{\beta}^1 \frac{dz}{z} \, C_{2/L, i} \Big( \frac{\beta}{z} \Big)\, f_i^D (z, Q^2; x_{\pom}, t)\,,
\end{eqnarray}
where the sum runs over quarks and gluons.

Considering QCD factorization theorem, various hard scattering diffractive processes are calculable by means of 
diffractive PDFs, such as the diffractive jet production in DIS. The concept of QCD hard factorization of the diffractive PDFs as well as the validity of the assumption of QCD hard factorization have been theoretically predicted to hold in diffractive DIS processes~\cite{Collins:1997sr}.
We should mentioned here that the hard QCD factorization has been tested at HERA in various diffractive processes. In recent H1 analyses the validity
of the hard factorization has been successfully examined for open charm production in photoproduction and DIS with D$^\star$ mesons~\cite{Aktas:2006up,H1:2017bnb} 
and in diffractive production of dijets in DIS~\cite{Aktas:2007bv,Aaron:2011mp,Andreev:2014yra,Andreev:2015cwa}. These studies support the validity of QCD hard scattering factorization in diffractive DIS.

We should notice here that in DGLAP NLO QCD global fits, NLO contributions to the splitting functions governing the evolution of unpolarized nonsinglet and singlet combinations of quark densities are the same as in fully inclusive DIS. Hence, the diffractive parton densities satisfy the same (DGLAP) evolution equations as
the usual parton distributions in inclusive DIS~\cite{Berera:1995fj,Martin:2006td,Kunszt:1996pj}.
The Wilson coefficient functions $C_{2}$ and $C_{L}$ in Eq.~\eqref{eq:factorisation} are also the same as in inclusive DIS and calculable in perturbative QCD~\cite{Vermaseren:2005qc}. The diffractive PDFs $f_i^D (\beta, Q^2; x_{\pom},t)$ are universal and non-perturbative quantities, which can be obtained from the QCD fit to the inclusive diffractive data.
Note that diffractive PDFs can be defined in terms of matrix elements of quark and gluon operators;
the renormalization of divergencies at next-to-leading order is carried out similarly to the inclusive case and leads to the DGLAP evolution equations.

In {\tt GKG18-DPDFs} analysis, the proton vertex factorization~\cite{Ingelman:1984ns} is assumed, where the $x_{\pom}$ and $t$ dependencies
of the diffractive PDFs factorize from the dependencies on $\beta$ and $Q^2$. In this framework, the diffractive PDFs can be written as,
\begin{eqnarray}\label{eq:fD}
f_{i/p}^D(\beta, Q^2; x_{\pom}, t)= f_{{\pom}/p}(x_{\pom}, t) f_{i/{\pom}}(\beta, Q^2) + f_{{\reg}/p}(x_{\pom}, t) f_{i/{\reg}}^{\reg}(\beta, Q^2) \,,
\end{eqnarray}
where $f_{i/{\pom}}(\beta, Q^2)$ and $f_{i/{\reg}}^{\reg}(\beta, Q^2)$ are the partonic structures of Pomeron and Reggeon, respectively. 
The emission of Pomeron and Reggeon from the proton can be described by the flux-factors of $f_{{\pom}/p}(x_{\pom}, t)$ and $f_{{\reg}/p}(x_{\pom}, t)$.
The detail discussion on the parametrization of the diffractive PDFs in Eq.~\eqref{eq:fD} will be presented in a separate section.

\subsection{Heavy flavour contributions to the diffractive DIS structure function}\label{sec:heavy-flavour}

In this section, we discuss a general framework for the inclusion of heavy quark contributions to diffractive DIS structure functions.
The correct treatment of heavy quark flavours in an analysis of diffractive PDFs is essential for precision measurements at DIS colliders as well as for the LHC phenomenology.
As an example, the cross section for the $W$-boson production at the LHC depends crucially on precise knowledge of the charm 
quark distribution. A detailed discussion on the impact of the heavy quark mass treatments 
in the parton distributions as well as the determination of the their uncertainty due to uncertainty in the heavy quark masses can be found in Ref.~\cite{Ball:2011mu}.

Like to the case of inclusive DIS, the treatment of heavy flavours has an
important impact on the diffractive PDFs extracted from the global analysis of diffractive DIS, due to the heavy flavour contribution to the total structure
function at small values of $z$. Recall that there are various choices that can be used to consider the heavy quark contributions. 
These are the so-called variable flavour number scheme (VFNS), fixed flavour number scheme (FFNS) and general-mass variable-flavor-number scheme (GM-VFNS).

In the case of FFNS, $Q^2 \simeq m_c^2, m_b^2$, the massive quark may be regarded as being only produced in the final state and not as partons within the nucleon. Hence, the light up-, down- and strange-quarks are active partons and the number of flavours is fixed to $n_f=3$. However one can also consider charm or bottom quark as light quark at high scales. It has been shown that the accuracy of the FFNS becomes increasingly uncertain as $Q^2$ increases above 
the heavy quark mass threshold $m_H^2$~\cite{Martin:2009iq}.
In the zero-mass VFNS, the massive quarks behave like massless partons for $Q^2 \gg m_c^2, m_b^2$. The ZM-VFNS misses out ${\cal O} (m_H^2/Q^2)$ contributions completely in the perturbative expansion, and hence, this scheme is not accurate enough to be used in a QCD analysis. One can also see a
discontinuity in the parton distributions and total structure function at $Q^2 = m^2_H$ in ZM-VFNS~\cite{Martin:2009iq}.

The GM-VFNS is the appropriate scheme to interpolate between these two regions and could correct FFNS at low $Q^2$ and ZM-VFNS at high $Q^2 \to \infty$, and hence, could 
improve the smoothness of the transition region where the number of active flavours is changed by one~\cite{Martin:2009iq}. 
Therefore,  for a precise analysis of structure functions and other inclusive DIS or hadron colliders data, one can use the GM-VFNS,
which smoothly connects the two well-defined scheme of VFNS and FFNS~\cite{Martin:2009iq}. This scheme is that most commonly approach in variety of global fits.
In H1-DPDFs-2006~\cite{Aktas:2006hy} and ZEUS-DPDFs-2010~\cite{Chekanov:2009aa} diffractive PDFs analyses, the heavy quark structure functions have been computed using the FFNS and general-mass variable-flavor-number scheme of Thorne and Roberts (TR GM-VFNS), respectively.
Our approach is based on the TR GM-VFNS~\cite{Harland-Lang:2014zoa,Thorne:1997ga,Thorne:2006qt} which extrapolates smoothly from the FFNS at low $Q^2$ to the ZM-VFNS
at high $Q^2$ and produces a good description of the effect of heavy quarks
on structure functions over the whole range of $Q^2$.

In our analysis, we follow the MMHT14 PDFs analysis and adopt their default values for the heavy quark masses as $m_c = 1.40$ and $m_b = 4.75$ GeV~\cite{Harland-Lang:2015qea}. In Ref.~\cite{Harland-Lang:2015qea}, the variation in the MMHT14 PDFs when the heavy
quark masses $m_c$ and $m_b$ were varied away from their default values of $m_c = 1.40$ and $m_b = 4.75$ GeV has been investigated.
The dependence of the MMHT14 PDFs and the quality of the comparison to analyzed data, under variations of the heavy quark masses away from their default values has been studied. 
It has been shown that the effects of varying $m_c$ and $m_b$ in the predictions of cross sections for standard processes at the LHC are small and the uncertainties on PDFs due
to the variation of quark masses are not hugely important~\cite{Harland-Lang:2015qea}.

%
\section{ The method of diffractive PDFs global QCD analysis }\label{sec:statistical}
%

In the following, we present the method of {\tt GKG18-DPDFs} global QCD analysis. This section also includes our parametrizations of the diffractive PDFs, the detailed discussion of the description of different data sets included in our global fit, and the method of minimization and uncertainties of our resulting diffractive PDFs.

\subsection{ {\tt GKG18-DPDFs} parametrizations of the diffractive PDFs }\label{sec:Parametrizations}

As we already mentioned, the scale dependence of the distributions $f_{i=q,g}(\beta, Q^2)$ of the quarks and gluons can be obtained by the DGLAP evolution equations, provided the diffractive PDFs are parametrized as functions of $\beta$ at some starting scale $Q^2_0$.   
In our analysis, the diffractive PDFs are modelled at the starting scale $Q^2_0 = 1.8 \, {\rm GeV}^2$
(below the charm threshold) in terms of quark $z  f_q (z, Q_0^2)$, and gluon $z  f_g (z, Q_0^2)$ distributions. Here, $z$ is the longitudinal momentum fraction of the struck parton, which enters the hard subprocess, with respect to the diffractive exchange.
Considering the lowest-order quark-parton model process, we have $z = \beta$, while the inclusion of higher-order processes leads to $0 < \beta < z$.
For the quark distributions we assume that all light-quarks and their antiquarks distributions are equal, $f_u = f_d = f_s = f_{\bar u} = f_{\bar d} = f_{\bar s}$. The heavy quark distributions $f_{q(=c,b)}$ are generated dynamically at the scale $Q^2$ > $m_{c,b}^2$ above the corresponding mass threshold in the TR GM-VFN scheme.

Due to the significantly smaller amount of data for inclusive diffractive DIS data than for the total DIS cross section, we adopt a slightly less
flexible, more economical functional form to parametrize the nonperturbative diffractive PDFs at the initial scale $Q^2_0 = 1.8 \, {\rm GeV}^2$.
Our standard parametrizations for the quarks and gluon diffractive PDFs are as follows:
\begin{eqnarray}
&& \label{DPDFsq-Q0} z  f_q (z, Q_0^2) = \alpha_q \, z^{\beta_q}(1 - z)^{\gamma_q} (1 + \eta_q \sqrt{z}) \,,   \\
&& \label{DPDFsg-Q0} z  f_g (z, Q_0^2) = \alpha_g \, z^{\beta_g}(1 - z)^{\gamma_g} (1 + \eta_g \sqrt{z}) \,.
\end{eqnarray}

An additional factor of $e^{-\frac{0.001}{1 - z}}$ is included to ensure that the distributions vanish for $z \to 1$. Therefore, the parameters $\gamma_q$ and
$\gamma_g$ have the freedom to take negative as well as positive values in the fit.
We have tested that Eqs.~\eqref{DPDFsq-Q0} and \eqref{DPDFsg-Q0} nevertheless yield a very satisfactory description of the analyzed diffractive DIS data. We 
found that the two parameters $\eta_q$ and $\eta_g$ had to be fixed to zero since the data do not constrain 
them well enough. These simple functional forms with significantly fewer parameters have the additional benefit of greatly facilitating the fitting procedure.

The $x_{\pom}$ dependence of diffractive PDFs $f_{i/p}^D(z, Q^2; x_{\pom}, t)$ in Eq.~\eqref{eq:fD} is parametrized by the Pomeron and Reggeon flux factors
\begin{equation}\label{flux}
f_{\pom, \reg}(x_{\pom}, t) = A_{\pom, \reg} \, \frac{e^{B_{\pom, \reg} \, t}}{x_{\pom}^{2 \alpha_{\pom, \reg}(t) - 1}}\,,
\end{equation}
where the trajectories are assumed to be linear, $\alpha_{\pom, \reg}(t) = \alpha_{\pom, \reg}(0) + \alpha_{\pom, \reg}^\prime t$.
The Pomeron and Reggeon intercepts, $\alpha_{\pom}(0)$ and $\alpha_{\reg}(0)$, and the normalization of the Reggeon term, $A_{\reg}$, are free parameters and should be extracted from the fit to data. Note that the value of the normalization parameter $A_{\pom}$ is absorbed in $\alpha_q$ and $\alpha_g$.

The Reggeon parton densities $f_{i/{\reg}}^{\reg}(z, Q^2)$ presented in Eq.~\eqref{eq:fD} are obtained from the GRV parametrization derived from a fit to pion structure function data~\cite{Gluck:1991ng}. The values of the parameters, which are fixed in {\tt GKG18-DPDFs} fit, are the following:
\begin{eqnarray}
&& \alpha^\prime_{\pom} = 0.0 \,,  \nonumber \\
&& \alpha^\prime_{\reg} = 0.90 \, {\text{GeV}}^{-2} \,, \nonumber \\
&& B_{\pom} = 7.0 \, {\text{GeV}}^{-2} \,, \nonumber \\
&& B_{\reg} = 2.0 \, {\text{GeV}}^{-2} \,. \nonumber
\end{eqnarray}
These values are taken from the following experimental measurements~\cite{Chekanov:2008fh,Donnachie:1992ny},
\begin{eqnarray}
&& \alpha^\prime_{\pom} = -0.01 \pm 0.06 \, (\text{stat.})^{+0.04}_{-0.08} \, (\text{syst.}) \pm 0.04 \, (\text{model}) \, {\text{GeV}}^{-2} \,,  \nonumber \\  
&& \alpha^\prime_{\reg} = 0.90 \pm 0.10  \, {\text{GeV}}^{-2}\,, \nonumber \\
&& B_{\pom}       = 7.1 \pm 0.7 \, (\text{stat.})^{+1.4}_{-0.7} \, (\text{syst.}) \, {\text{GeV}}^{-2}\,, \nonumber \\
&& B_{\reg}       = 2.0 \pm 2.0 \, {\text{GeV}}^{-2}\,. \nonumber
\end{eqnarray}
In total, 9 free parameters are left in {\tt GKG18-DPDFs} QCD analysis, which are $\alpha_q$, $\beta_q$, $\gamma_q$, $\alpha_g$, $\beta_g$, $\gamma_g$, $\alpha_{\pom}(0)$, $\alpha_{\reg}(0)$, and $A_{\reg}$.

\subsection{ Diffractive DIS data sets used in {\tt GKG18-DPDFs} fits }\label{sec:data}

In this section, we present the new experimental data and their treatment in {\tt GKG18-DPDFs} diffractive PDFs analysis.
After reviewing the analyzed data sets, which include the recent H1 and ZEUS combined data, we discuss each of the new data sets in turn.
We finally review the way in which the total diffractive DIS data sets are constructed and, in particular, which data and which cuts are included.

A list of all diffractive DIS data points used in {\tt GKG18-DPDFs} global analysis is presented in Tables~\ref{tab:DDISdata-FitA} and \ref{tab:DDISdata-FitB}. These tables 
correspond to our two different scenarios for including inclusive diffractive DIS data in {\tt GKG18-DPDFs} global analyses, namely {\tt Fit A} and {\tt Fit B}.

For each data set presented in these tables, we have provided the corresponding references, the kinematical coverage of $\beta$, $x_{\pom}$, and $Q^2$ and the number of data points.
We strive to include as much of the available diffractive DIS experimental data as possible in our diffractive PDF analysis. However, some cuts have to be applied in order to ensure that only proper data are included in the analysis. 

The first data set we have used in our QCD analysis is the inclusive diffractive DIS data from H1-LRG-11, which
were taken with the H1 detector in the years 2006 and 2007. These data 
correspond to three different center-of-mass energies of $\sqrt{s} = 225$, $252$ and $319$ GeV~\cite{Aaron:2012zz}. In this measurement, the reduced cross sections have been measured in the range of photon virtualities $4.0 \leq Q^2 \leq 44.0 \, {\text{GeV}}^2$ and of the longitudinal momentum fraction of the diffractive exchange
$5 \times 10^{-4} \leq x_{\pom} \leq 3 \times 10^{-3}$.

In addition to the H1-LRG-11 data set, we have used for the first time the H1-LRG-12 data, where
the diffractive process $e p \to e X Y$ with $M_Y < 1.6 \, {\text{GeV}}$ and $|t| < 1 \, {\text{GeV}}^2$ has been studied with the H1 experiment at HERA~\cite{Aaron:2012ad}. This high statistics measurement covering the data taking periods 1999-2000 and 2004-2007, has been combined with previously published results~\cite{Aktas:2006hy} and covers the range of $3.5 < Q^2  < 1600 \, {\text{GeV}}^2$, $0.0017 \leq \beta \leq 0.8$, and $0.0003 \leq x_{\pom} \leq 0.03$.

Finally, for the first time, we have used the recent and up-to-date H1/ZEUS combined data set for the reduced diffractive cross sections, $\sigma_r^{D(3)} (e p \to e p X)$~\cite{Aaron:2012hua}. This measurement used samples
of diffractive DIS $ep$ scattering data at a centre-of-mass energy of $\sqrt{s} = 318 \, {\text{GeV}}$ and combined the previous 
the H1 FPS HERA I~\cite{Aktas:2006hx}, H1 FPS HERA II~\cite{Aaron:2010aa}, ZEUS LPS 1~\cite{Chekanov:2004hy} and ZEUS LPS 2~\cite{Chekanov:2008fh} data sets. This combined data cover the photon virtuality range of  $2.5 < Q^2 < 200 \, {\text{GeV}}^2$,
$3.5 \times 10^{-4} < x_{\pom} < 0.09$ in proton fractional momentum loss, $0.09 < |t| < 0.55 \, {\text{GeV}}^2$ in squared four-momentum transfer
at the proton vertex, and $1.8 \times 10^{-3} < \beta < 0.816$.

While all H1-LRG data are given for the range $|t|<1 \, {\text{GeV}}^2$, the combined H1/ZEUS diffractive DIS, 
which is based upon proton-tagged samples, are restricted to the range $0.09<|t|<0.55 \, {\text{GeV}}^2$, so one needs to use a global normalization factor between those two measurement regions. Assuming an exponential $t$ dependence of the inclusive diffractive cross section, the extrapolation from $0.09<|t|<0.55$ GeV$^2$ to $|t|<1$ GeV$^2$ has been done using the H1 value of exponential slope parameter $b \simeq 6 \, {\rm GeV}^{-2}$~\cite{Aaron:2010aa,Aaron:2012hua}.
The slope parameter can be extracted from fits to the reduced cross section $x_{\pom} \sigma_r^{D(4)}$. With the above choice of constant slope parameter, a good description of the data
over the full $x_{\pom}$, $Q^2$ and $\beta$ range is obtained~\cite{Aaron:2010aa,Aktas:2006hx}.

In addition to the extrapolation discussed above, distinct methods have been employed by the H1 and ZEUS experiments, and hence, cross sections are not always given with the corrections for proton dissociation background. The different contributions from proton dissociation in the different data sets should be considered by application of different global factors. Proton dissociation is simulated using an approximate $\frac{d \sigma}{d M^{2}_{Y}} \propto \frac{1}{M^{2}_{Y}}$ dependence~\cite{Aktas:2006hy,Monfared:2011xf}. The combined H1/ZEUS diffractive DIS are corrected by a global factor of 1.21 to account for such contributions.

It should be noted that the two data normalization factors, which we described above, bring a small systematic uncertainty to the 
fitted data. However, since the extrapolation in $|t|$ is rather modest and the slope parameter $b$ is experimentally 
determined with better than $10$\% accuracy~\cite{Aktas:2006hx} and the factor due to proton dissociation is rather well-constrained
phenomenologically and experimentally, this uncertainty is at the level of a few percent. Hence, it can be safely neglected compared to the
total experimental error of the H1/ZEUS combined data~\cite{Aaron:2012hua}.  

As in the case of H1-DPDFs-2006~\cite{Aktas:2006hy} and ZEUS-DPDFs-2010~\cite{Chekanov:2009aa} fits, we apply a cut on $M_X$, $\beta$ and ${Q}^2$. 
To determine our diffractive PDFs, we apply $\beta \leq 0.80$ over the data sets. 
The data with $M_X > 2 \, {\text{GeV}}$ are included in the fit and the data with ${Q}^2 < {Q}^2_{\rm min}$ are excluded to avoid regions,
 which are most likely to be influenced by higher twist (HT) corrections or other problems with the chosen theoretical framework.

%
\begin{table*}[htb]
\caption{\small List of all diffractive DIS data points used in {\tt Fit A} global analysis. For each
dataset we have provided the references, the kinematical coverage of $\beta$, $x_{\pom}$, and $Q^2$ and the number of data points. The details of kinematic cuts explained in the text.} \label{tab:DDISdata-FitA}
\begin{tabular}{l c c c c c c}
Experiment & Observable & [$\beta^{\text{min}}, \beta^{\text{max}}$] & [$x_{\pom}^{\text{min}}, x_{\pom}^{\text{max}}$]  & $Q^2\,[{\text{GeV}}^2]$  & \# of points
\tabularnewline
\hline\hline
H1-LRG-11 $\sqrt{s} = 225$~\cite{Aaron:2012zz} & $\sigma_r^{D(3)}$ & [$0.089$--$0.699$]   & [$5.0 \times 10^{-4}$ -- $3.0 \times 10^{-3}$] & 11.5--44 & \textbf{13}   \\
H1-LRG-11 $\sqrt{s} = 252$~\cite{Aaron:2012zz} & $\sigma_r^{D(3)}$ & [$0.089$--$0.699$]   & [$5.0 \times 10^{-4}$ -- $3.0 \times 10^{-3}$] & 11.5--44 & \textbf{12}   \\
H1-LRG-11 $\sqrt{s} = 319$~\cite{Aaron:2012zz} & $\sigma_r^{D(3)}$ & [$0.089$--$0.699$]   & [$5.0 \times 10^{-4}$ -- $3.0 \times 10^{-3}$] & 11.5--44 & \textbf{12}   \\	
H1-LRG-12~\cite{Aaron:2012ad} & $\sigma_r^{D(3)}$ & [$0.0067$--$0.80$]   & [$3.0 \times 10^{-4}$ -- $3.0 \times 10^{-2}$] & 12--1600 & \textbf{165}   \\	 \hline	
H1/ZEUS combined~\cite{Aaron:2012hua} & $\sigma_r^{D(3)}$  &   [$0.0056$--$0.562$] & [$9.0 \times 10^{-4}$ -- $9.0 \times 10^{-2}$] & 15.3--200 & \textbf{96}  \\			
\hline \hline
\multicolumn{1}{c}{\textbf{Total data}} ~~  & ~~ & ~~ & ~~ & ~~ \textbf{298}  \\  \hline
\end{tabular}
\end{table*}
%
%

%
\begin{table*}[htb]
\caption{\small List of all diffractive DIS data points used in {\tt Fit B} global analysis. See the caption of Table.~\ref{tab:DDISdata-FitA} for more details.} \label{tab:DDISdata-FitB}
\begin{tabular}{l c c c c c c}
Experiment & Observable & [$\beta^{\text{min}}, \beta^{\text{max}}$] & [$x_{\pom}^{\text{min}}, x_{\pom}^{\text{max}}$]  & $Q^2\,[{\text{GeV}}^2]$  & \# of points
\tabularnewline
\hline\hline
H1-LRG-11 $\sqrt{s} = 225$~\cite{Aaron:2012zz} & $\sigma_r^{D(3)}$ & [$0.089$--$0.699$]   & [$5.0 \times 10^{-4}$ -- $3.0 \times 10^{-3}$] & 11.5--44 & \textbf{13}   \\
H1-LRG-11 $\sqrt{s} = 252$~\cite{Aaron:2012zz} & $\sigma_r^{D(3)}$ & [$0.089$--$0.699$]   & [$5.0 \times 10^{-4}$ -- $3.0 \times 10^{-3}$] & 11.5--44 & \textbf{12}   \\
H1-LRG-11 $\sqrt{s} = 319$~\cite{Aaron:2012zz} & $\sigma_r^{D(3)}$ & [$0.089$--$0.699$]   & [$5.0 \times 10^{-4}$ -- $3.0 \times 10^{-3}$] & 11.5--44 & \textbf{12}   \\	
H1-LRG-12~\cite{Aaron:2012ad} & $\sigma_r^{D(3)}$ & [$0.0067$--$0.80$]   & [$3.0 \times 10^{-4}$ -- $3.0 \times 10^{-2}$] & 12--1600 & \textbf{165}   \\ 	 \hline 	
H1/ZEUS combined~\cite{Aaron:2012hua} & $\sigma_r^{D(3)}$  &   [$0.0056$--$0.562$] & [$9.0 \times 10^{-4}$ -- $9.0 \times 10^{-2}$] & 26.5--200 & \textbf{70}  \\			
\hline \hline
\multicolumn{1}{c}{\textbf{Total data}} ~~  & ~~ & ~~ & ~~ & ~~ \textbf{272}  \\  \hline
\end{tabular}
\end{table*}
%
%

To ensure the validity of the DGLAP evolution equations, we have to impose certain cuts on the above mentioned data sets.
In order to finalize the cut on $Q^2$, the sensitivity of $\chi^2$ to variations in $Q^2 > Q_{\text{min}}^2$ is investigated for data used in the analysis.
Considering these $\chi^2$ scans, our full diffractive PDFs fits are repeated for each different $Q^2 > Q_{\text{min}}^2$ cut.

\begin{figure*}[htb]
\vspace{0.50cm}
\includegraphics[clip,width=0.70\textwidth]{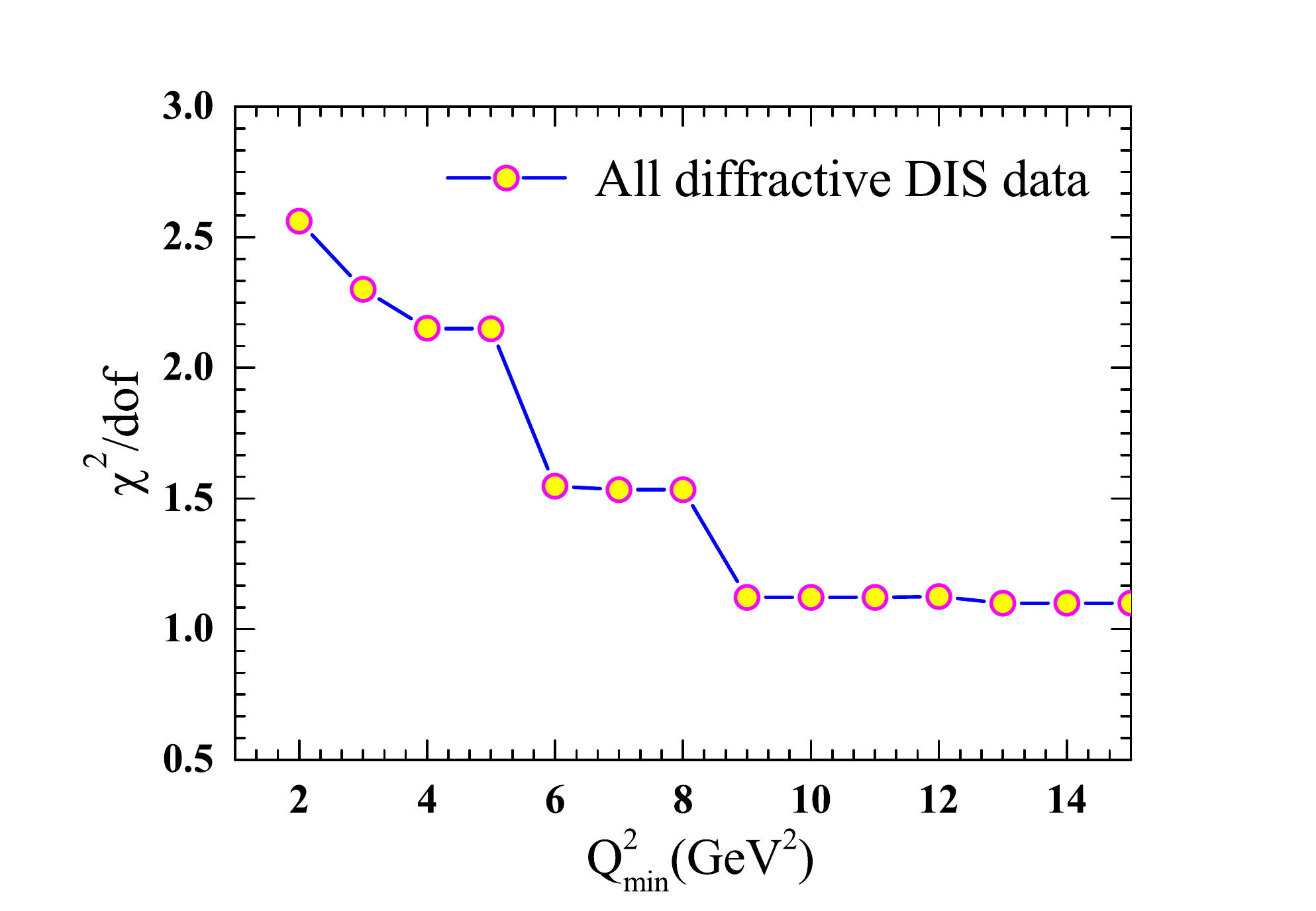}
\begin{center}
\caption{{\small Dependence of $\chi^2/{\text{dof}}$ on the minimum cut value of $Q_{\text{min}}^2$ for all data sets used in the analysis.  \label{fig:chi2_total}}}
\end{center}
\end{figure*}

\begin{figure*}[htb]
\vspace{0.50cm}
\includegraphics[clip,width=0.70\textwidth]{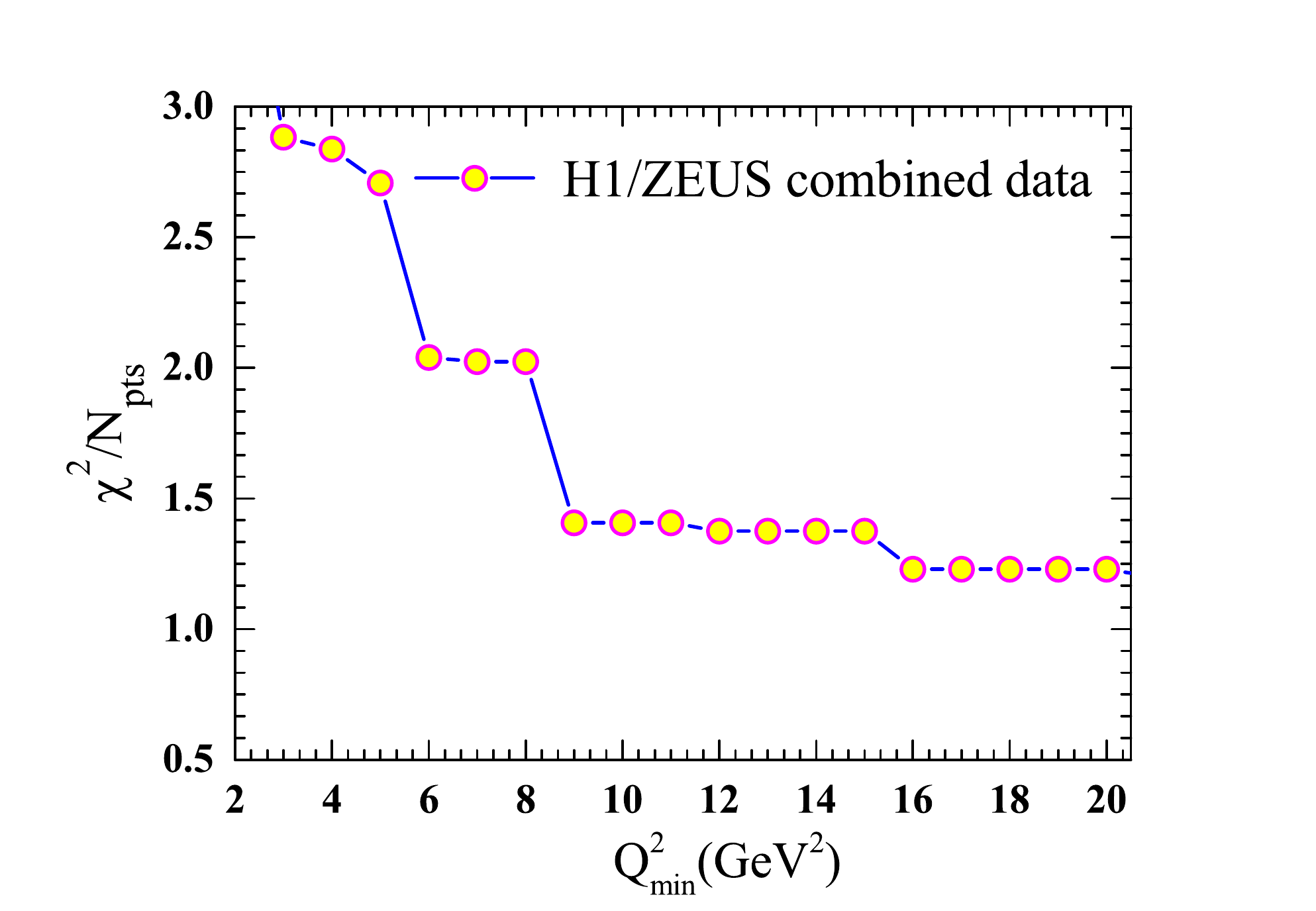}
\includegraphics[clip,width=0.70\textwidth]{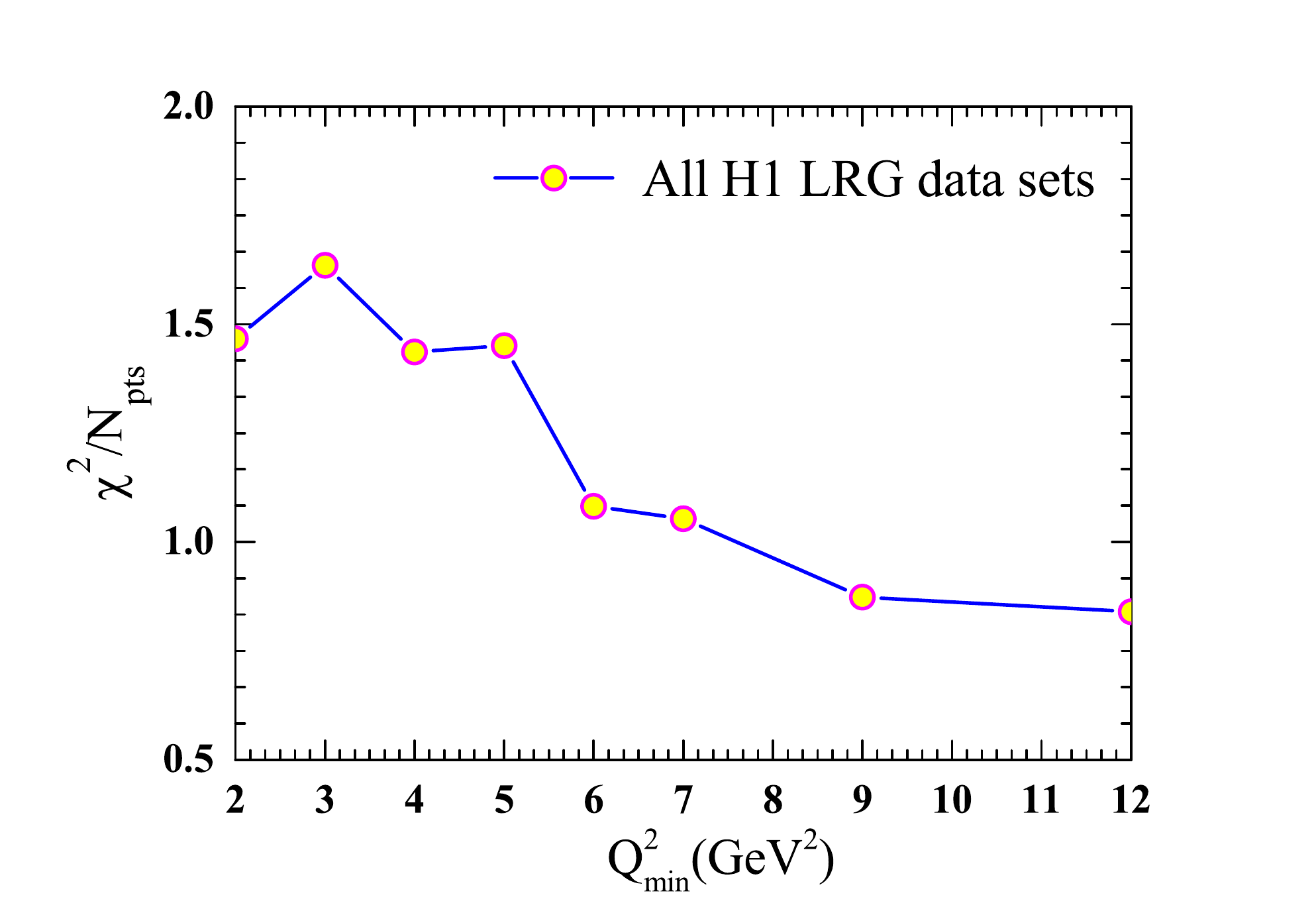}
\begin{center}
\caption{{\small Dependence of $\chi^2/N_{\text{pts}}$ on the minimum cut value of $Q_{\text{min}}^2$ for H1/ZEUS combined data (up) and all H1 LRG data sets (down). \label{fig:chi2_Combin}}}
\end{center}
\end{figure*}

In Fig.~\ref{fig:chi2_total}, the dependence of $\chi^2$ per number of degrees of freedom, $\chi^2/{\text{dof}}$, 
on the minimum cut value of $Q^2$ has been presented as a function of $Q_{\rm min}^2$ for all inclusive diffractive DIS data sets used in {\tt GKG18-DPDFs} (see Table.~\ref{tab:DDISdata-FitA}).
The $Q_{\text{min}}^2$ dependence is reflected from this plot and no further improvement on $\chi^2/{\text{dof}}$ can be expected for larger value of  $Q^2 > Q_{\text{min}}^2 = 9 \, \text{GeV}^2$. Therefore, the lowest $Q^2$ data are omitted from our QCD fit and $Q_{\text{min}}^2 \geq 9 \, \text{GeV}^2$ is applied to the diffractive DIS data sets. We refer this fit to {\tt Fit A}.

However, this choice is somewhat different from the cut used in Refs.~\cite{Aktas:2006hy,Chekanov:2009aa} ($ Q_{\text{min}}^2 > 8.5 \, \text{GeV}^2$).
Since this issue can be related to the possible tension between the H1-LRG-11 and H1-LRG-12 data sets with the  H1/ZEUS combined data in  low-$Q^2$ bins, some further investigations are required. To resolve this issue,  we also present similar plots for the H1/ZEUS combined data as well as for all H1 LRG data sets.
As one can see from the upper panel of Fig.~\ref{fig:chi2_Combin}, an improvement on $\chi^2$ per number of data points, $\chi^2/N_{\text{pts}}$, can be expected for larger value of  $Q^2 > Q_{\text{min}}^2 = 16 \, \text{GeV}^2$ for the H1/ZEUS combined data. 
In Fig.~\ref{fig:chi2_Combin}, we have also shown the same plot for the H1 LRG data sets. This plot clearly shows that the appropriate choice for the case of H1 LRG data sets is $Q_{\text{min}}^2 > 9 \, \text{GeV}^2$. This fact indicates that the choice of $Q_{\text{min}}^2 > 9 \, \text{GeV}^2$ is still suitable for all data sets excluding the H1/ZEUS combined data.
Hence, we repeated our analysis by applying an additional cuts on $Q_{\text{min}}^2 \geq 16 \, \text{GeV}^2$ for the H1/ZEUS combined and keeping $Q_{\text{min}}^2 \geq 9 \, \text{GeV}^2$ for other H1-LRG-11 and H1-LRG-12 data sets. We refer this fit to {\tt Fit B}. The number of data points after all cuts for both {\tt Fit A} and {\tt Fit B} are summarized in Tables~\ref{tab:DDISdata-FitA} and \ref{tab:DDISdata-FitB}, respectively. Note that since higher twist (HT) can be potentially large is inclusive diffractive DIS~\cite{Motyka:2012ty}, the choice of larger $Q_{\text{min}}^2$ also tends to reduce the HT influence.

\subsection{The method of minimization and diffractive PDF uncertainties}\label{sec:minimization}

As we already discussed, {\tt GKG18-DPDFs} diffractive PDFs are provided at NLO in perturbative QCD and the data used in our fits cover a wide range of  $\beta$, $x_{\pom}$ and $Q^2$ kinematics.
In order to achieve an accurate theoretical descriptions of both the diffractive PDFs evolution and the hard scattering cross sections, a well-tested software package is necessary. In {\tt GKG18-DPDFs} analysis, we have used the {\tt xFitter}~\cite{Alekhin:2014irh} which is a standard package for performing the global QCD analysis of PDFs. Fortunately, the necessary tools for making theoretical predictions of the diffractive DIS observables have been implemented in the {\tt xFitter}, allowing one to perform also a global analysis of diffractive PDFs.
For the minimization, $\chi^2$ definition and treatment of experimental uncertainties, we used the methodology implemented in {\tt xFitter} to determine the unknown parameters of diffractive PDFs.

The QCD fit strategy follows closely the one adopted for the determination of the PDFs in the HERAPDF methodology~\cite{Aaron:2009aa,Aaron:2012qi}. 
The QCD predictions for the inclusive diffractive cross section are obtained by solving the DGLAP evolution equations at NLO. As we mentioned, the heavy quark coefficient functions are calculated in the TR GM-VFNS~\cite{Thorne:1997ga,Harland-Lang:2014zoa} and the heavy quark masses for charm and beauty are chosen as $m_c = 1.40$ GeV and $m_b = 4.75$ GeV~\cite{Harland-Lang:2015qea}. The strong
coupling constant is fixed to the $\alpha_s(M_Z^2) = 0.1176$~\cite{Patrignani:2016xqp} which is close to the best-fit value of NNLO MMHT2014 global PDF analysis, $\alpha_s(M_Z^2) = 0.1172 + \pm 0.0013$~\cite{Harland-Lang:2015nxa}. The $\chi^2$ function is minimized using the CERN {\tt MINUIT} package~\cite{James:1975dr}. The form of the $\chi^2$ minimized during our QCD fits is expressed as follows~\cite{Aaron:2012qi},

\begin{eqnarray}\label{chi2}
\chi^2 (\{\xi_k\}) && = \sum_{i} \frac{\big[ \mu_i - T_i(\{\xi_k\}) (1-\sum_{j} \gamma_j^i b_j) \big]^{2}  }{\delta^2_{i, unc} T_i^2(\{\xi_k\})+ \delta^2_{i,stat} \mu_i T_i \big( 1- \sum_{j} \gamma^i_j b_k\big) }  \\ \nonumber 
&& + \sum_{j} b_j^2 + \sum_{i} \ln \frac{\delta^2_{i,unc} T_i^2(\{\xi_k\}) + \delta^2_{i,stat} \mu_i T_i  (\{\xi_k\}) } { \delta^2_{i,unc} \mu_i^2 + \delta^2_{i,stat} \mu^2_i } \,,
\end{eqnarray}
where $\mu_i$ is the the measured value of inclusive diffractive cross section at point $i$, and $T_i $ is the corresponding theoretical predictions. 
The parameters $\delta_{i,stat}$, $\delta_{i, unc}$, and $\gamma_j^i$ are the relative statistical, uncorrelated systematic, and correlated systematic uncertainties.
The nuisance parameters $b_j$ are associated to the correlated systematics which are determined simultaneously with the unknown parameters $\{\xi_k\}$ of our functional forms of Eq.~\eqref{DPDFsq-Q0} and \eqref{DPDFsg-Q0}. We minimize the above $\chi^2$ value with the $k=9$ unknown fit parameters $\{\xi_k\}$ of our diffractive PDFs.

Table~\ref{tab:chi2} contains the final results of $\chi^2/N_{\text{pts}}$ for our global fits. For each data set, the value of $\chi^2/N_{\text{pts}}$ has been presented for both {\tt Fit A} and {\tt Fit B}. In the last row of the table, the values of $\chi^2/{\text{dof}}$ have also been presented as well. These table illustrates the quality of our QCD fits to inclusive diffractive cross section at NLO accuracy in terms of the individual $\chi^2$ values obtained for each experiment. For {\tt Fit A} and {\tt Fit B}, we obtain $\chi^2$ of 322 and 280 with the total 289 and 263 data points, respectively. As one can see from this Table, a $Q_{\text{min}}^2 \geq 16 \, \text{GeV}^2$ cut on the H1/ZEUS combined data set significantly reduces the $\chi^2/N_{\text{pts}}$ from 128/96 to 85/70. Note also that the values of $\chi^2/N_{\text{pts}}$ for H1-LRG-11 data sets
at $\sqrt{s} = 225$ and 252 GeV do not change from {\tt Fit A} to {\tt Fit B} and just a very small reduction is observed for the H1-LRG-11 $(\sqrt{s} = 319~{\rm GeV})$ and H1-LRG-12 data sets.
In conclusion, the quality of {\tt Fit B} is slightly better than that of {\tt Fit A}, indicating a better description of the inclusive diffractive DIS data. A substantial part of the improvement in the description is driven by the H1/ZEUS combined data.

%
\begin{table*}[htb]
\caption{ \small The values of $\chi^2/N_{\text{pts}}$ for the data sets included in the global fits. } \label{tab:chi2}
\begin{tabular}{l c c c }
\hline \hline
& {\tt Fit A}             & {\tt Fit B}  \\ \hline
Experiment & $\chi^2/N_{\text{pts}}$  & $\chi^2/N_{\text{pts}}$ 
\tabularnewline
\hline
H1-LRG-11 $\sqrt{s} = 225$~GeV~\cite{Aaron:2012zz} & 11/13 &  12/13  \\	
H1-LRG-11 $\sqrt{s} = 252$~GeV~\cite{Aaron:2012zz} & 20/12 &  21/12  \\		
H1-LRG-11 $\sqrt{s} = 319$~GeV~\cite{Aaron:2012zz} & 6.5/12 &  6.2/12  \\			
H1-LRG-12~\cite{Aaron:2012ad}        & 135/165 &  138/165  \\	
H1/ZEUS combined~\cite{Aaron:2012hua} & 128/96 &  85/70  \\	 \hline
Correlated $\chi^2$  & 10  & 11      \\  \hline 
Log penalty $\chi^2$  & +11  & +6.9  \\ 	\hline \hline
\multicolumn{1}{c}{~\textbf{$\chi^2/{\text {dof}}$}~}  &    $~322/289=1.11~$ &    $~280/263=1.06~$  \\  \hline
\end{tabular}
\end{table*}
%
%

In order to obtain the uncertainties on the diffractive PDFs, we use the {\tt xFitter} framework, which includes both the experimental statistical and systematic errors on the data points and their correlations in the definition of the $\chi^2$ function. The uncertainties on the diffractive PDFs as well as the corresponding observables throughout our analysis are computed using the standard ``Hessian'' error propagation~\cite{Pumplin:2001ct,Nadolsky:2008zw,Martin:2009iq}.

%
\section{Results and discussions}\label{sec:results}
%

Key results of the current NLO diffractive PDFs fit compared to all previous analyses are the inclusion of
all new and up-to-date experimental diffractive DIS data, in particular, the H1/ZEUS combined data set~\cite{Aaron:2012hua}, 
and the error analysis of the extracted diffractive PDFs. Since these new data sets may have the potential to provide more information on the extracted diffractive PDFs, it is important to precisely study their impact on the diffractive PDFs as well as on their uncertainty bands.
The second significant addition is the first determination of the diffractive PDFs in the framework of {\tt xFitter}~\cite{Alekhin:2014irh}.

The diffractive PDFs in our fits are parameterized at the input scale $ Q_0^2 = 1.8 \, {\rm GeV}^2 $ according to Eqs.~\eqref{DPDFsq-Q0} and \eqref{DPDFsg-Q0}, which 
provide considerable flexibility. As we mentioned, the available diffractive DIS experimental data are not sufficient enough to constrain all parameters of such a flexible parameterization. However, due to more precise data from H1/ZEUS combined experiments, an enhanced flexibility is maybe allowed for the quark and gluon parameterizations compared to the {\tt H1-2006} and {\tt ZEUS-2010} fits. We investigated Eqs.~\eqref{DPDFsq-Q0} and \eqref{DPDFsg-Q0} in our analysis and found that relaxing $\eta_g$ and $\eta_q$ does not cause significant changes to the fit results. Therefore, in our {\tt Fit A} and {\tt Fit B} QCD analyses, we set these parameters to zero.
The details of the fits are summarized in Table~\ref{tab-dpdf}, which shows our best fit values of the free parameters.
In this table, the values of the fixed parameters of $\alpha_s(M_Z^2)$, $m_c$ and $m_b$ for our {\tt Fit A} and {\tt Fit B} QCD analyses are also listed.  

%
%
\begin{table}[ht]
\begin{center}
\caption{\small Parameters obtained with the different fits at the initial scale $Q_{0}^{2} = 1.8 \, {\text{GeV}}^2$ and their experimental uncertainties. Values marked with (*) are fixed in the fit. }
\begin{tabular}{ c | c | c }
\hline \hline
Parameters	  & {\tt Fit A}       & {\tt Fit B}        \\  \hline \hline
$\alpha_g$    & $1.01 \pm 0.16$ & $0.80 \pm 0.13$  \\ 
$\beta_g$     & $0.213 \pm 0.065$ & $0.166 \pm 0.072$  \\ 
$\gamma_g$    & $0.29 \pm 0.16$ & $0.11 \pm 0.15$   \\ 
$\eta_g$      & $0.0^*$           & $0.0^*$            \\   
$\alpha_q$    & $0.303 \pm 0.022$ & $0.283 \pm 0.021$  \\ 
$\beta_q$     & $1.464 \pm 0.069$ &  $1.514 \pm 0.075$  \\ 
$\gamma_q$    & $0.512 \pm 0.035$ & $0.512 \pm 0.036$  \\ 
$\eta_q$      & $0.0^*$           & $0.0^*$            \\     
$\alpha_{\pom}(0)$  & $1.0938 \pm 0.0032$  & $1.0988 \pm 0.0037$    \\
$\alpha_{\reg}(0)$  & $0.318 \pm 0.053$    & $0.382 \pm 0.057$      \\
$A_{\reg}$          & $21.5 \pm 5.7$       & $18.1 \pm 5.1$          \\  \hline
$\alpha_s(M_Z^2)$   & $0.1176^*$~\cite{Harland-Lang:2015nxa,Patrignani:2016xqp}           & $0.1176^*$~\cite{Harland-Lang:2015nxa,Patrignani:2016xqp}            \\
$m_c$               & $1.40^*$~\cite{Harland-Lang:2015qea}             & $1.40^*$~\cite{Harland-Lang:2015qea}               \\
$m_b$               & $4.75^*$~\cite{Harland-Lang:2015qea}             & $4.75^*$~\cite{Harland-Lang:2015qea}               \\ 	\hline \hline
\end{tabular}
\label{tab-dpdf}
\end{center}
\end{table}
%
%

\begin{figure*}[htb]
\begin{center}
\vspace{0.5cm}
\resizebox{0.480\textwidth}{!}{\includegraphics{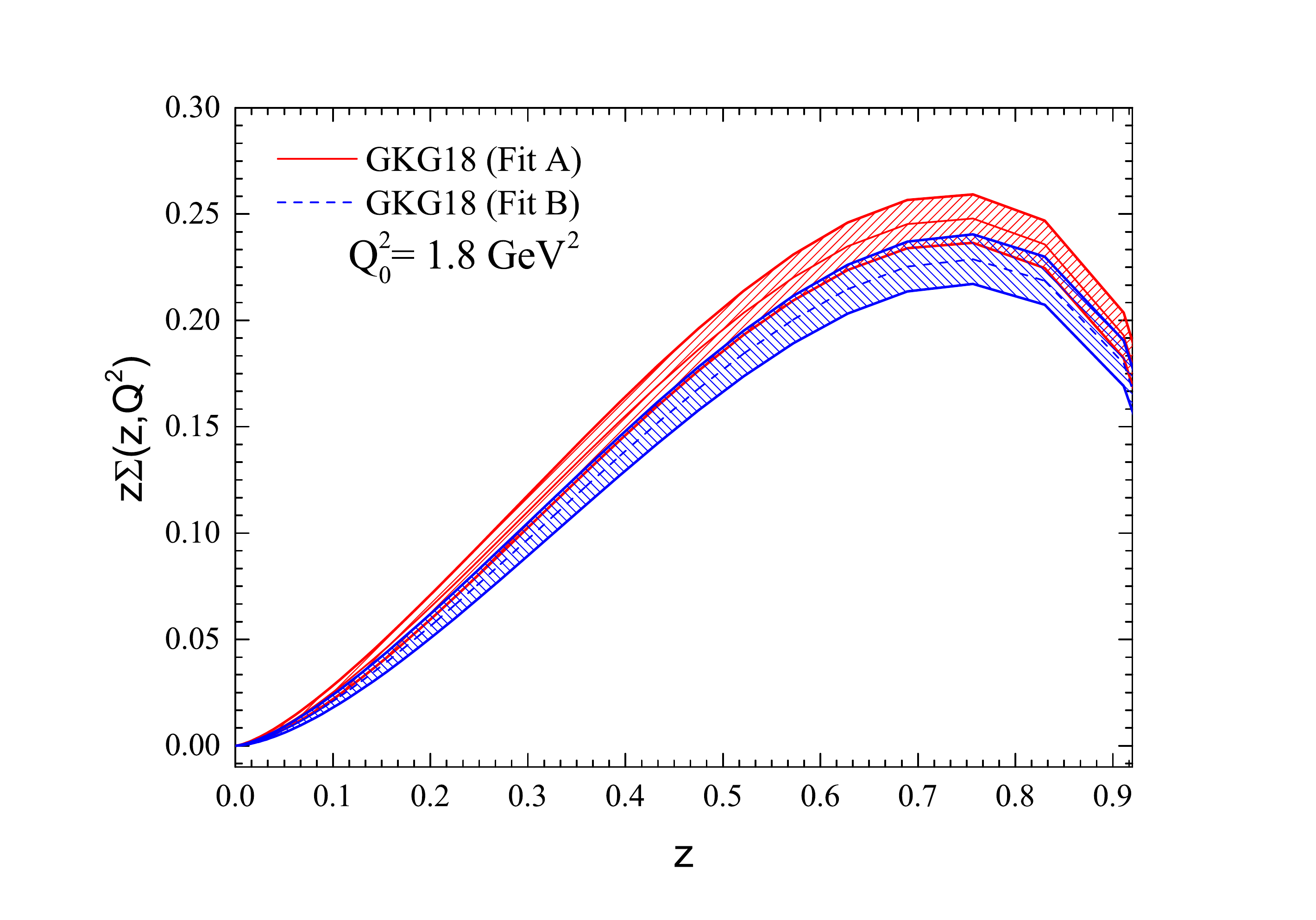}} \vspace{-0.5cm}  
\resizebox{0.480\textwidth}{!}{\includegraphics{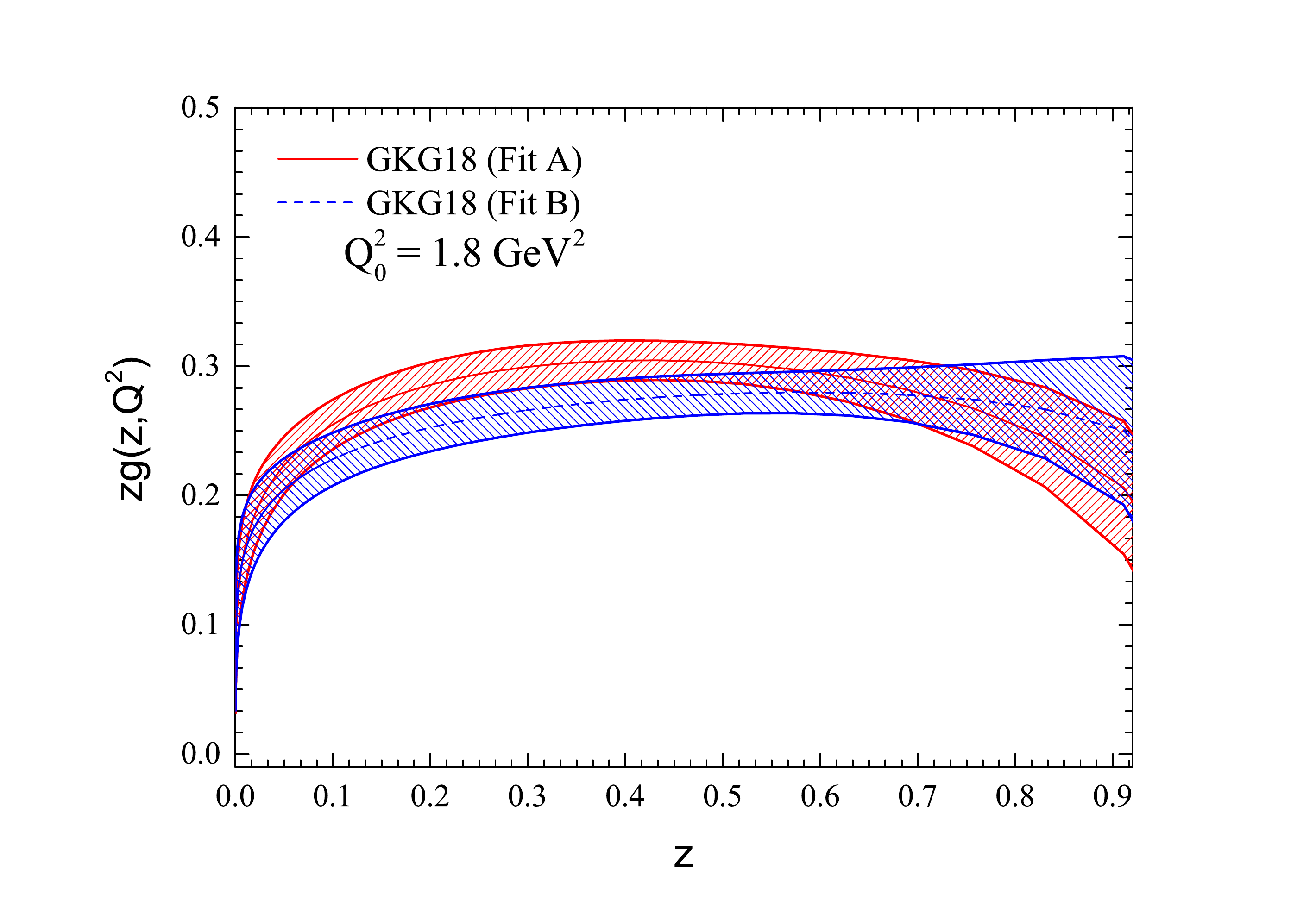}}   
\resizebox{0.480\textwidth}{!}{\includegraphics{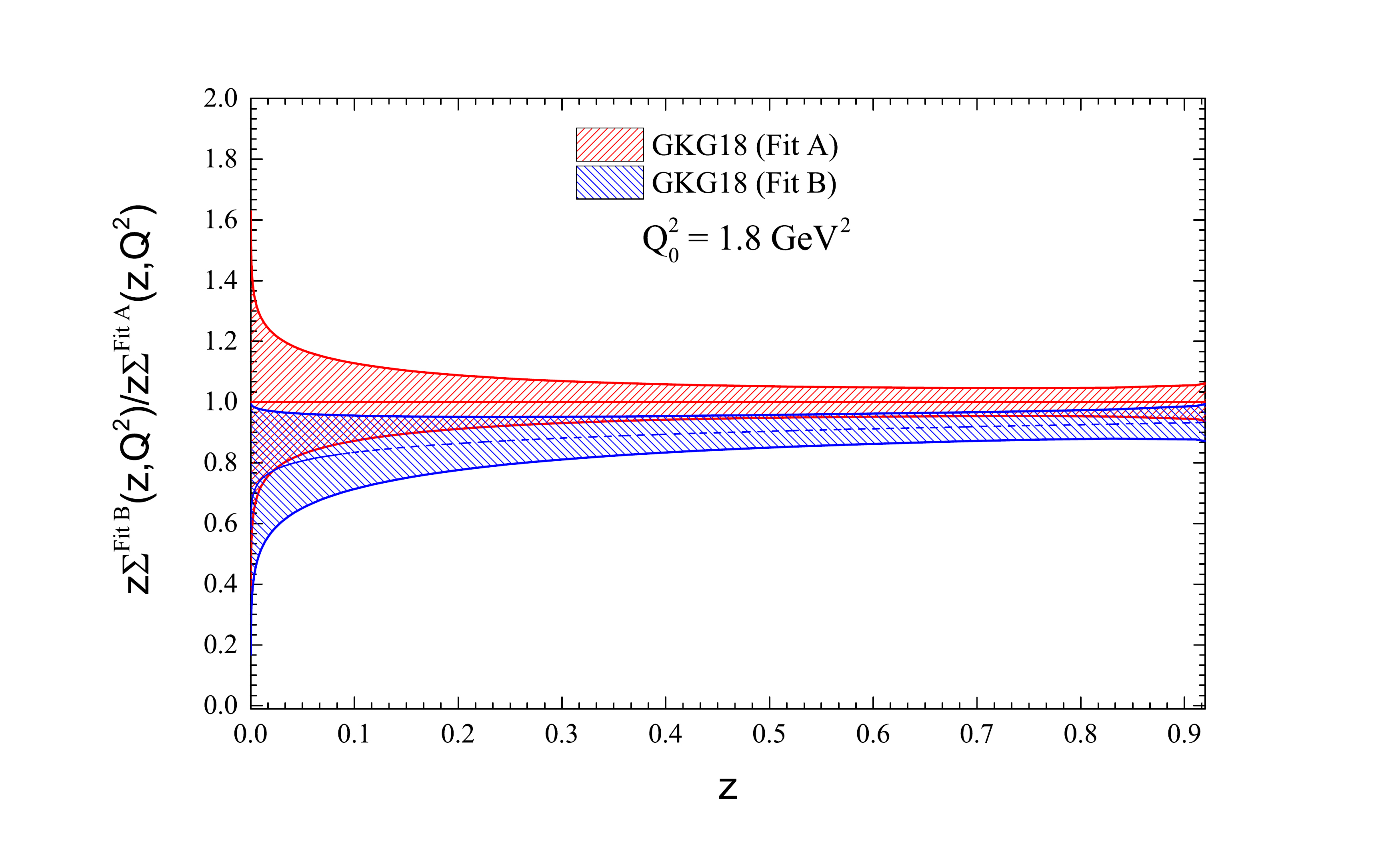}}   
\resizebox{0.480\textwidth}{!}{\includegraphics{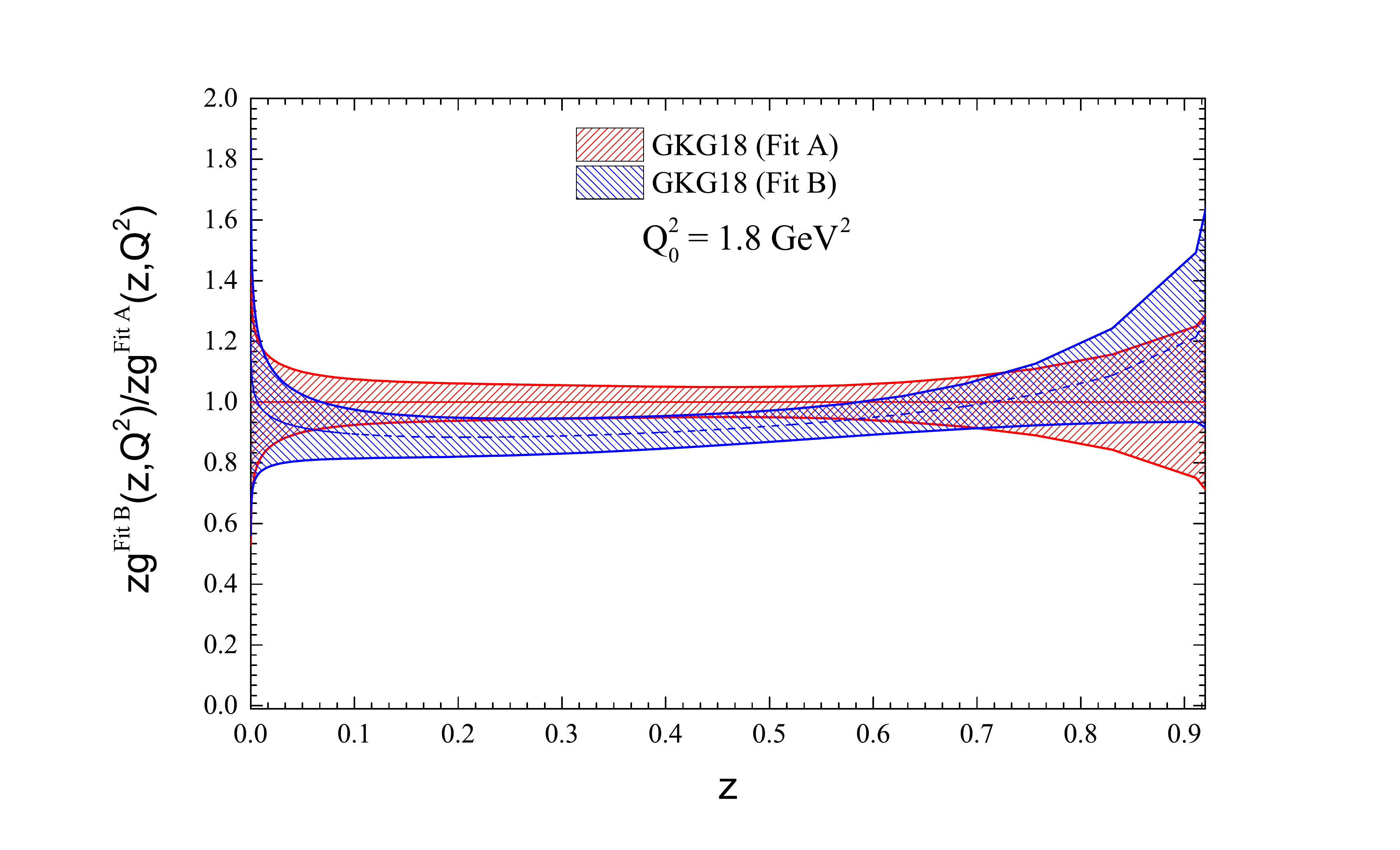}}   
\caption{\small (Color online) The total quark singlet $z\Sigma(z,Q_0^2)=\sum\limits_{q=u,d,s}z[q(z,Q_0^2)+\bar q(z,Q_0^2)]$ (left) and gluon $z g(z, Q_0^2)$ (right) distributions obtained from our NLO QCD fits, shown at the input scale ${\rm Q}_0^2 = 1.8 \, \text{GeV}^2$. The error bands correspond to the experimental uncertainties. }\label{fig:DPDFs-Q02}
\end{center}
\end{figure*}

The total quark singlet $z\Sigma(z, Q_0^2)=\sum\limits_{q=u,d,s}z[q(z,Q_0^2)+\bar q(z,Q_0^2)]$ and gluon densities $z g(z, Q_0^2)$, obtained from our QCD fits are shown with their
uncertainties in Fig.~\ref{fig:DPDFs-Q02} at the input scale of $Q_0^2 = 1.8 \, \text{GeV}^2$. 
As can be seen, in  the case of the quark singlet distribution (left panel), the result of {\tt Fit A} 
is on top of {\tt Fit B} for all kinematic ranges of $z$. Overall, one can conclude that there is 
a slight difference between {\tt Fit A} and {\tt Fit B} in the total quark singlet channel. However, in the case of the gluon distribution (right panel), the differences between the two analyses are noticeable almost for all kinematic ranges of $z$. This result can be considered as a evidence for the existence of a possible tension between the low $Q^2$ data points of the H1/ZEUS combined data. Note that in our {\tt Fit A} there are more lower-$Q^2$ data points of the H1/ZEUS combined data than in our {\tt Fit B}. Overall, it seems that {\tt Fit B} can be considered as a more conservative analysis because the tension between these data sets has been decreased as much as possible by imposing a more restrictive cut on the H1/ZEUS combined data. 

As a last point, we have shown the rations of $z\Sigma^{\tt Fit~B}(z,Q_0^2)/z\Sigma^{\tt Fit~A}(z,Q_0^2)$ and $zg^{\tt Fit~B}(z,Q_0^2)/zg^{\tt Fit~A}(z,Q_0^2)$ in Fig.~\ref{fig:DPDFs-Q02}. As illustrated in this figure, in view of the uncertainties of the obtained diffractive PDFs, there is no significant difference between {\tt Fit A} and {\tt Fit B}. Consequently, imposing a more restrictive cut on the H1/ZEUS combined data has a slight impact on the central values of the diffractive PDFs, though they do not reduce the uncertainty of the diffractive PDFs. However, from obtained $\chi^2/{\rm ndf}$, one can conclude that the {\tt GKG18} predictions describe these data very well, particularly for {\tt Fit B}.

In summary, despite slightly different central values, {\tt Fits A} and {\tt Fit B} have overlapping uncertainty bands and, hence, are compatible. 
The difference comes from the inclusion of the lower-$Q^2$ region of the combined H1/ZEUS data and thus reflects the overall compatibility of the used data sets.
It is in turn related to a few-percent systematic uncertainty in the relative normalization of the data sets, see our discussion above.

The uncertainties on diffractive PDFs need to be improved in the future for very high precision predictions at present and future hadron colliders. Like the total DIS cross section, the diffractive DIS cross section is directly sensitive to the diffractive quark density, whilst the gluon density is only indirectly 
constrained through scaling violations.
Since the gluons directly contribute to the jet production through the boson-gluon fusion process~\cite{Andreev:2015cwa,Andreev:2014yra,H1:2017bnb,Abramowicz:2015vnu,Aaron:2011mp}, one can use the measurements of dijet production in diffractive DIS to further constrain the diffractive gluon PDF. 
As an example of the inclusion of dijet production data in the QCD analysis of the diffractive PDFs, one can refer to the ZEUS analysis~\cite{Chekanov:2009aa}.

\subsection{$Q^2$ evolution and comparison to other diffractive PDFs}\label{sec:comparison-to-other-DPDFs}

Having the optimised values of the free parameters, we study next the shape and behaviour of {\tt GKG18-DPDFs} diffractive PDFs extracted from {\tt Fit A} and {\tt Fit B} analyses
with an increase of $Q^2$ and also compare our results  with those of other collaborations, in particular with the {\tt ZEUS-2010 Fit SJ} and {\tt H1-2006 Fit B} parton sets.

\begin{figure*}[htb]
\begin{center}
		\vspace{0.5cm}
		\resizebox{0.480\textwidth}{!}{\includegraphics{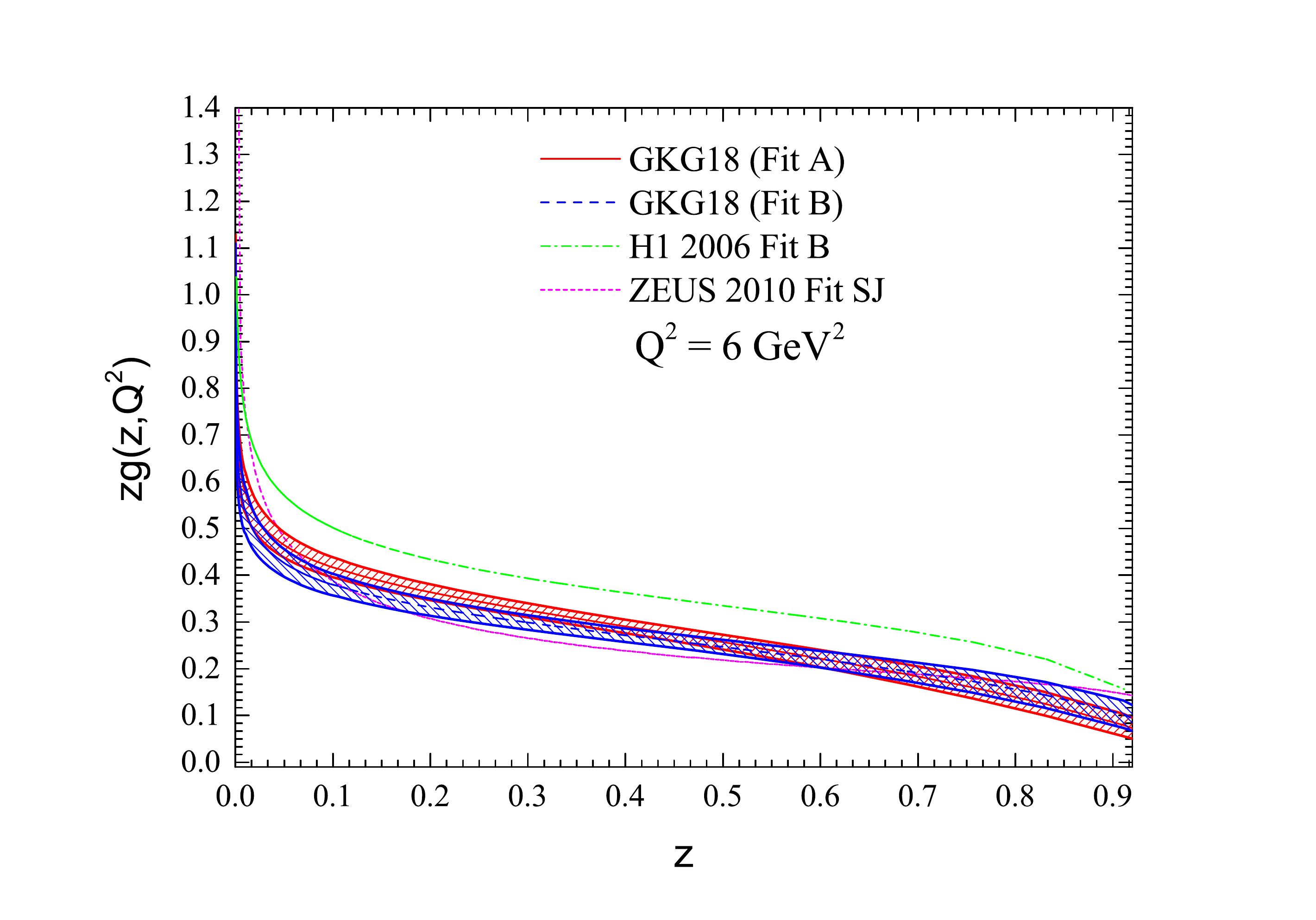}}   
		\resizebox{0.480\textwidth}{!}{\includegraphics{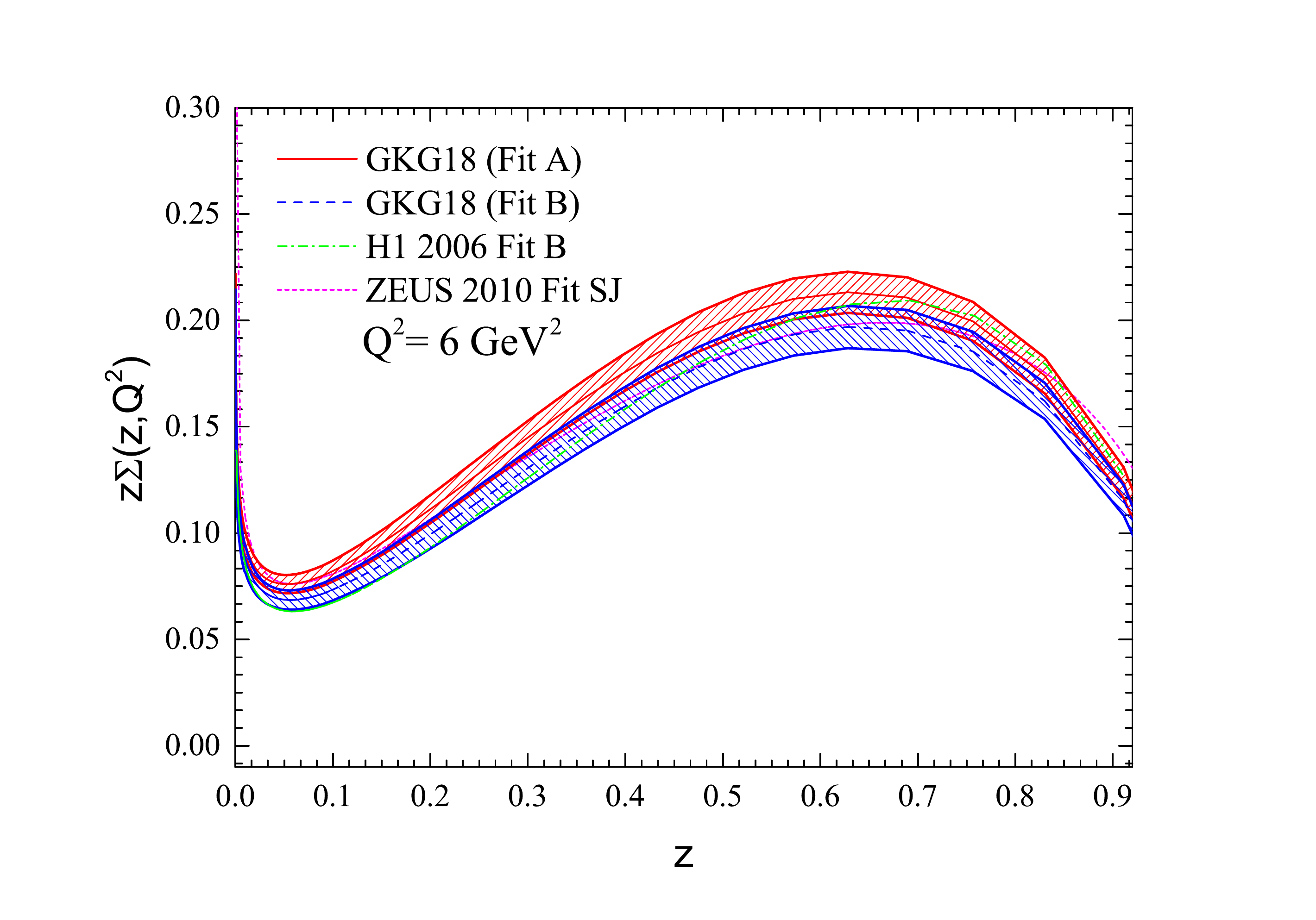}}   
		\resizebox{0.480\textwidth}{!}{\includegraphics{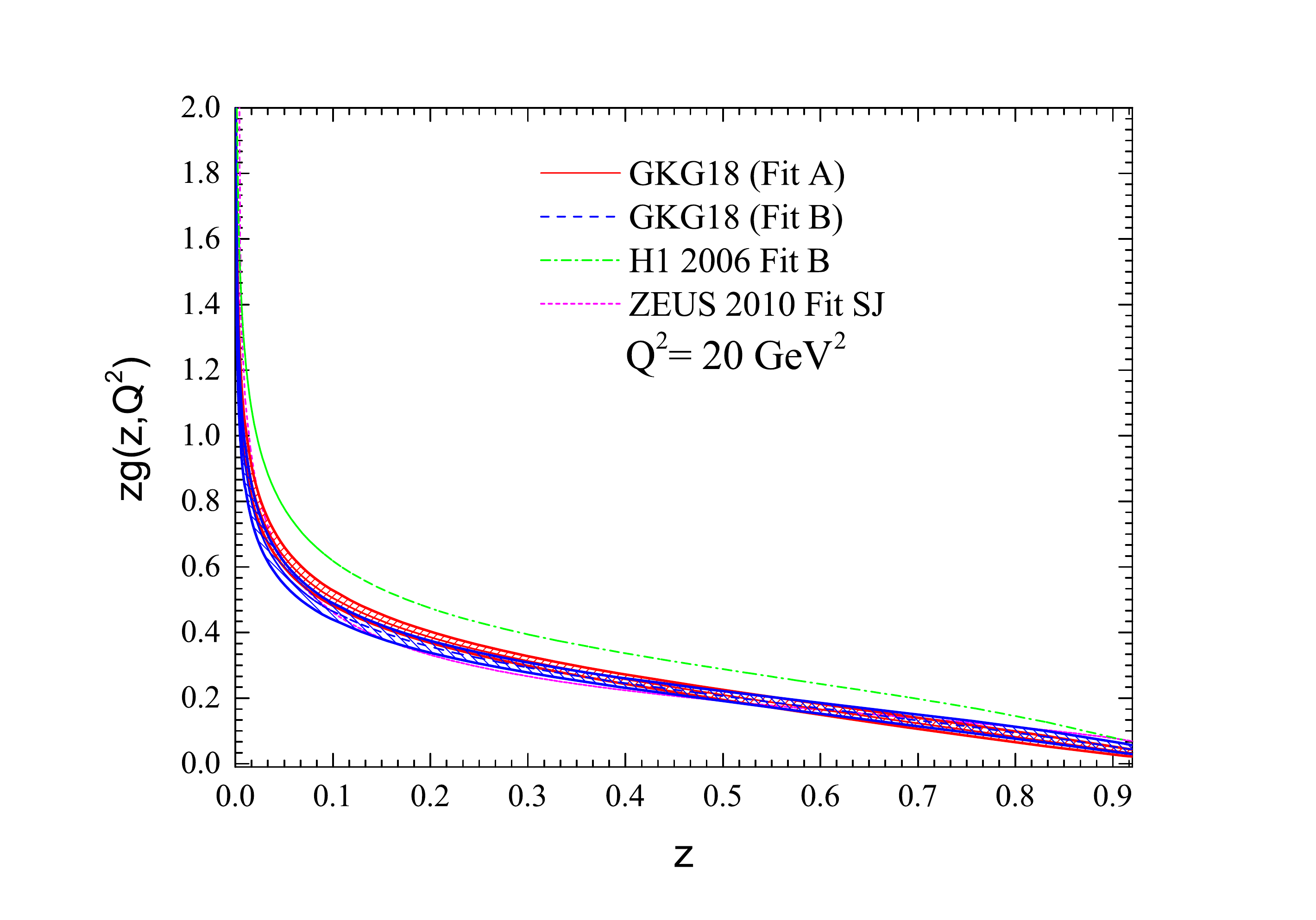}}   
		\resizebox{0.480\textwidth}{!}{\includegraphics{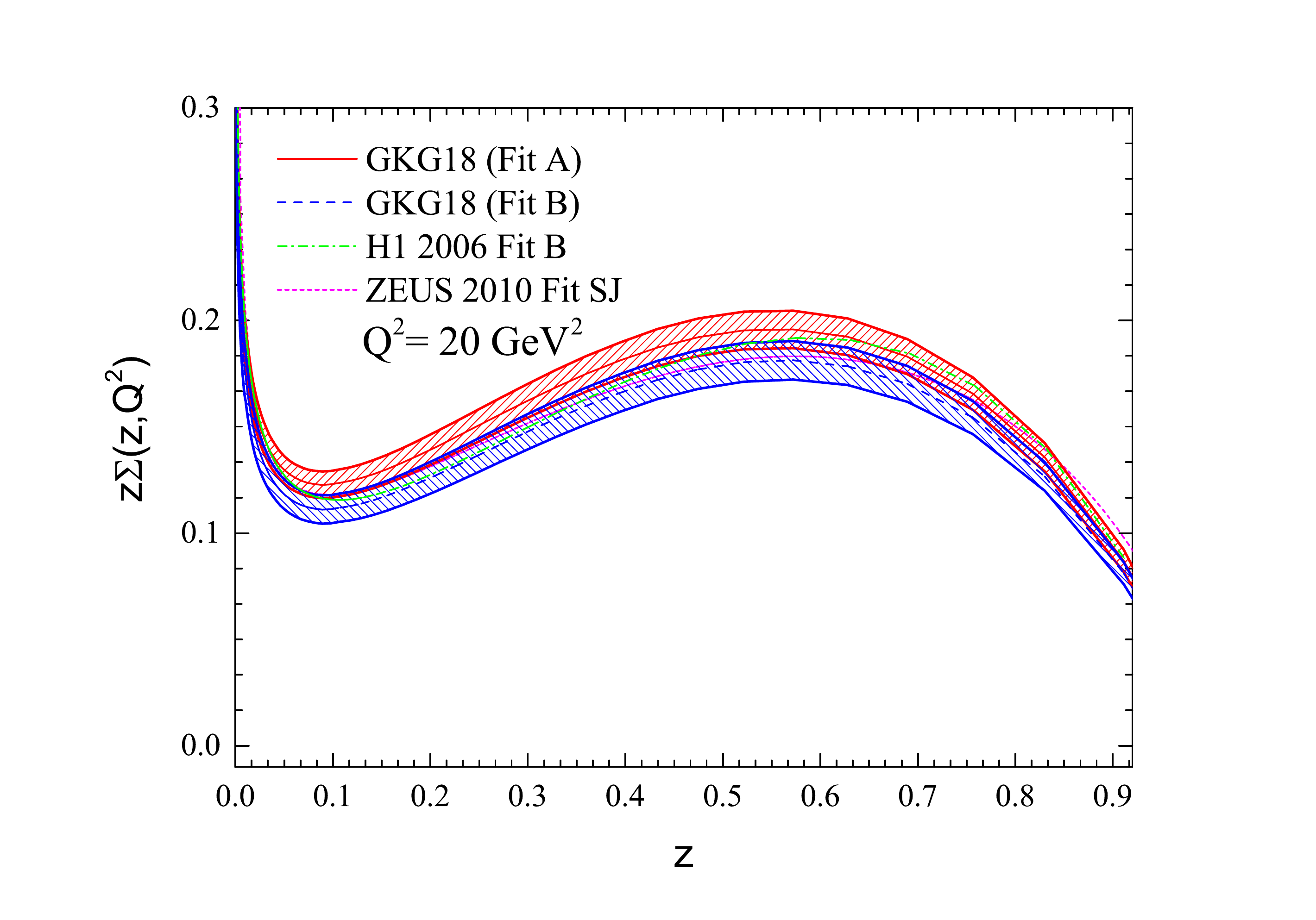}}   
		\resizebox{0.480\textwidth}{!}{\includegraphics{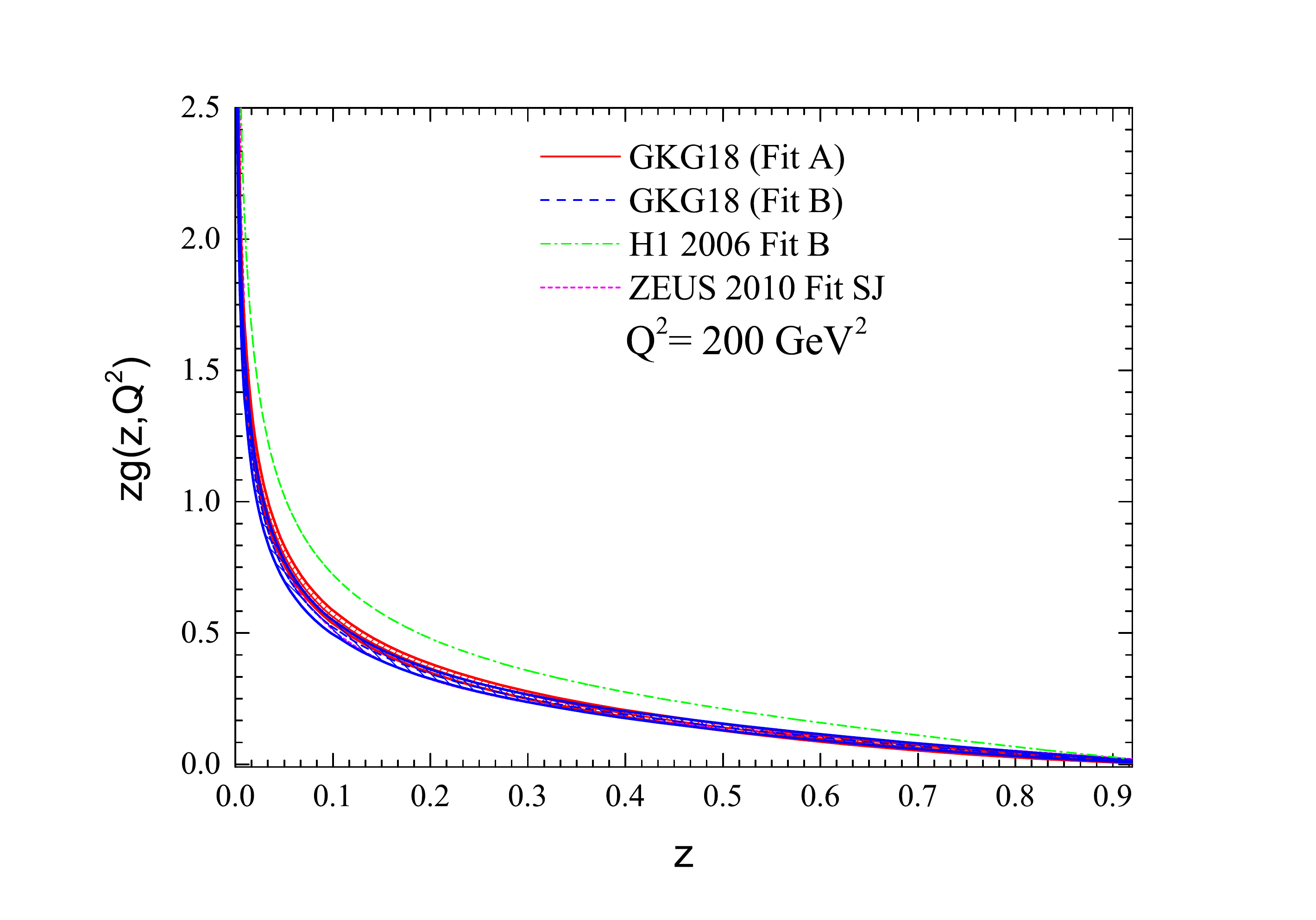}}   
		\resizebox{0.480\textwidth}{!}{\includegraphics{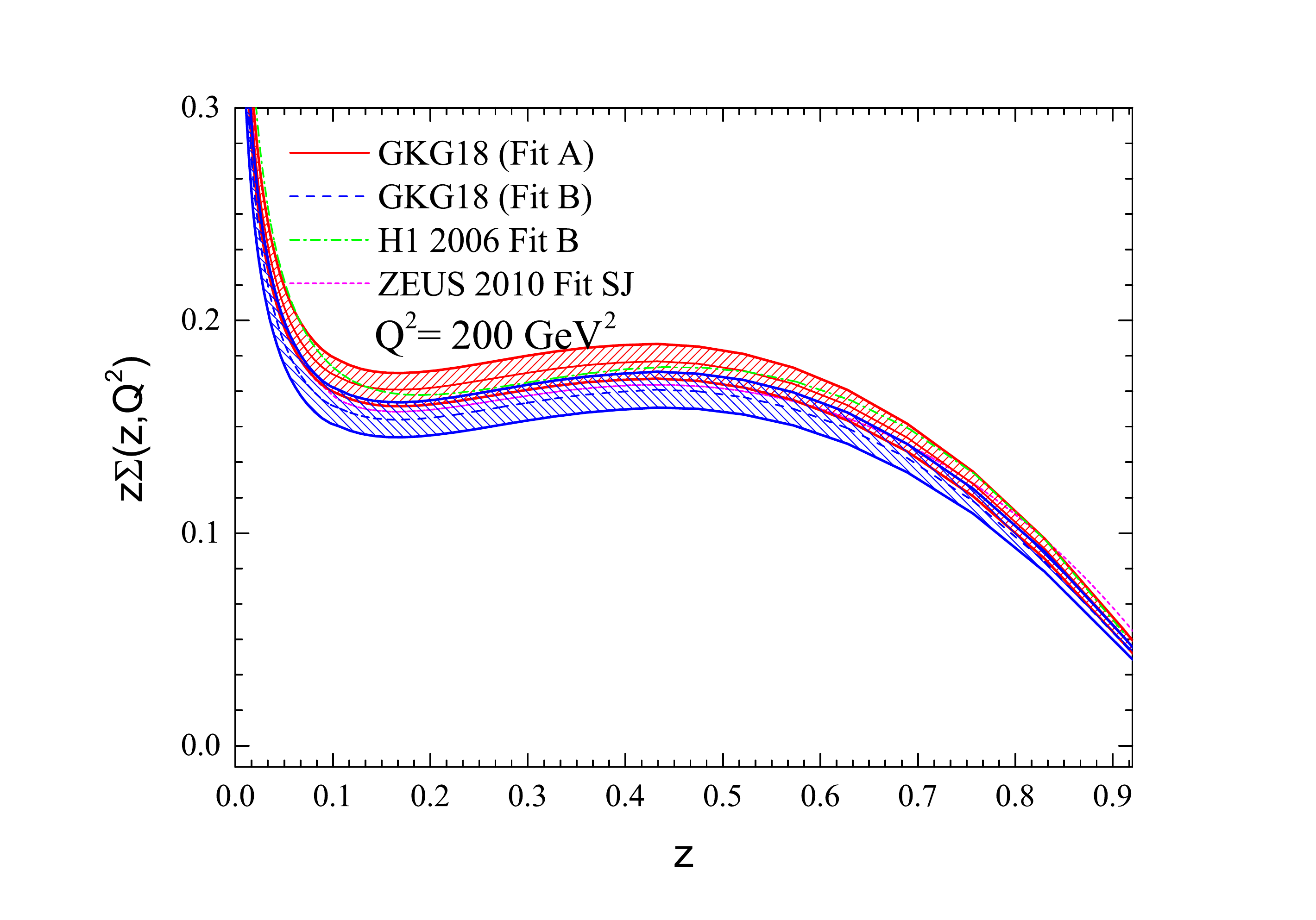}}   
\caption{\small (Color online) The total quark singlet $z \Sigma(z, Q^2)$ (right) and gluon $z g(z, Q^2)$ (left) distributions obtained from our NLO QCD fits
at selected $Q^2$ value of $Q^2 = 6, 20$ and 200 ${\rm GeV}^2$. The error bands correspond to the fit uncertainties. }\label{fig:DPDFs-AllQ2}
\end{center}
\end{figure*}

In order to study the scale dependence of diffractive PDFs, in Fig.~\ref{fig:DPDFs-AllQ2} we show
the obtained total quark singlet $z \Sigma(z, Q^2)$ and gluon $z g(z, Q^2)$ densities with their uncertainties at some selected $Q^2$ values of $Q^2 = 6, \, 20$ and 200 $\text{GeV}^2$. These plots also contain the related results of two previous analyses of diffractive PDFs from H1~\cite{Aktas:2006hy} and ZEUS~\cite{Chekanov:2009aa} Collaborations. Note that for the H1 analysis we have used the result of their {\tt H1-2006 Fit B}, while for the ZEUS analysis, their standard analysis of {\tt ZEUS-2010 Fit SJ} has been considered for comparison.

As can be seen from Fig.~\ref{fig:DPDFs-AllQ2}, due to the evolution effects, both the quark singlet and gluon distributions are undergone an enhancement at low values of $z$. For large value of $z$, one can see a reduction of the diffractive PDFs
with an increase of $Q^2$. For the gluon distributions (left panels), the results of our {\tt Fit A} and {\tt Fit B} are
in good agreements with the {\tt ZEUS-2010 Fit SJ} analysis. However, there are some deviations between our results
and the H1 ones, especially at smaller and larger values of $z$. To summarize, the agreement between our results for the gluon diffractive PDFs and the {\tt ZEUS-2010 Fit SJ}  is somewhat better than for {\tt H1-2006 Fit B}. The discrepancy between our results and H1 fit can be directly attributed to the inclusion of the H1-LRG-12 and H1/ZEUS combined data sets which is not used in the H1 analysis. 
	
For the total quark singlet distributions (right panels), there are no significant differences between our results and both the H1 and ZEUS analyses, almost at all values of $z$. As one can see from Fig.~\ref{fig:DPDFs-AllQ2}, in all region of $z$, the total quark singlet distributions of {\tt H1-2006 Fit B} and {\tt ZEUS-2010 Fit SJ} are inside the error bands of the two {\tt Fit A} and {\tt Fit B} total quark singlet distributions. Overall, we have obtained comparable singlet distribution in comparison to the other groups.  According to the obtained results, one can conclude that the preliminary impact of these new data sets on the extracted diffractive PDFs is mostly on the behavior of the quark diffractive PDFs.

\begin{figure*}[htb]
\begin{center}
		\vspace{0.5cm}
		\resizebox{0.480\textwidth}{!}{\includegraphics{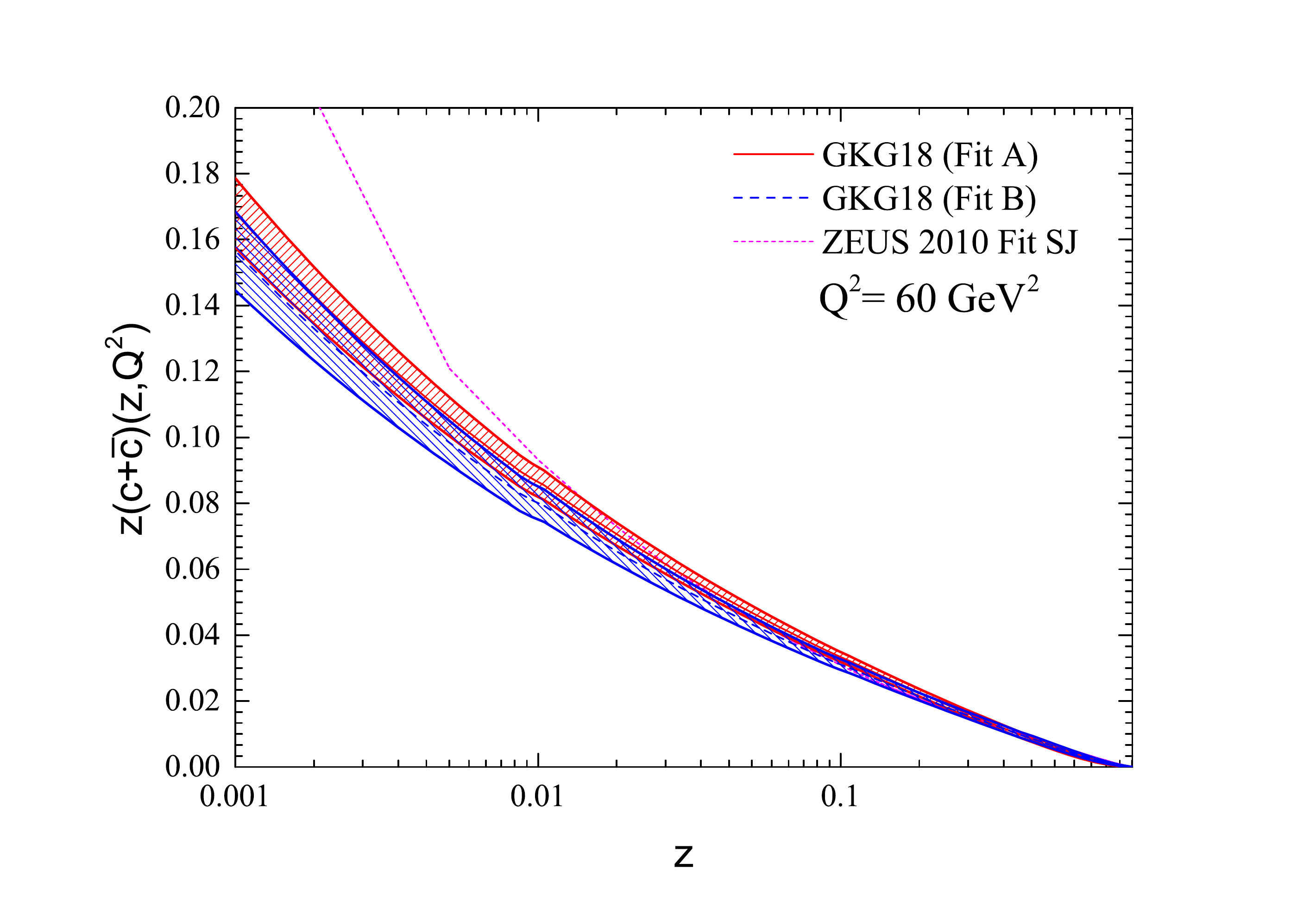}}   
		\resizebox{0.480\textwidth}{!}{\includegraphics{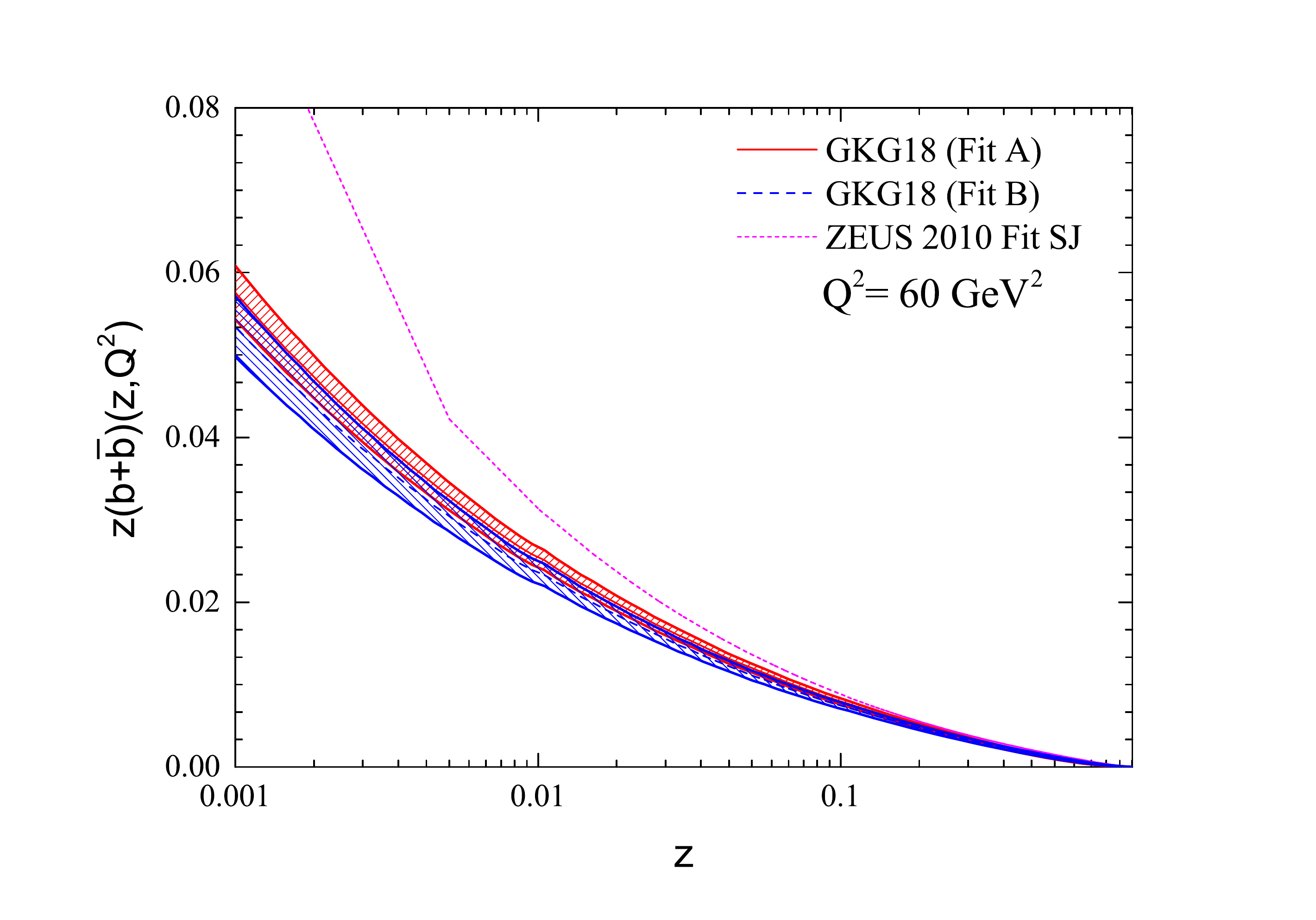}}   
		\resizebox{0.480\textwidth}{!}{\includegraphics{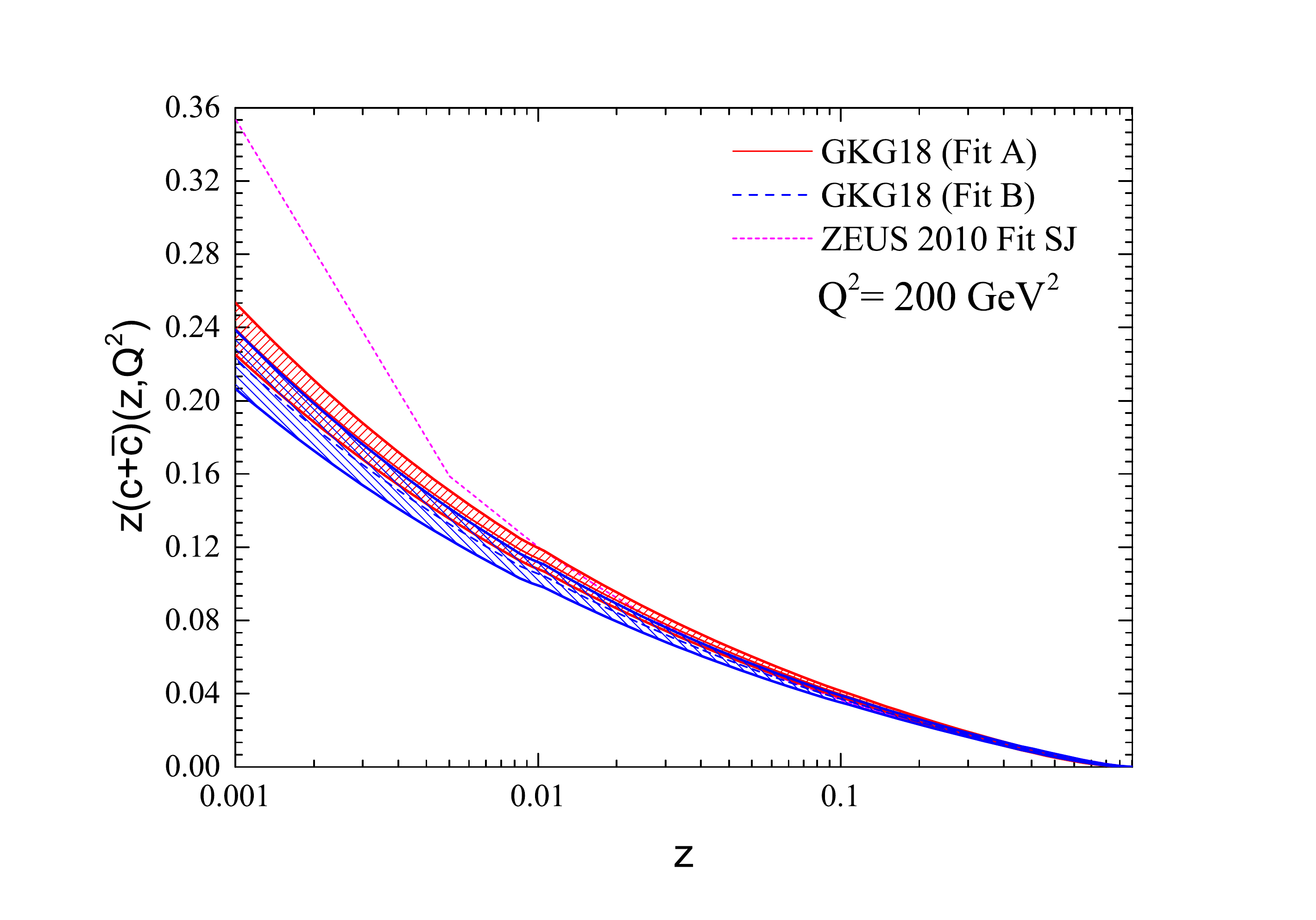}}   
		\resizebox{0.480\textwidth}{!}{\includegraphics{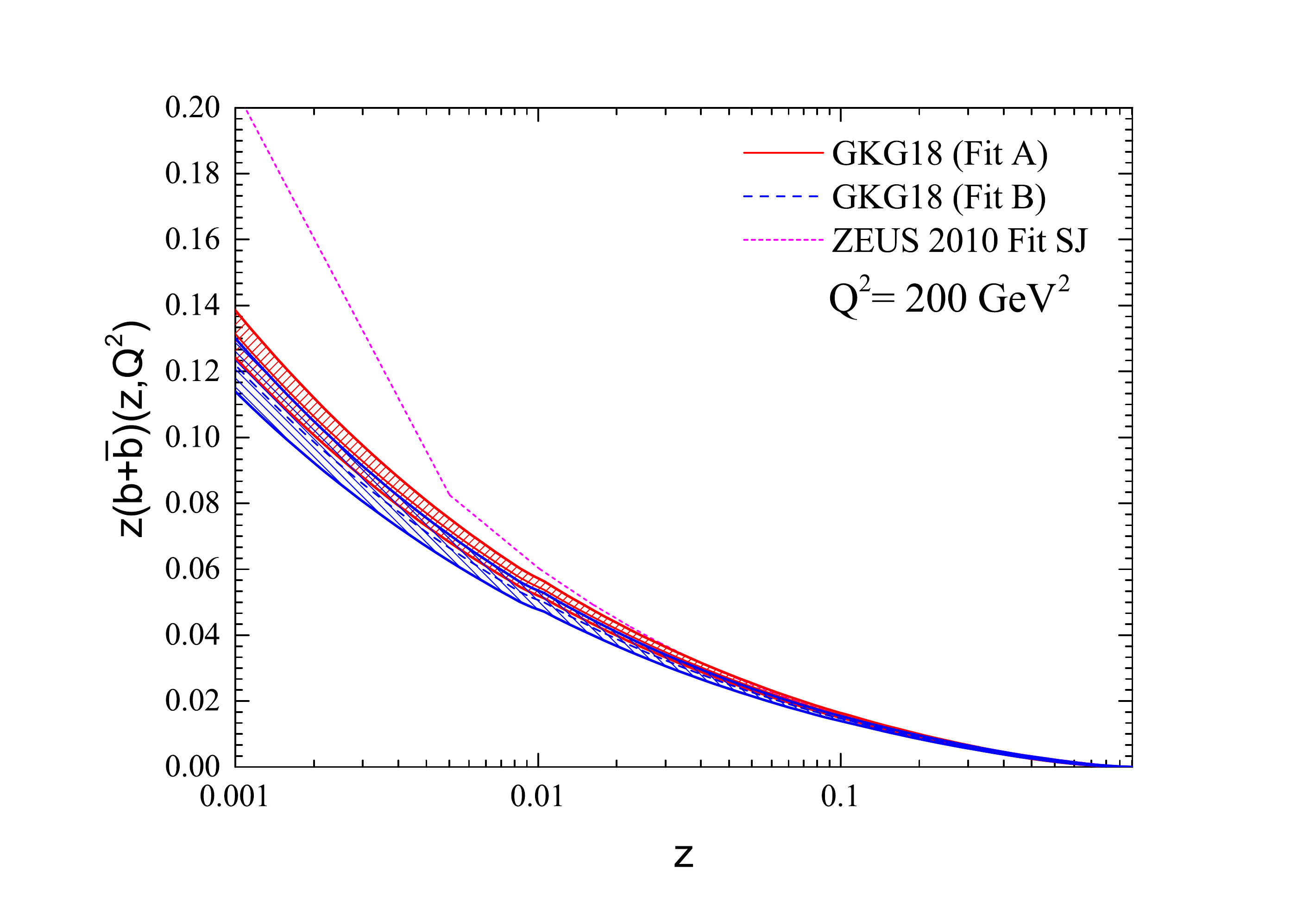}}   
\caption{\small (Color online) The charm  $z (c + \bar c)(z, Q^2)$ (left) and bottom  $z (b + \bar b )(z, Q^2)$ (right) quark diffractive PDFs obtained from our NLO QCD fits
at selected $Q^2$ value of $Q^2 = 60$ and 200 ${\rm GeV}^2$. The error bands correspond to the fit uncertainties. }\label{fig:DPDFs-HQ}
\end{center}
\end{figure*}

We conclude this section by presenting the heavy quark diffractive PDFs determined in this analysis in the TR GM-VFNS.
In Fig.~\ref{fig:DPDFs-HQ}, the charm $z (c + \bar c)(z, Q^2)$ (left) and bottom $z (b + \bar b )(z, Q^2)$ (right) quark diffractive PDFs obtained from our NLO QCD fits have been shown at selected $Q^2$ value of $Q^2 = 60$ and 200 GeV$^2$. The error bands correspond to the fit uncertainties derived only from the experimental input. The results from {\tt ZEUS-2010 Fit SJ} also presented for comparison. As one can see from these plots, 
only insignificant differences between our results and {\tt ZEUS-2010 Fit SJ} can be found for all heavy quark diffractive PDFs at low values of $z$; $z < 0.01$. 

\subsection{Comparison to the diffractive DIS data}\label{sec:comparison-to-DDISdata}

\begin{figure*}[htb]
\vspace{1.0cm}
\includegraphics[clip,width=1.02\textwidth]{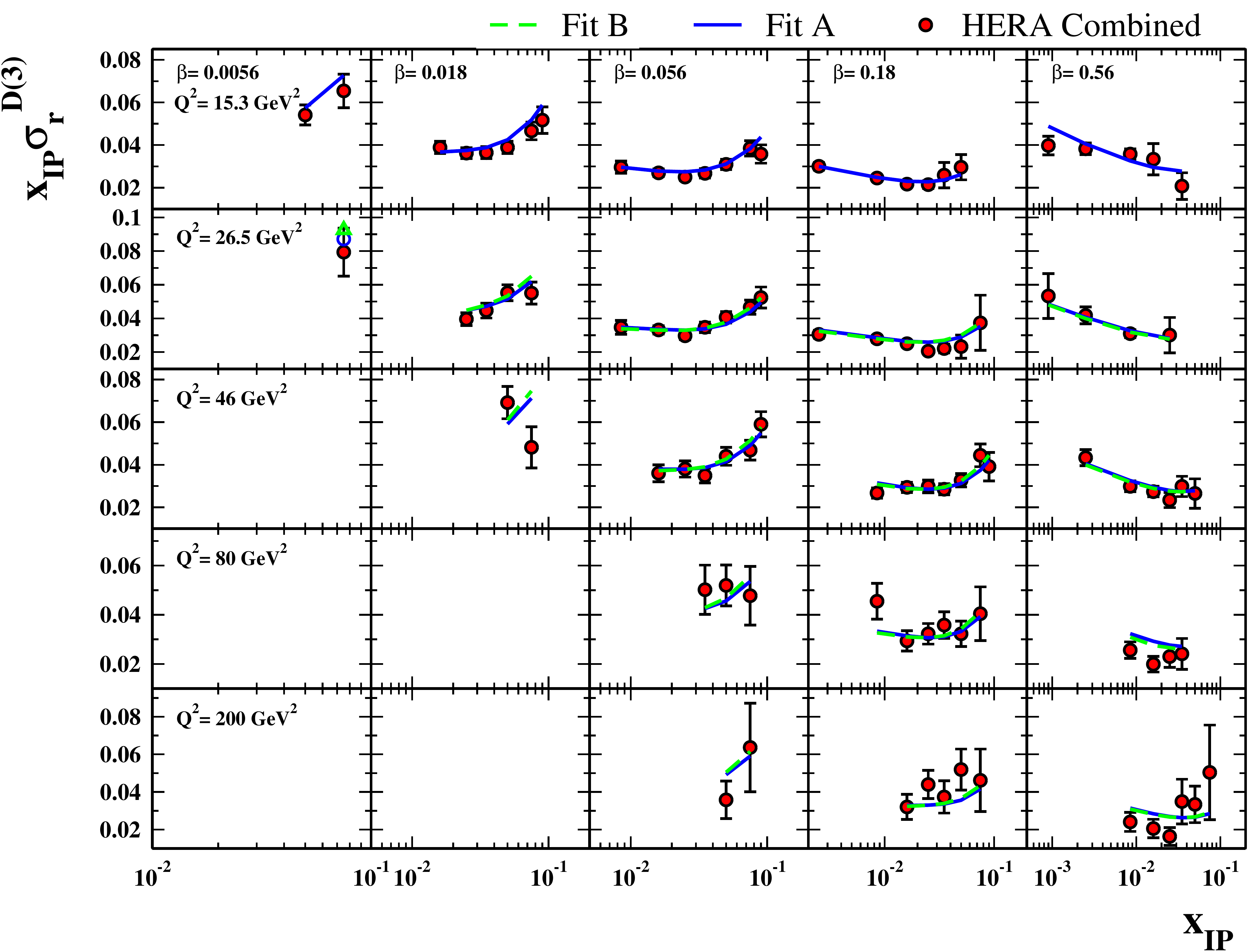}
\begin{center}
\caption{ \small A comparison between the results of {\tt Fit A} and {\tt Fit B} and the HERA combined reduced diffractive cross section $x_{\pom} \sigma_r^{D(3)}$~\cite{Aaron:2012hua}
as a function of $x_{\pom}$ for different values of $\beta$ and $Q^2$. The vertical error
bars indicate the statistical, systematic and procedural uncertainties added in quadrature. The combined H1/ZEUS diffractive DIS data are corrected by a global factor of 1.21 to consider the contributions of proton dissociation processes and also corrected by a global normalization factor to extrapolate from $0.09<|t|<0.55$ GeV$^2$ to $|t|<1$ GeV$^2$  as described in the text.} \label{Combined_2012}
\end{center}
\end{figure*}

This section presents a detailed comparison of the theoretical predictions based on our diffractive PDFs extracted from the analyses {\tt Fit A} and {\tt Fit B} with the experimental data used in these analyses. Note that for all figures, the error bars shown on the experimental data points correspond to the statistical and systematic errors added in quadrature. 
Figure~\ref{Combined_2012} presents a detailed comparison between the results of {\tt Fit A} and {\tt Fit B} and the HERA combined reduced diffractive cross sections $x_{\pom} \sigma_r^{D(3)}$~\cite{Aaron:2012hua} as a function of $x_{\pom}$ for different values of $\beta$ and ${Q}^2$. The plots clearly show that our pQCD fits describe the diffractive DIS data well for all ranges of $\beta$ from 0.0056 to 0.56 and ${Q}^2$ from 15.3 to 200 GeV$^2$. There are only some small deviations at larger values of $\beta$ and ${Q}^2$. It should be noted here that the data points excluded from the analysis with $Q^2 \leq Q_{min}^2=9$ GeV$^2$, due to the requirement cuts mentioned in Sec.~\ref{sec:data}, 
are not shown in the figures in this section. In addition, note that the HERA combined data are corrected by a global factor of 1.21 to consider the contributions of proton dissociation processes as described in Sec.~\ref{sec:data}.  As we discussed in Sec.~\ref{sec:data}, while all H1-LRG data sets have been given for the range of $|t|<1 \, {\text{GeV}}^2$, the combined H1/ZEUS diffractive DIS data are restricted to the  $0.09<|t|<0.55 \, {\text{GeV}}^2$ range. Hence all the combined H1/ZEUS diffractive DIS data sets are corrected by a global normalization factor to extrapolate from $0.09<|t|<0.55$ GeV$^2$ to $|t|<1$ GeV$^2$.

\begin{figure*}[htb]
\vspace{1.0cm}
\includegraphics[clip,width=0.30\textwidth]{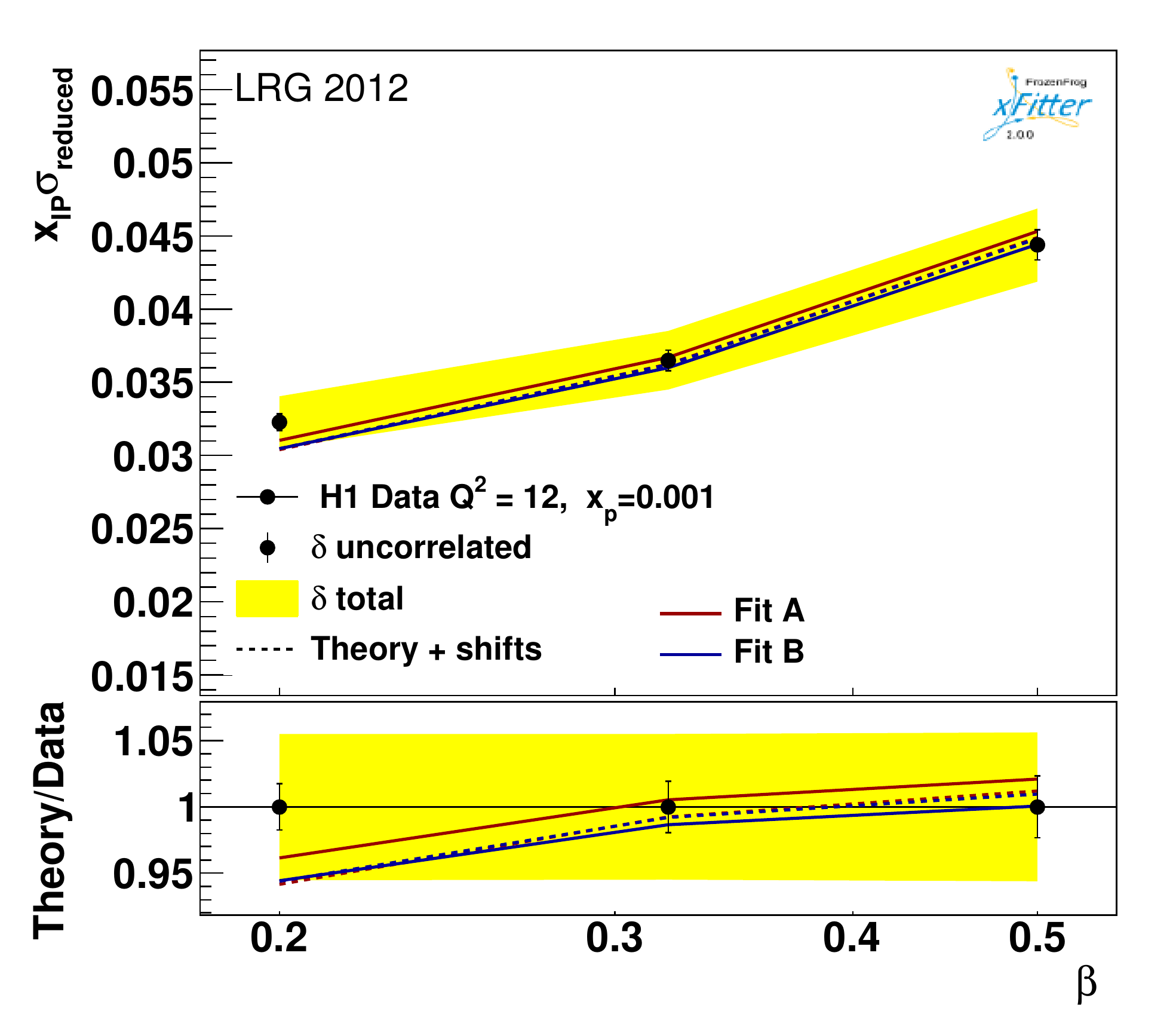}
\includegraphics[clip,width=0.30\textwidth]{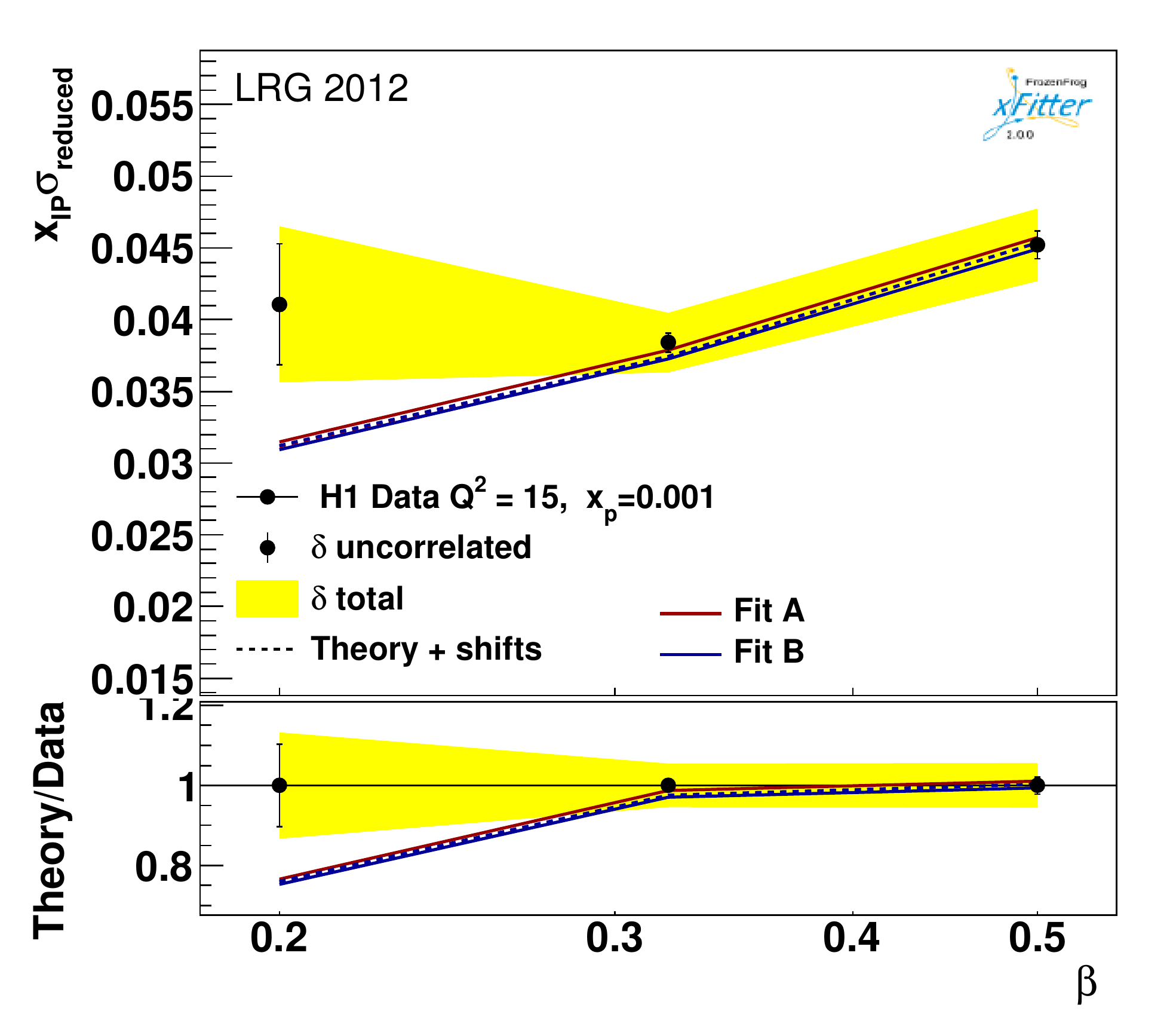}
\includegraphics[clip,width=0.30\textwidth]{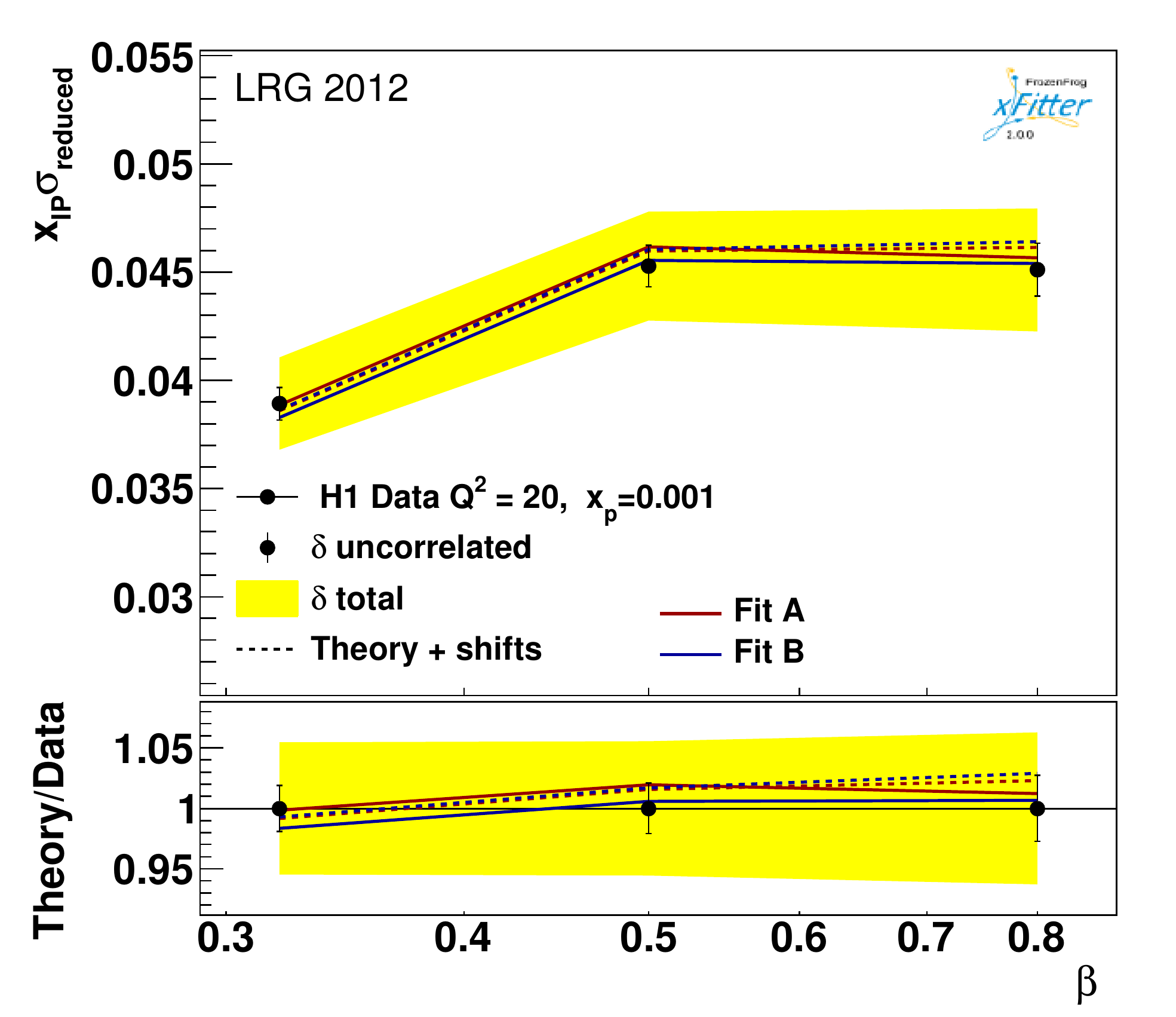}
\includegraphics[clip,width=0.30\textwidth]{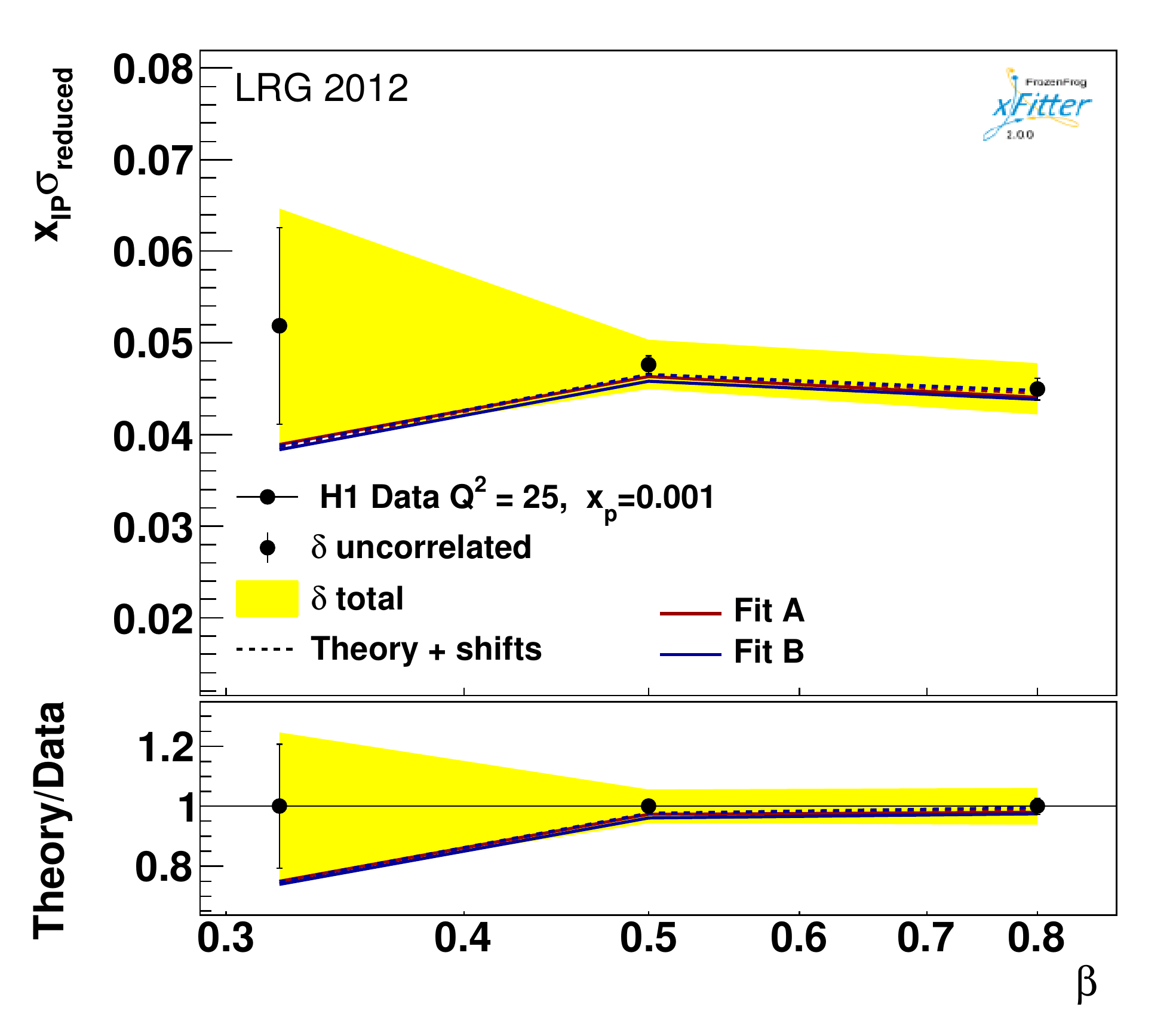}
\includegraphics[clip,width=0.30\textwidth]{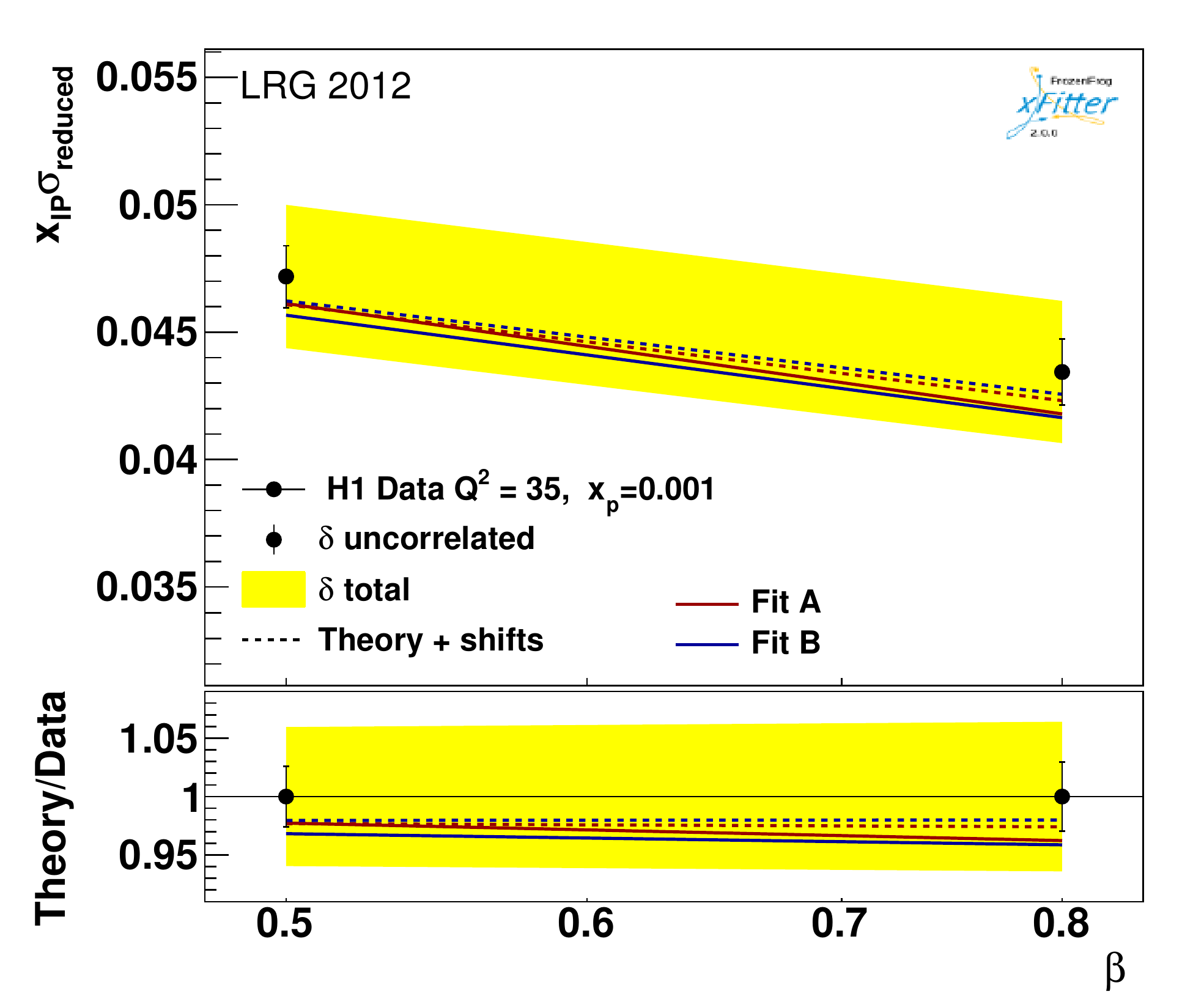}
\includegraphics[clip,width=0.30\textwidth]{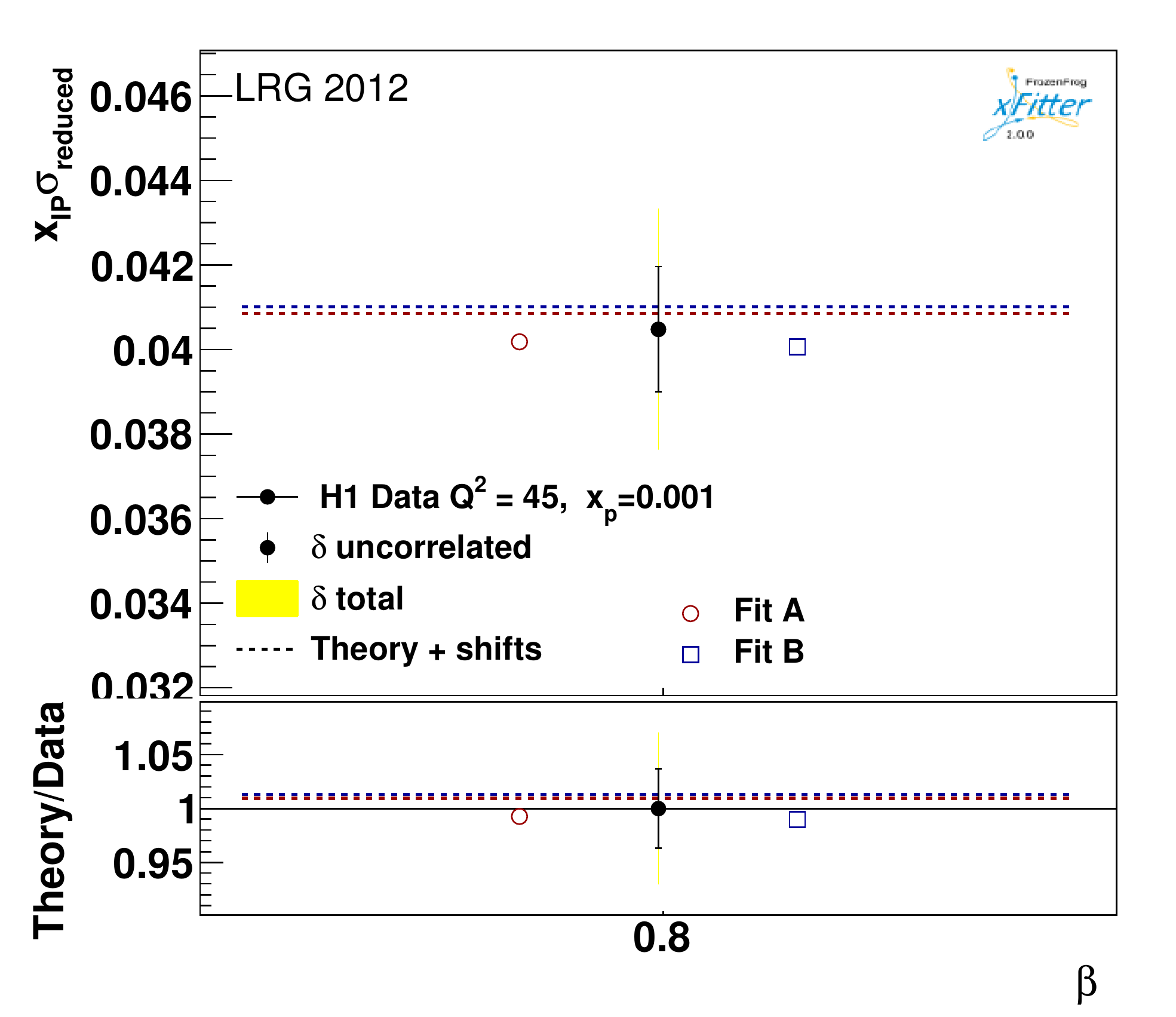}
\begin{center}
\caption{\small The results of our NLO pQCD fit based on {\tt Fit B} for the reduced diffractive cross section $x_{\pom} \sigma_r^{D(3)}$ as a function of $\beta$ for $x_{\pom} = 0.001$ in comparison with H1-LRG-2012 data~\cite{Aaron:2012ad}. The error bars on the data points represent the uncorrelated uncertainties and the yellow bands represent the total uncorrelated and correlated uncertainties. } \label{LRG-2012-xp001}
\end{center}
\end{figure*}

\begin{figure*}[htb]
\vspace{1.0cm}
\includegraphics[clip,width=0.30\textwidth]{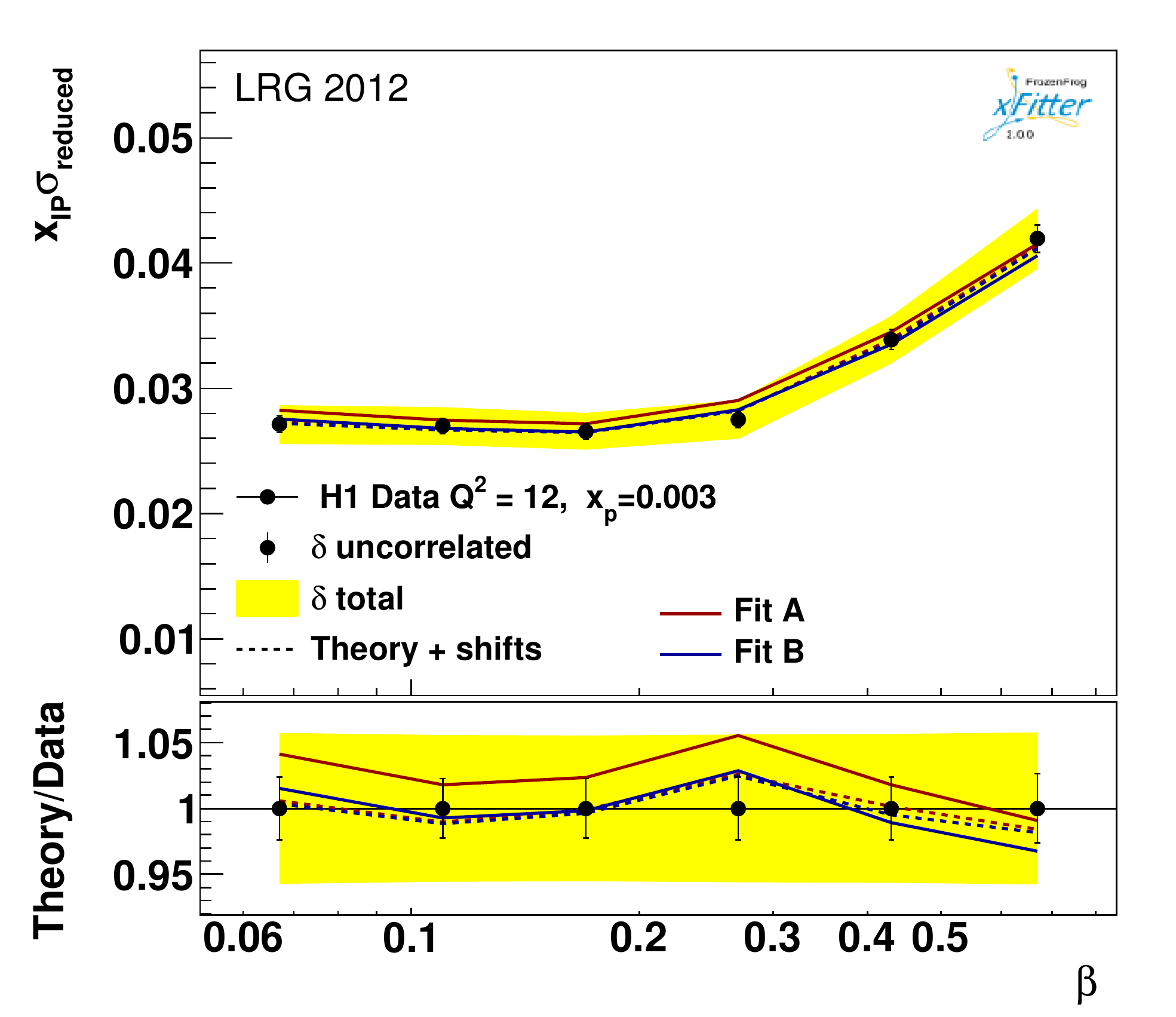}
\includegraphics[clip,width=0.30\textwidth]{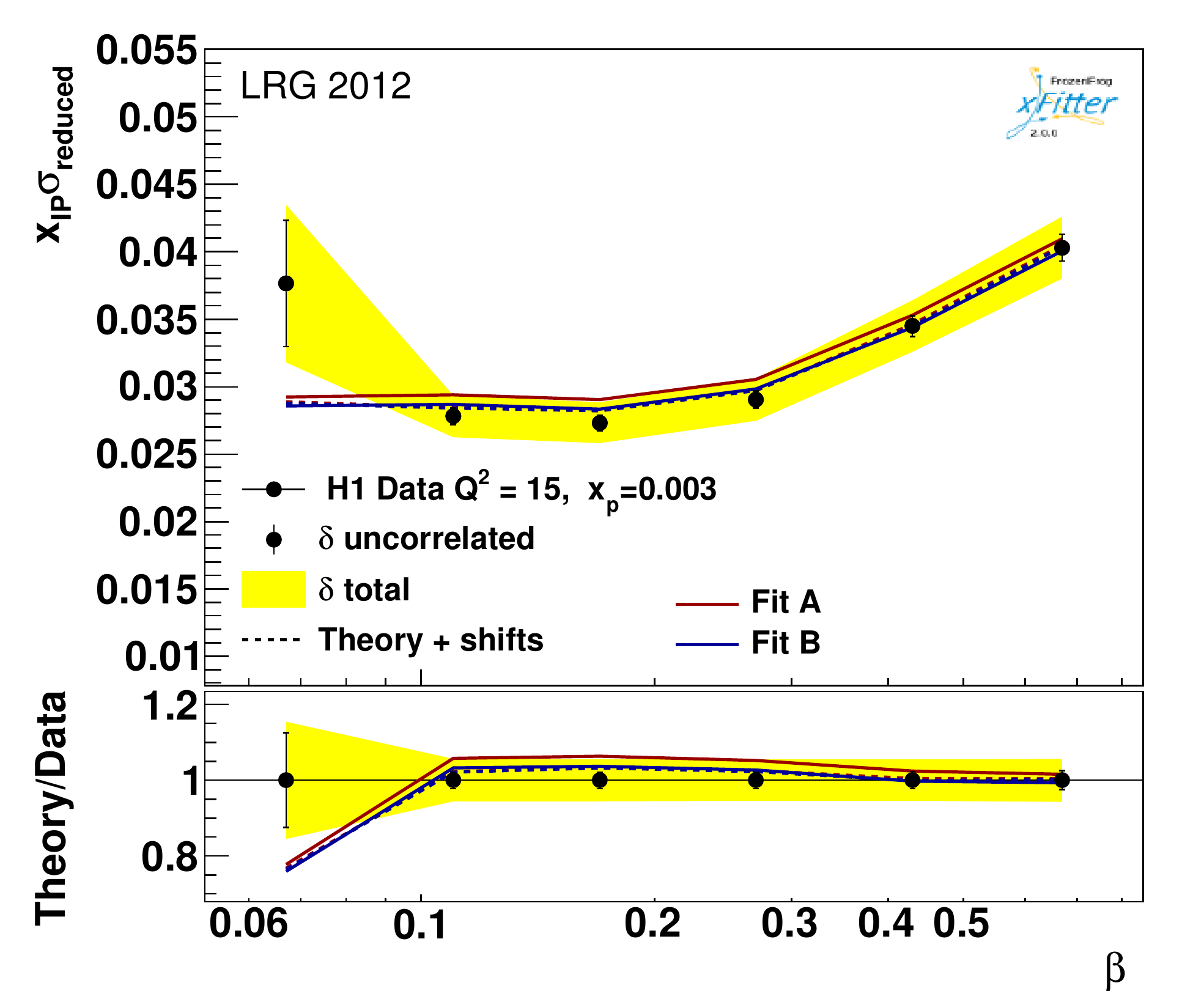}
\includegraphics[clip,width=0.30\textwidth]{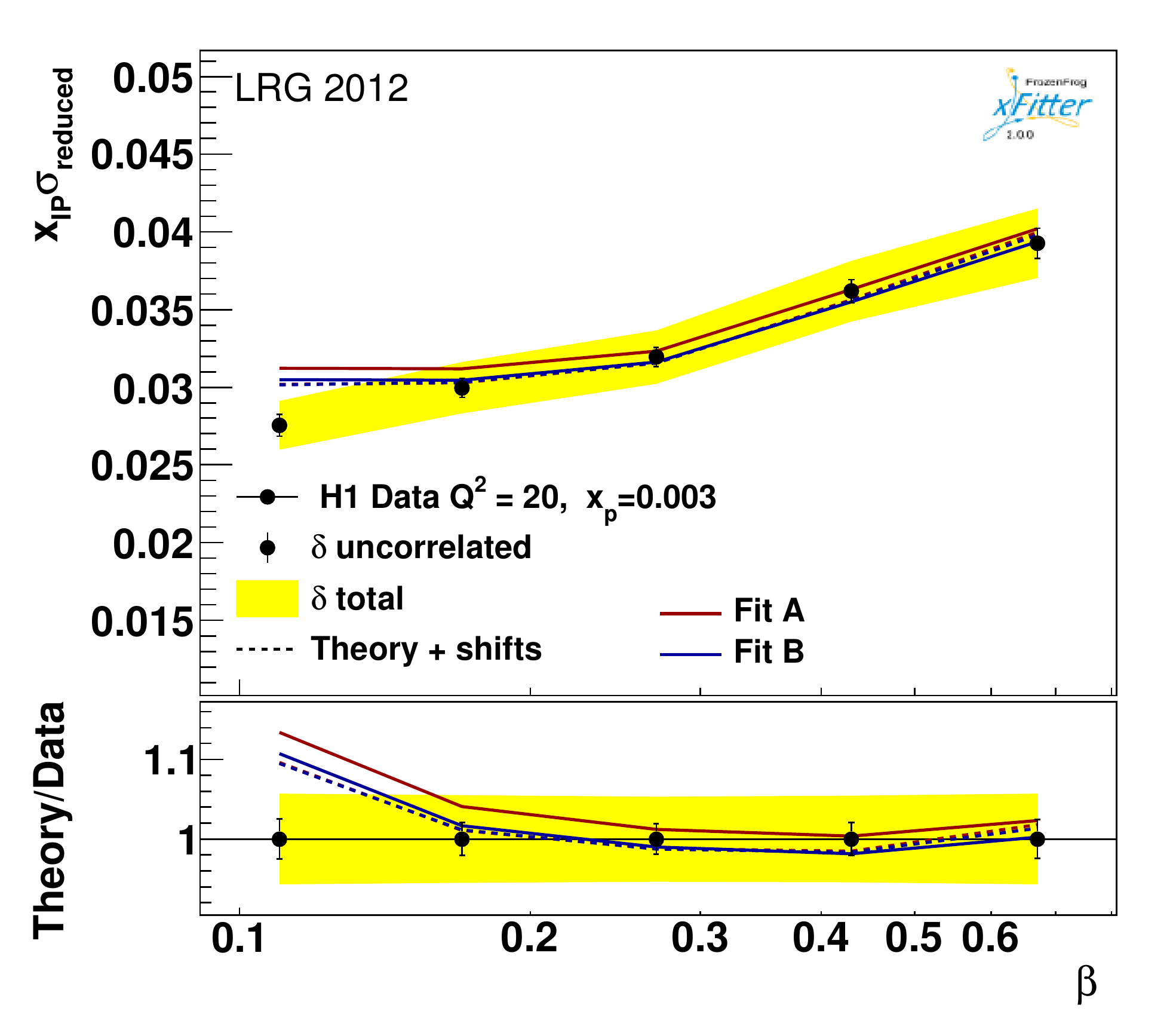}
\includegraphics[clip,width=0.30\textwidth]{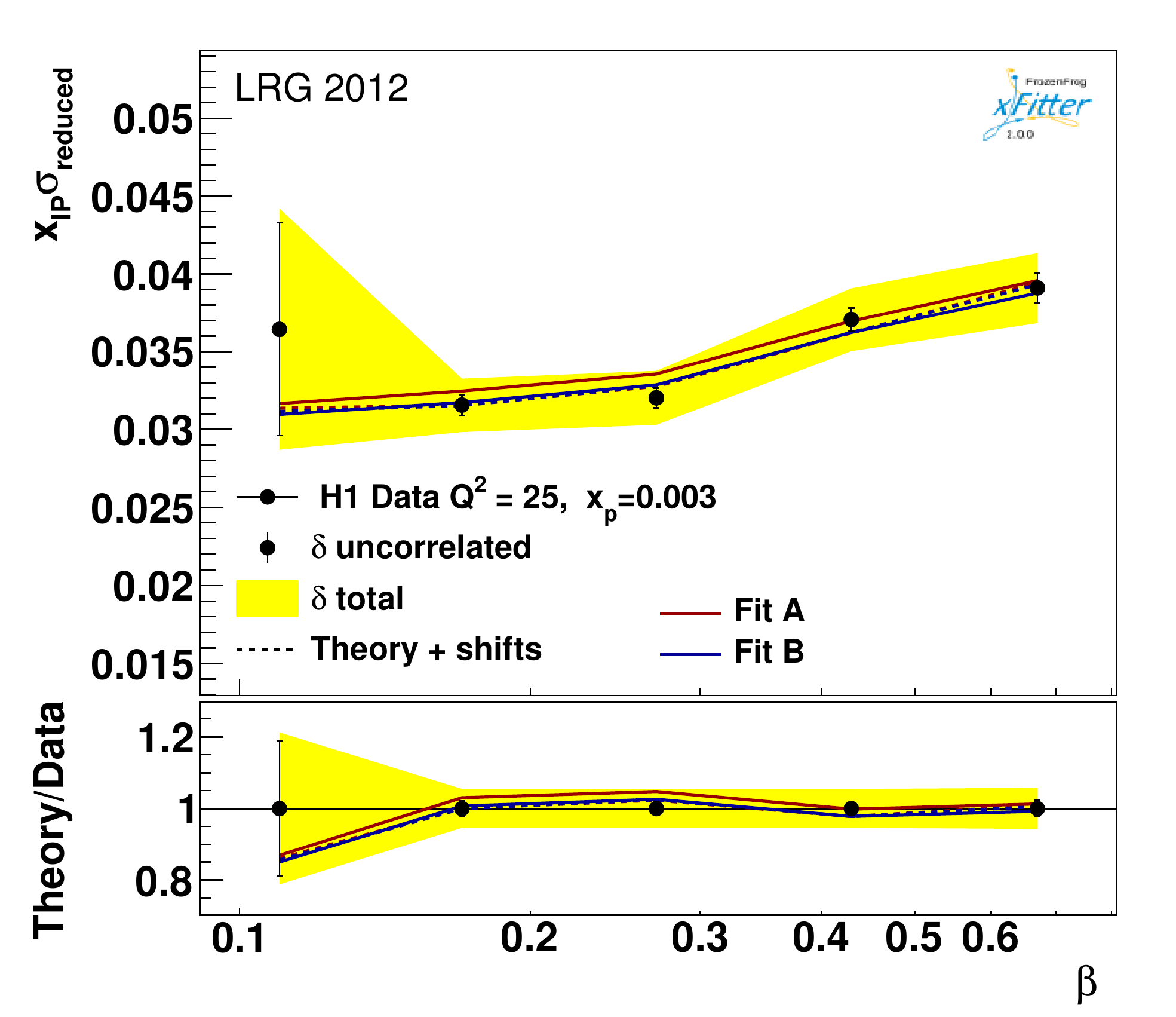}
\includegraphics[clip,width=0.30\textwidth]{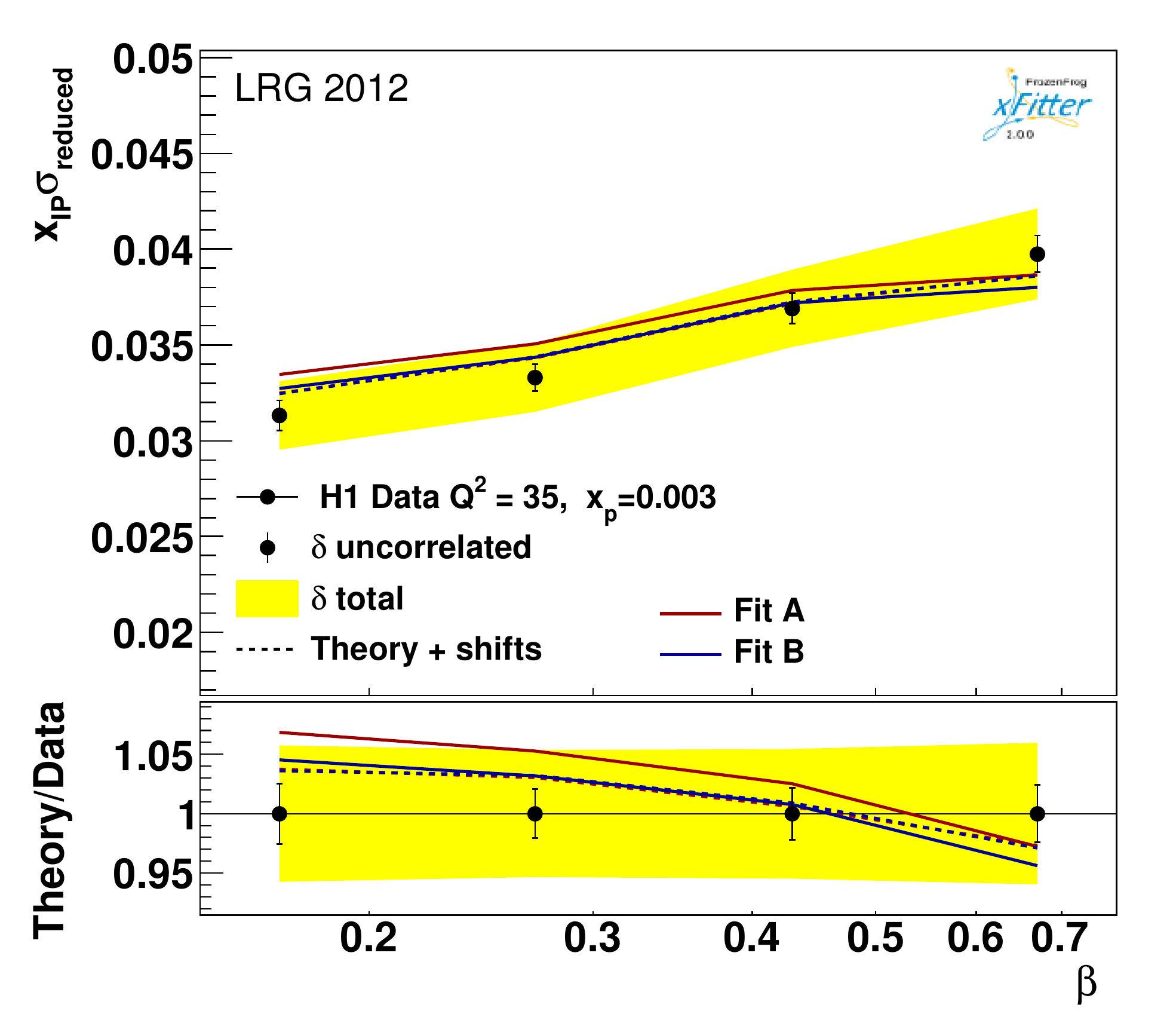}
\includegraphics[clip,width=0.30\textwidth]{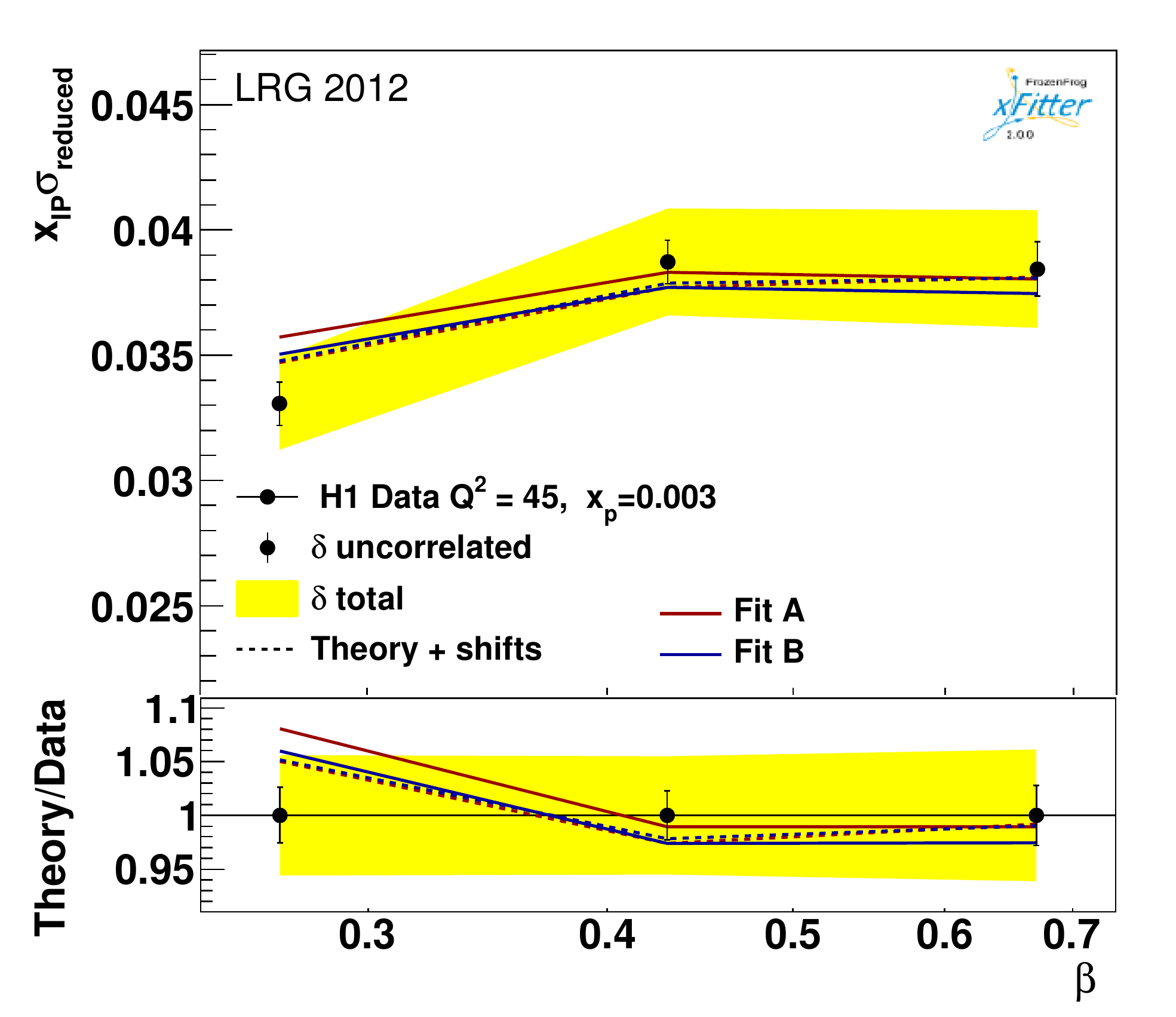}
\includegraphics[clip,width=0.30\textwidth]{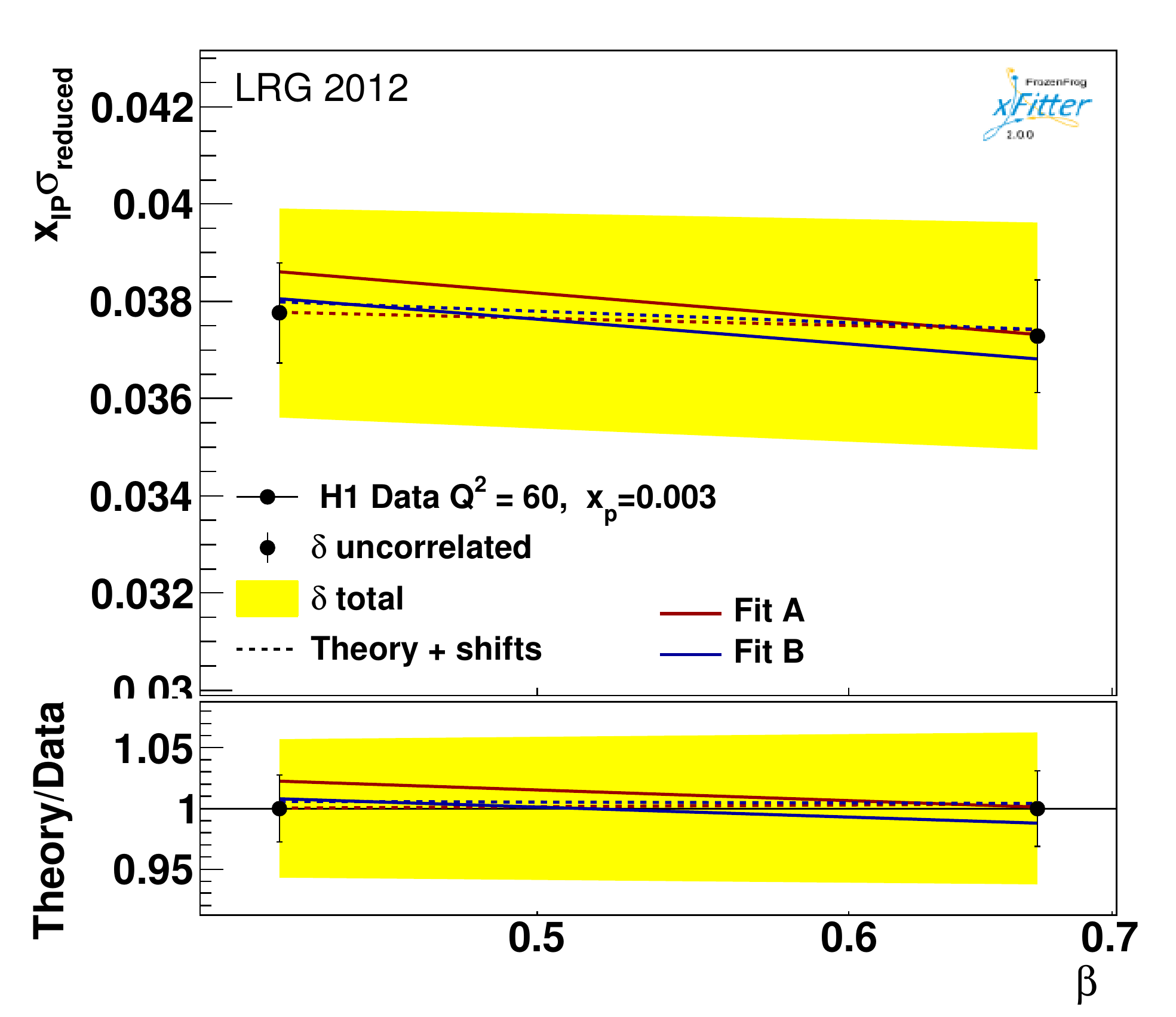}
\includegraphics[clip,width=0.30\textwidth]{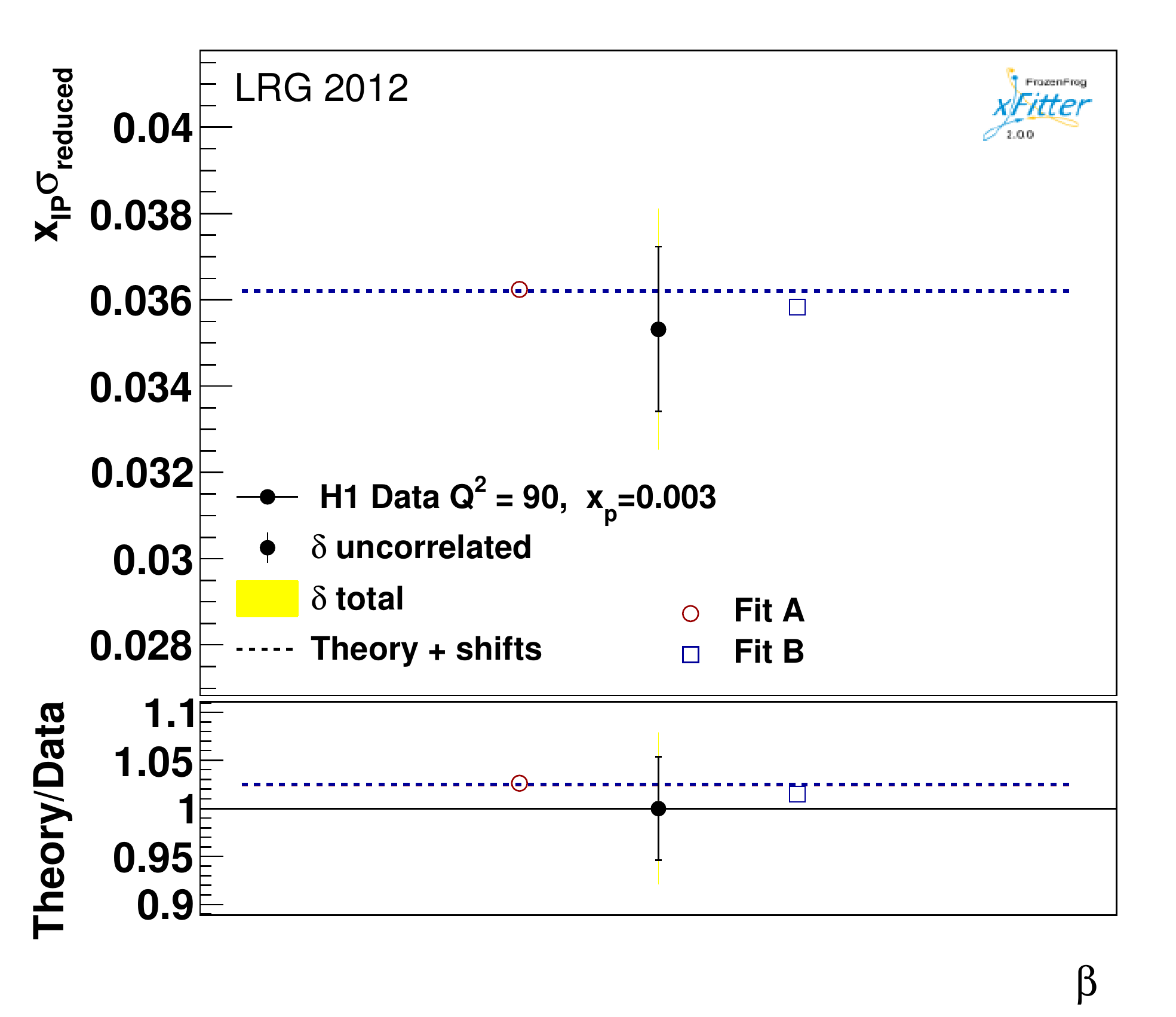}
\begin{center}
\caption{\small The results of our NLO pQCD fit based on {\tt Fit B} for the reduced diffractive cross section $x_{\pom} \sigma_r^{D(3)}$ as a function of $\beta$ for $x_{\pom} = 0.003$ in comparison with H1-LRG-2012 data~\cite{Aaron:2012ad}. See the caption of Fig.~\ref{LRG-2012-xp001} for further details. } \label{LRG-2012-xp003}
\end{center}
\end{figure*}

\begin{figure*}[htb]
\vspace{1.0cm}
\includegraphics[clip,width=0.30\textwidth]{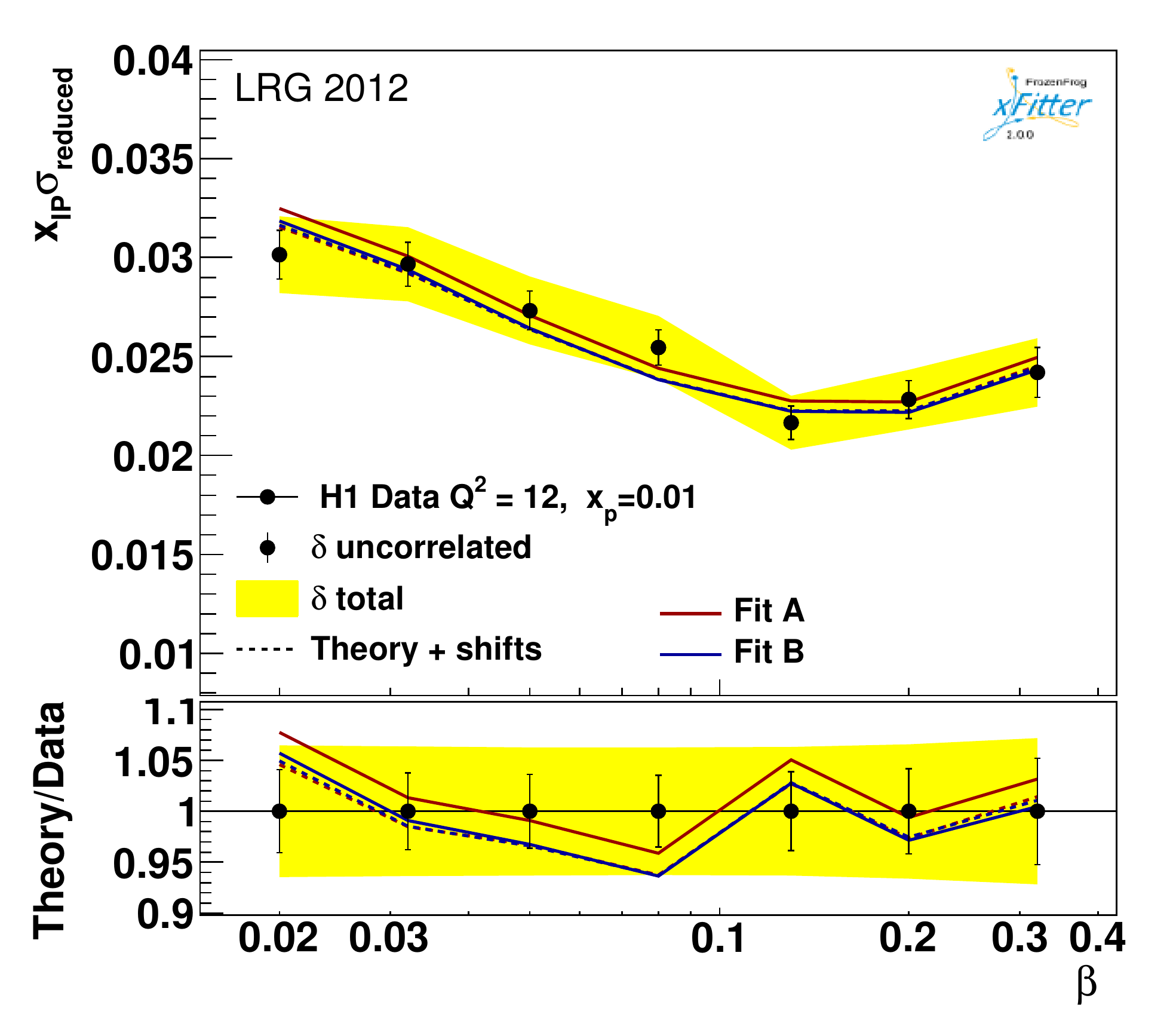}
\includegraphics[clip,width=0.30\textwidth]{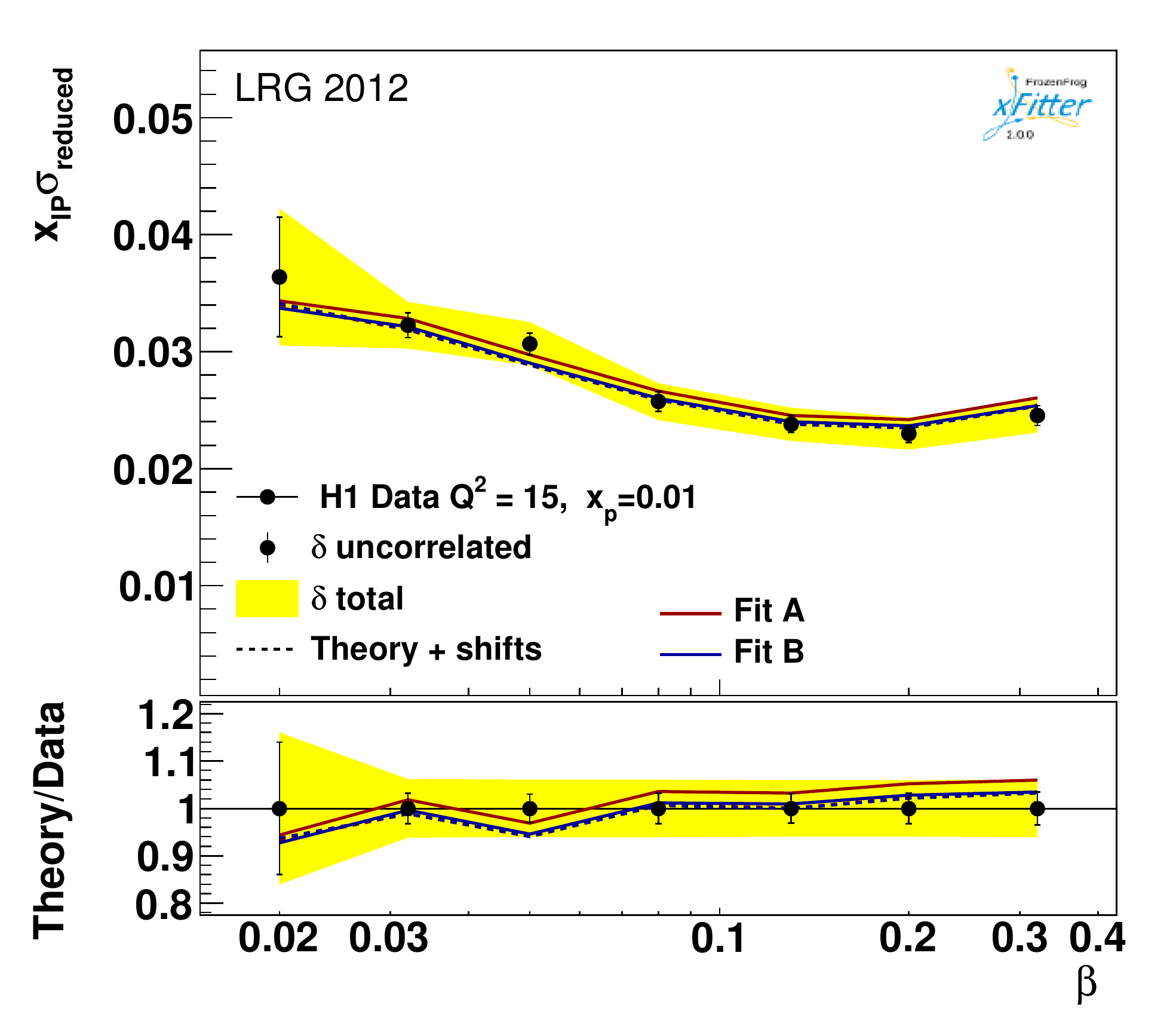}
\includegraphics[clip,width=0.30\textwidth]{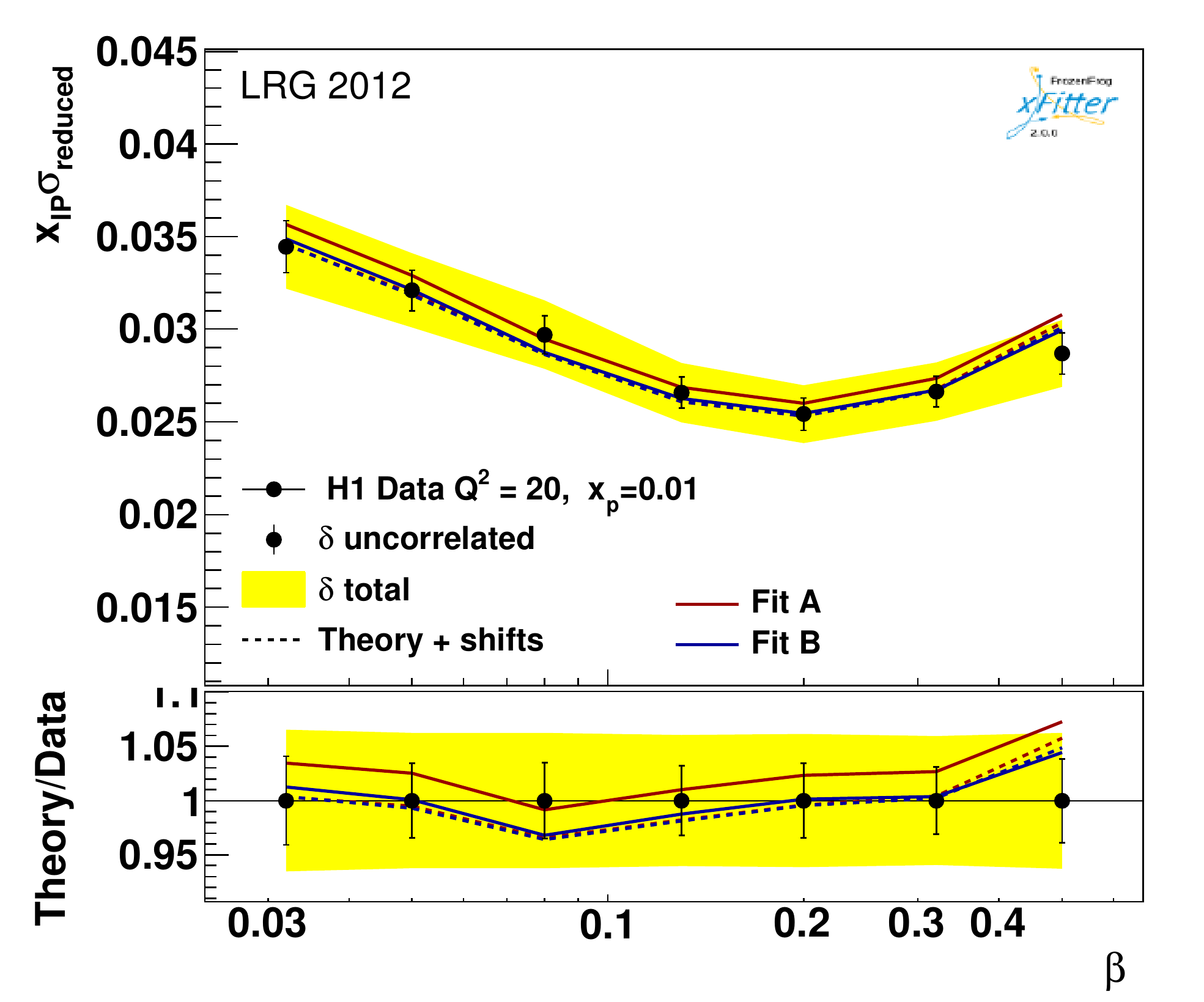}
\includegraphics[clip,width=0.30\textwidth]{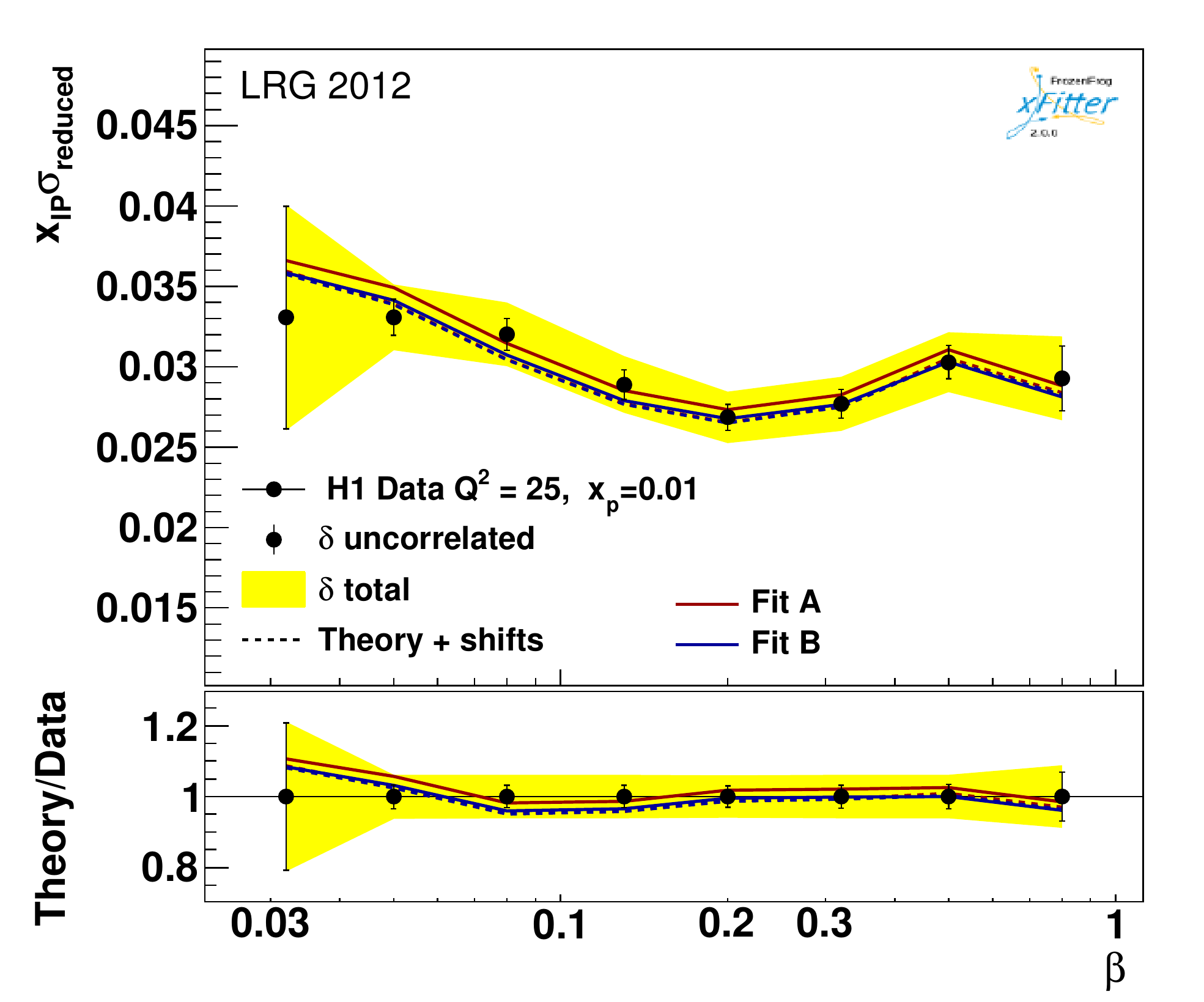}
\includegraphics[clip,width=0.30\textwidth]{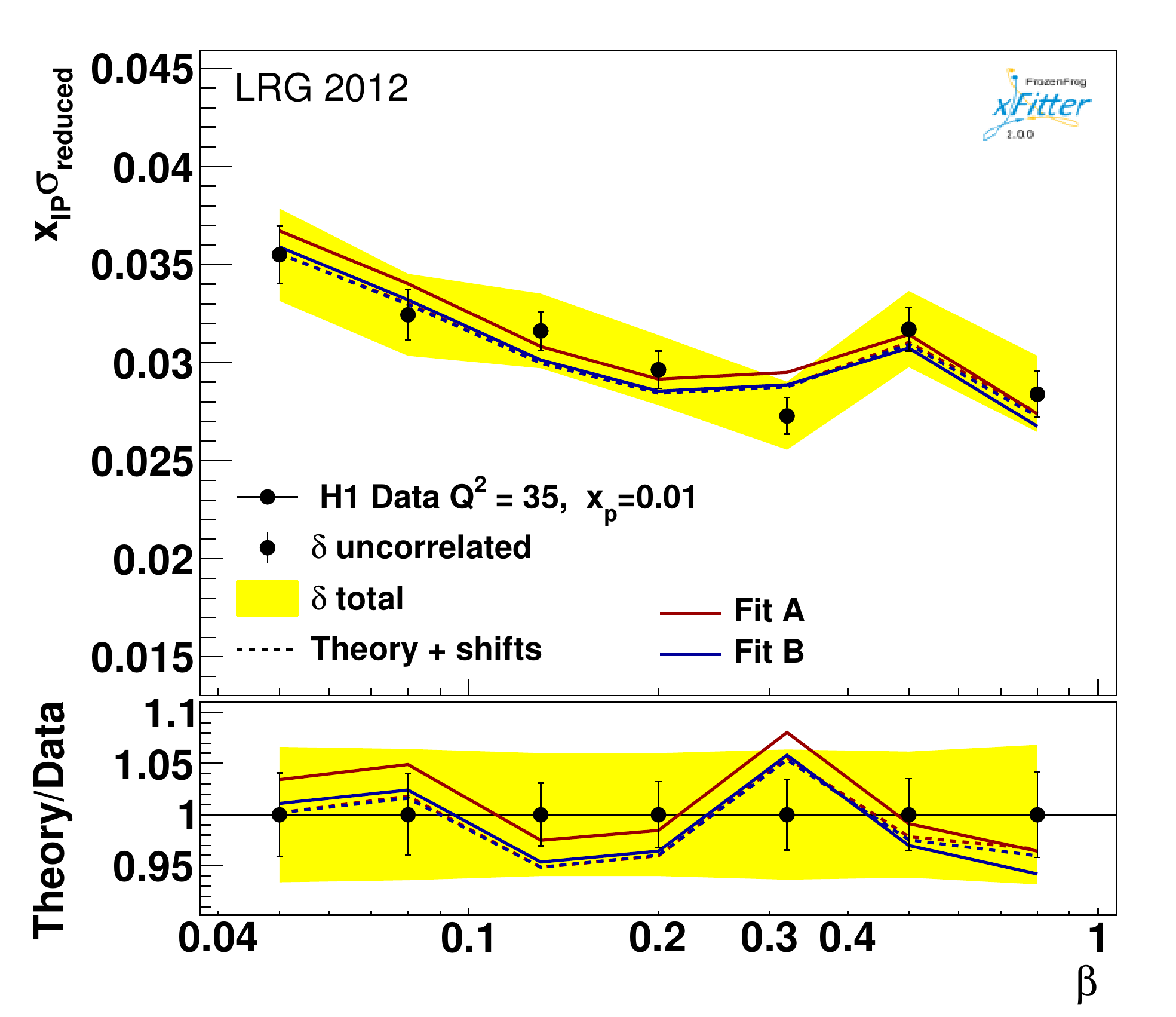}
\includegraphics[clip,width=0.30\textwidth]{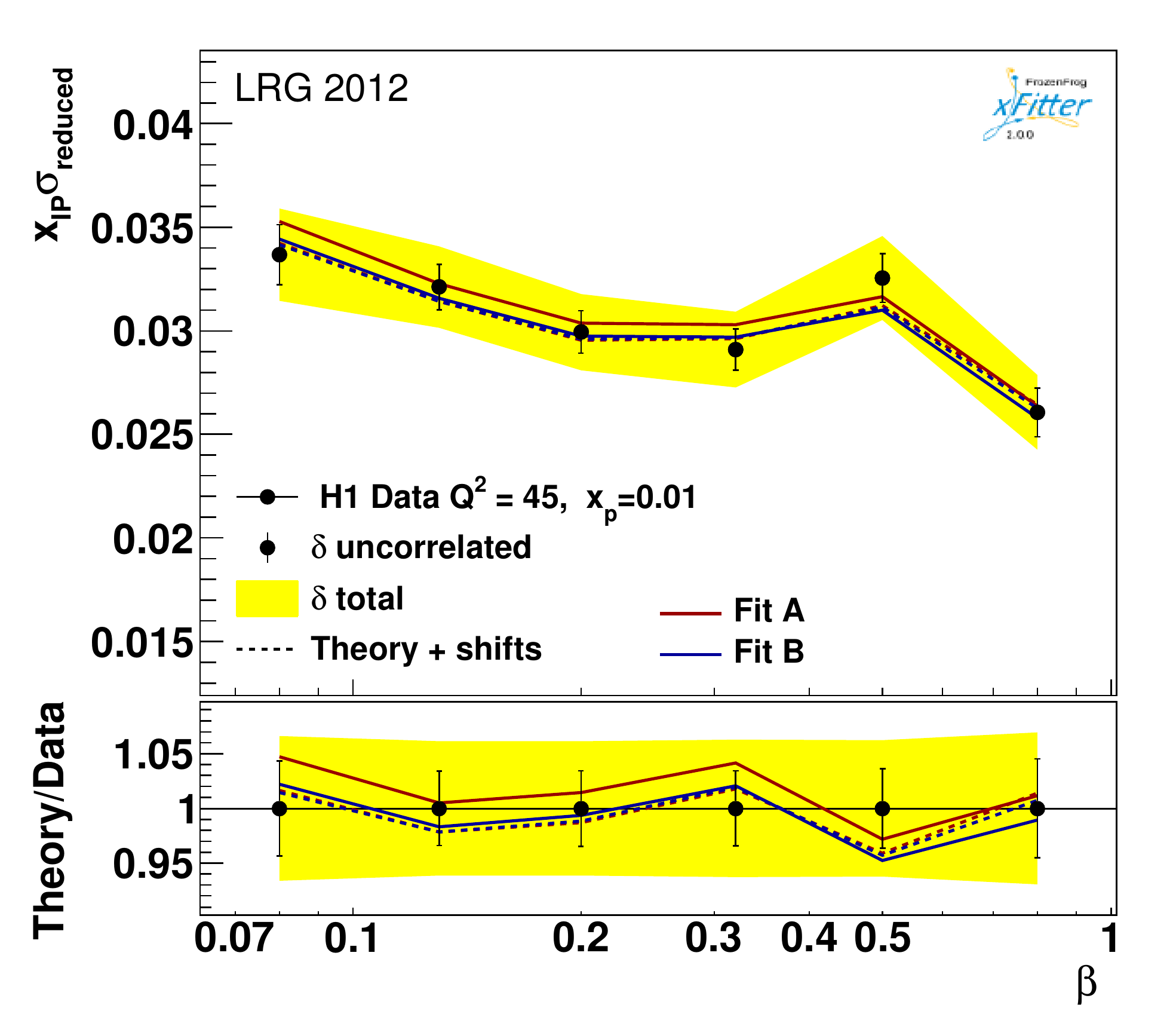}
\includegraphics[clip,width=0.30\textwidth]{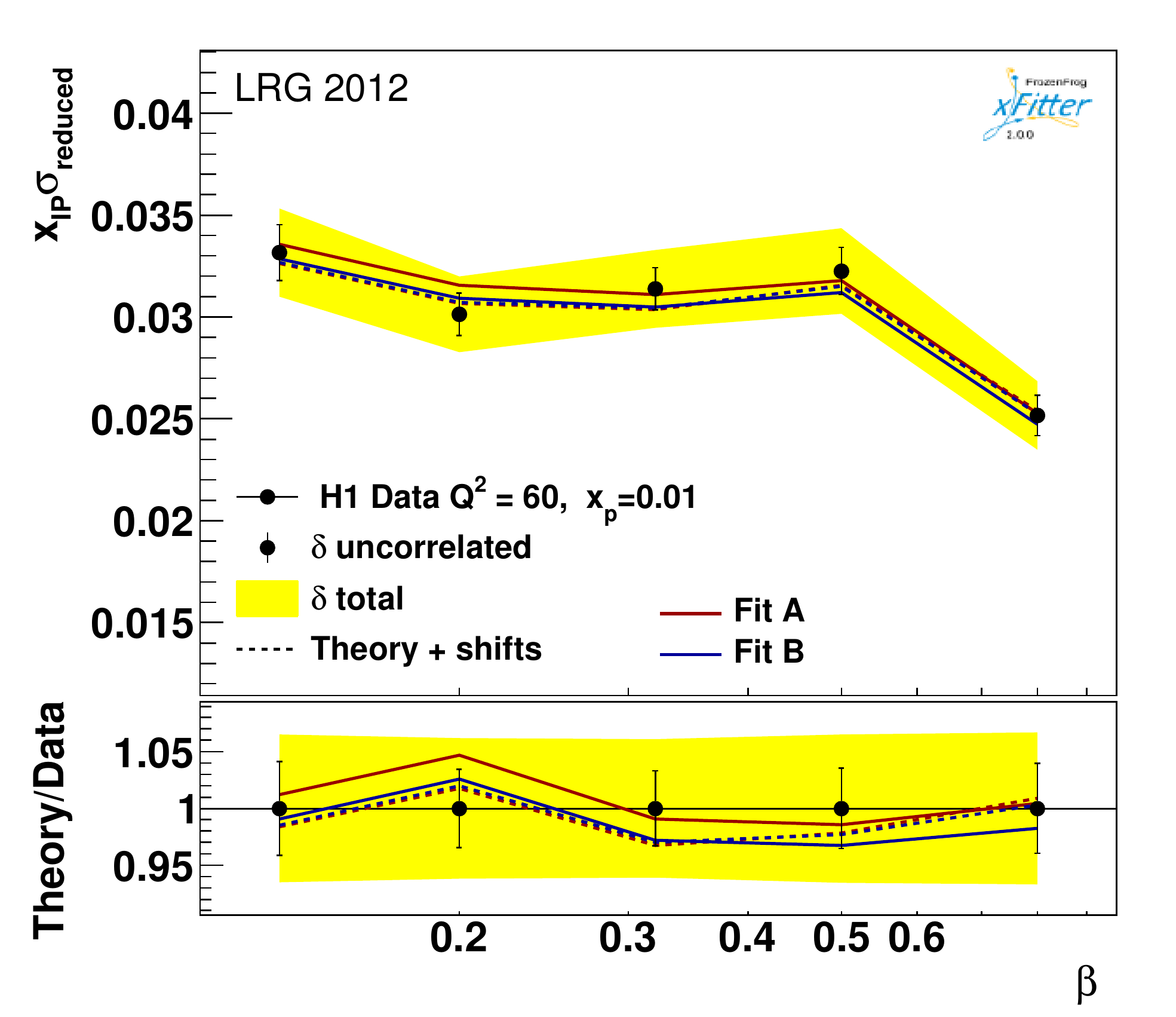}
\includegraphics[clip,width=0.30\textwidth]{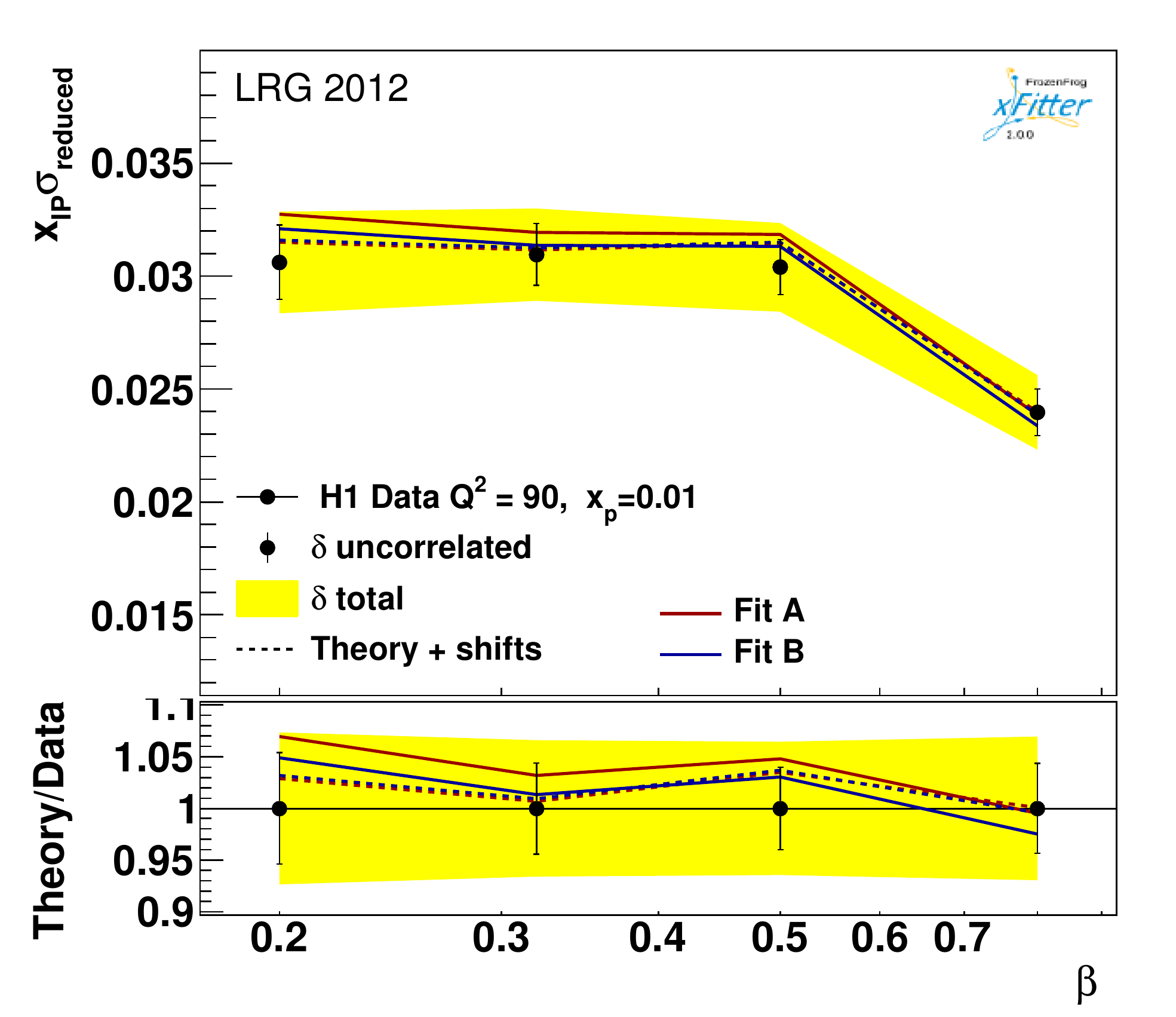}
\includegraphics[clip,width=0.30\textwidth]{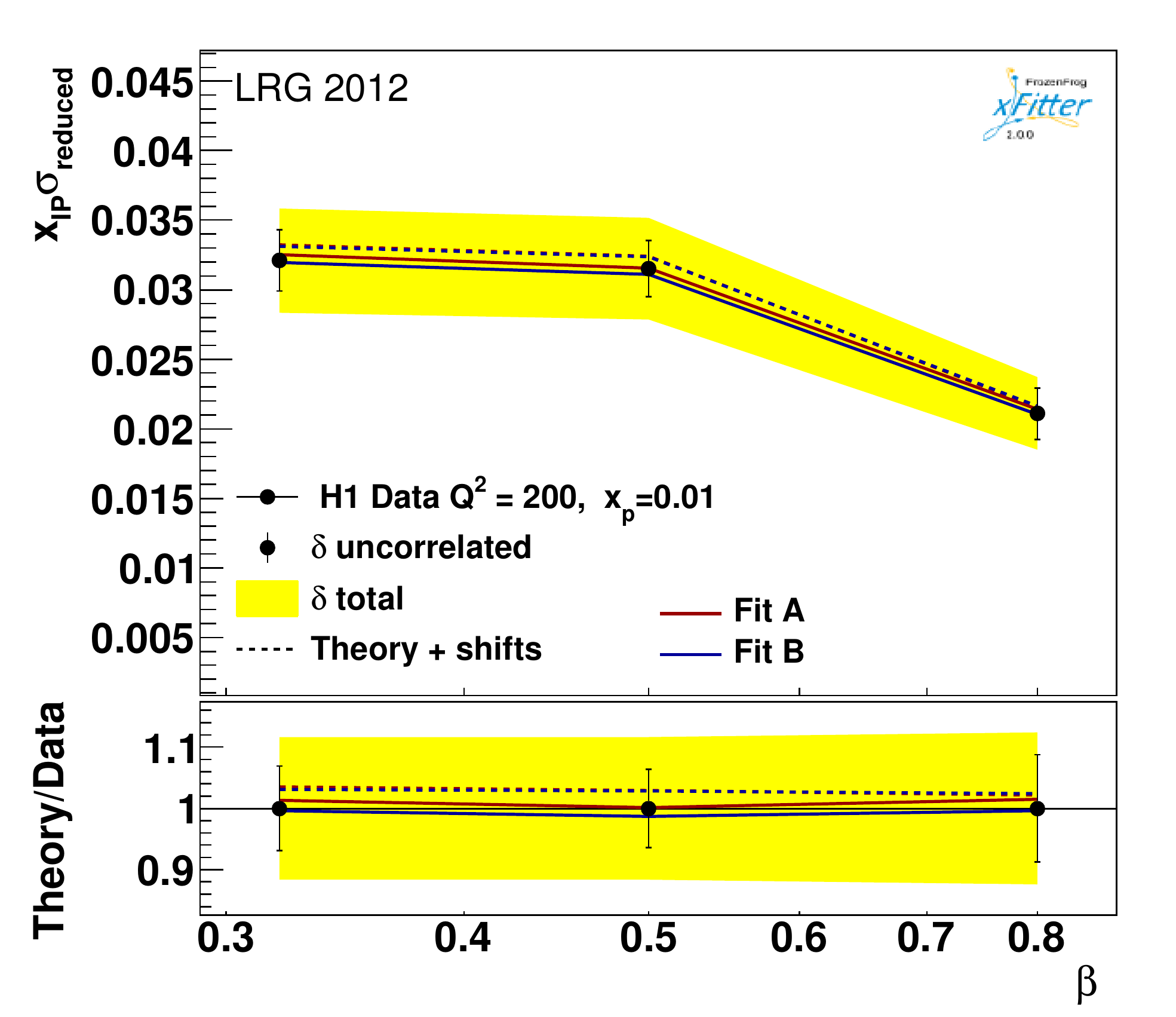}
\includegraphics[clip,width=0.30\textwidth]{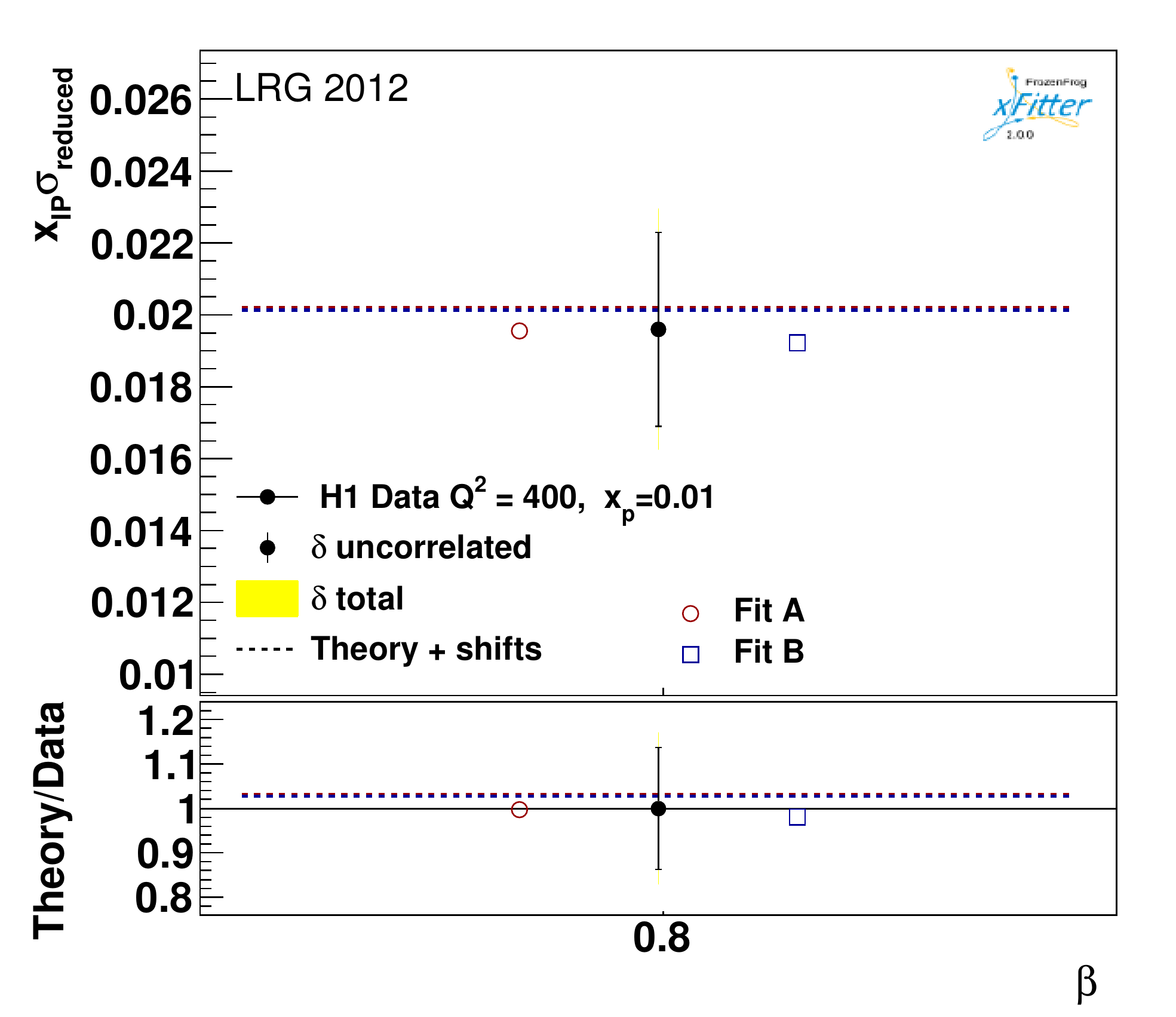}
\begin{center}
\caption{\small The results of our NLO pQCD fit based on {\tt Fit B} for the reduced diffractive cross section $x_{\pom} \sigma_r^{D(3)}$ as a function of $\beta$ for $x_{\pom} = 0.01$ in comparison with H1-LRG-2012 data~\cite{Aaron:2012ad}. See the caption of Fig.~\ref{LRG-2012-xp001} for further details. } \label{LRG-2012-xp01}
\end{center}
\end{figure*}

\begin{figure*}[htb]
\vspace{1.0cm}
\includegraphics[clip,width=0.30\textwidth]{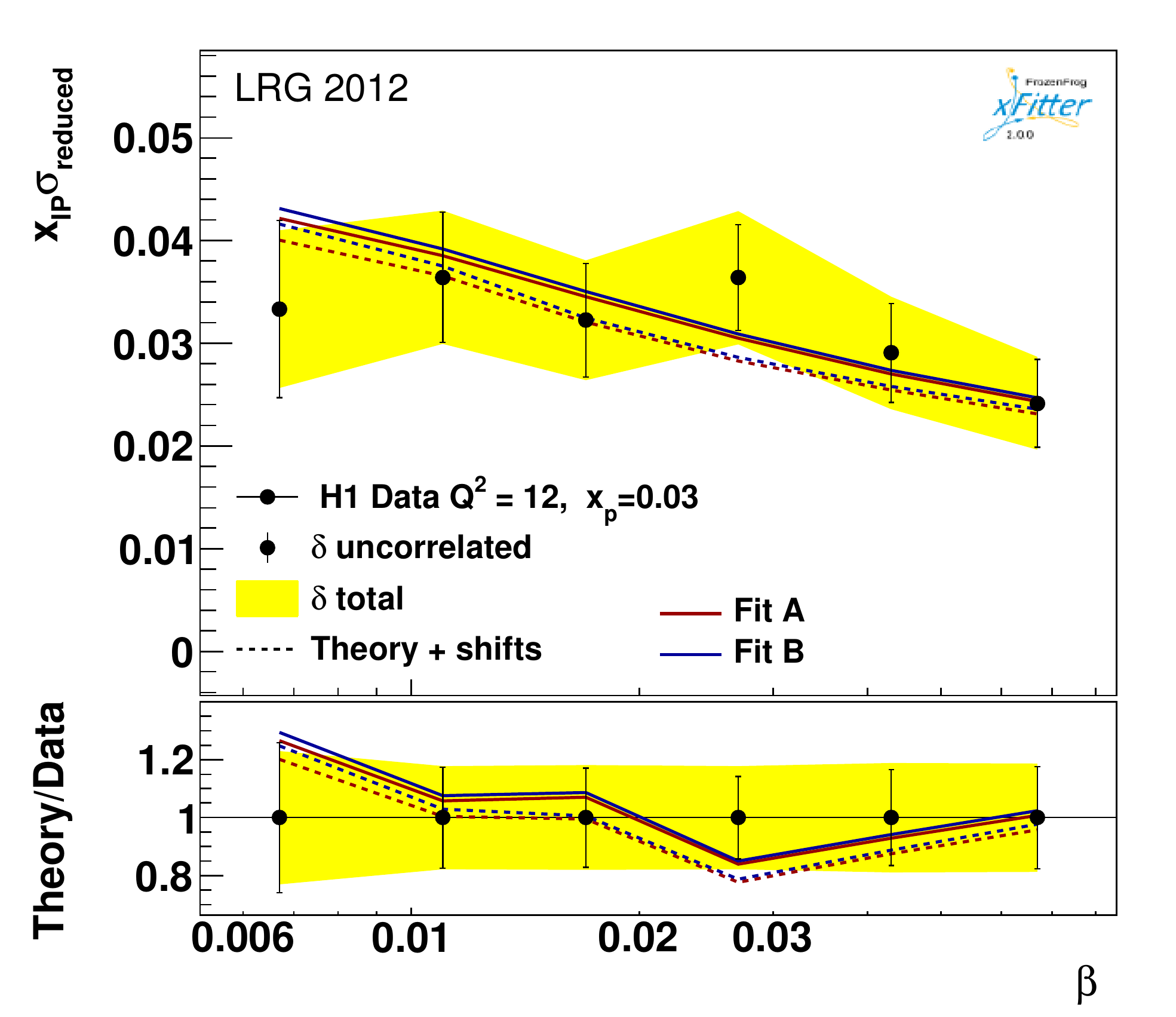}
\includegraphics[clip,width=0.30\textwidth]{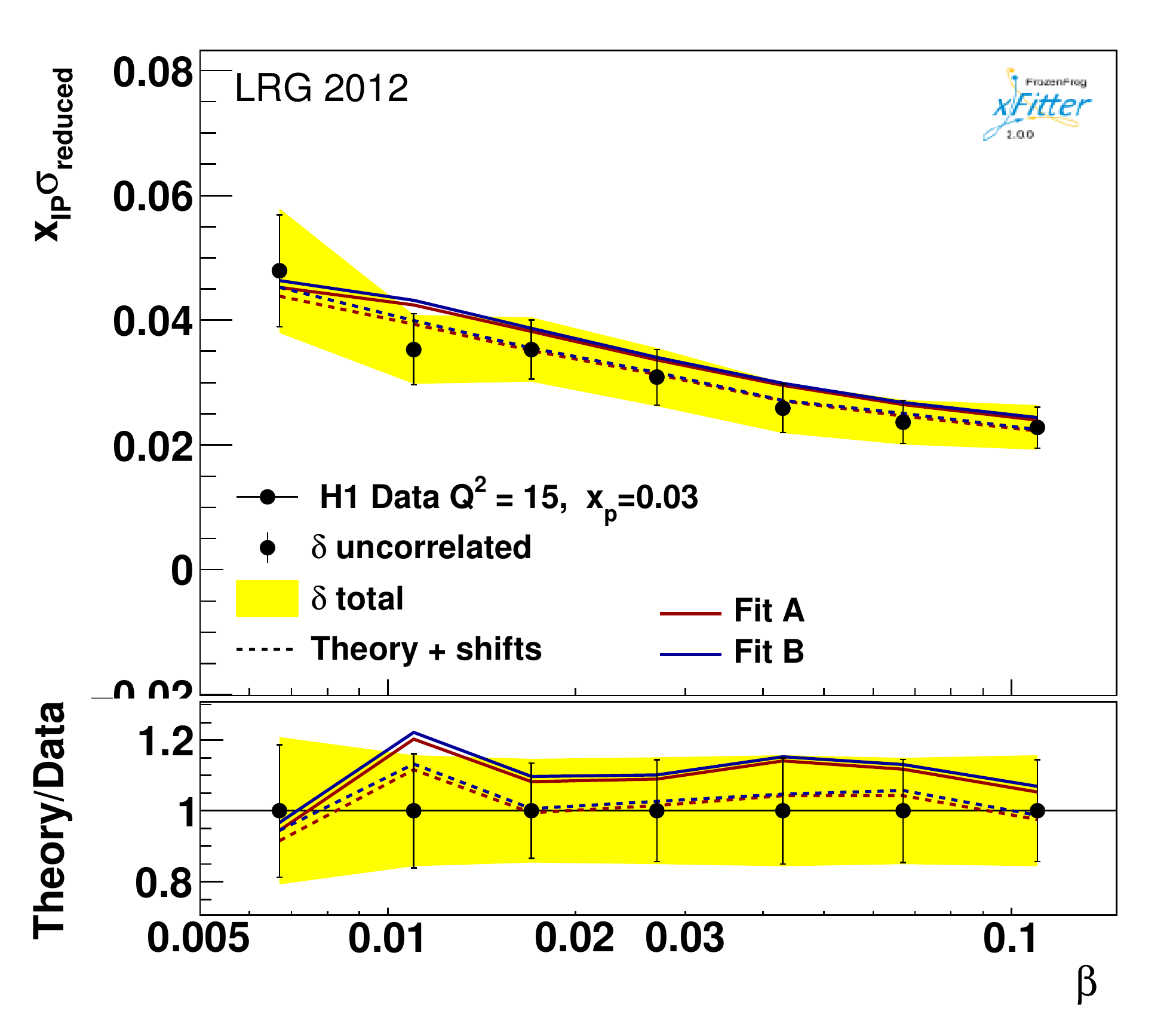}
\includegraphics[clip,width=0.30\textwidth]{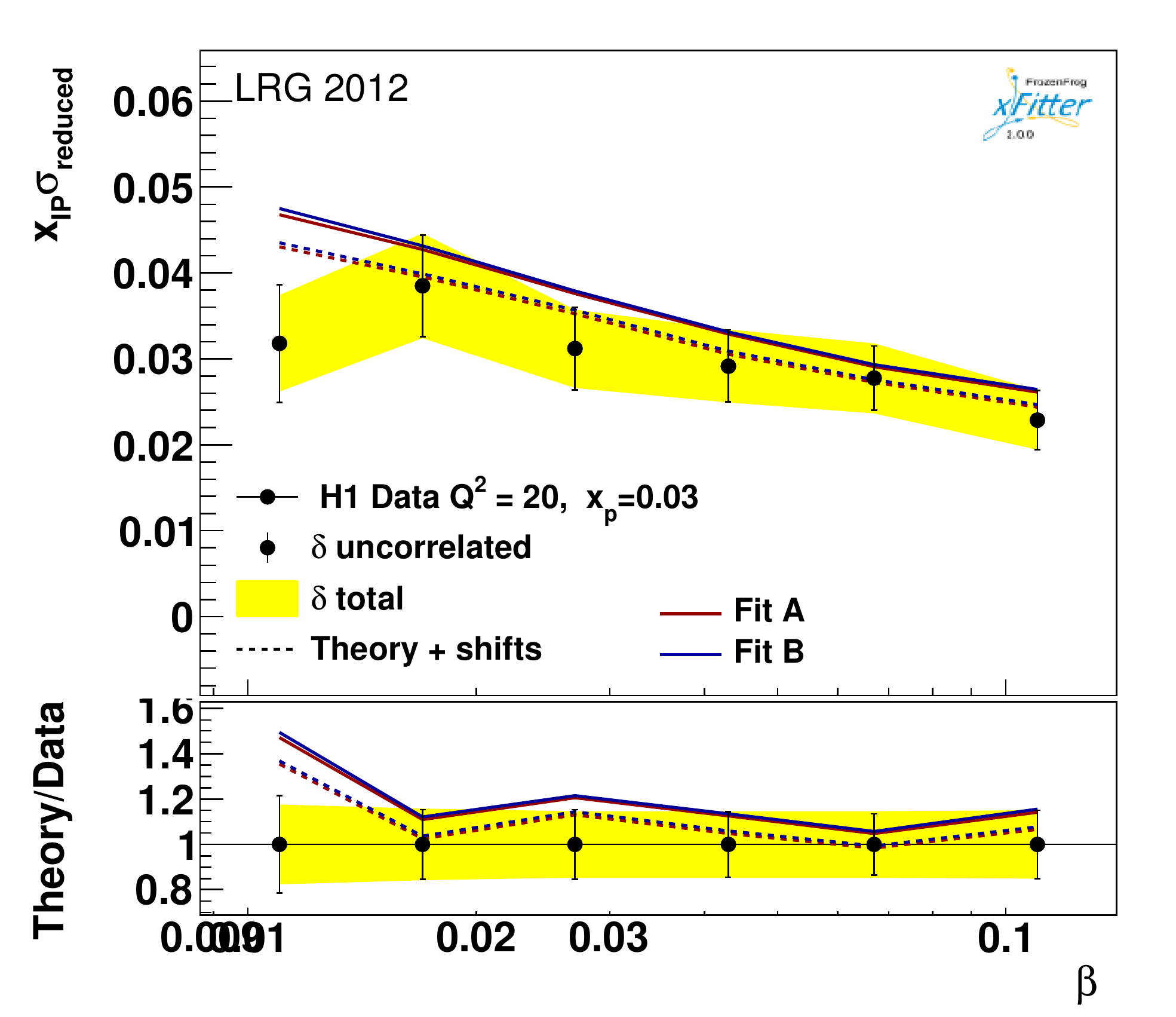}
\includegraphics[clip,width=0.30\textwidth]{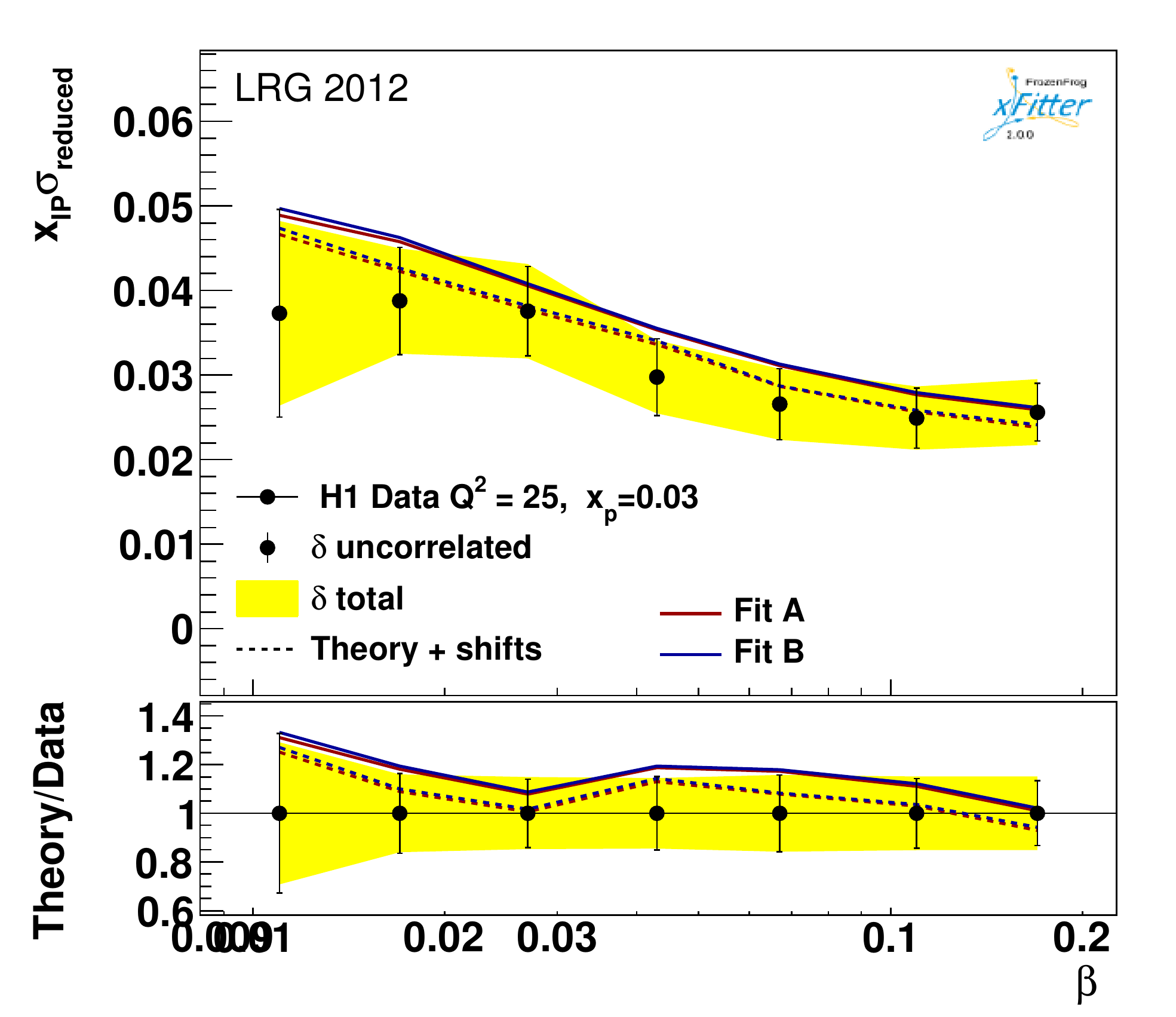}
\includegraphics[clip,width=0.30\textwidth]{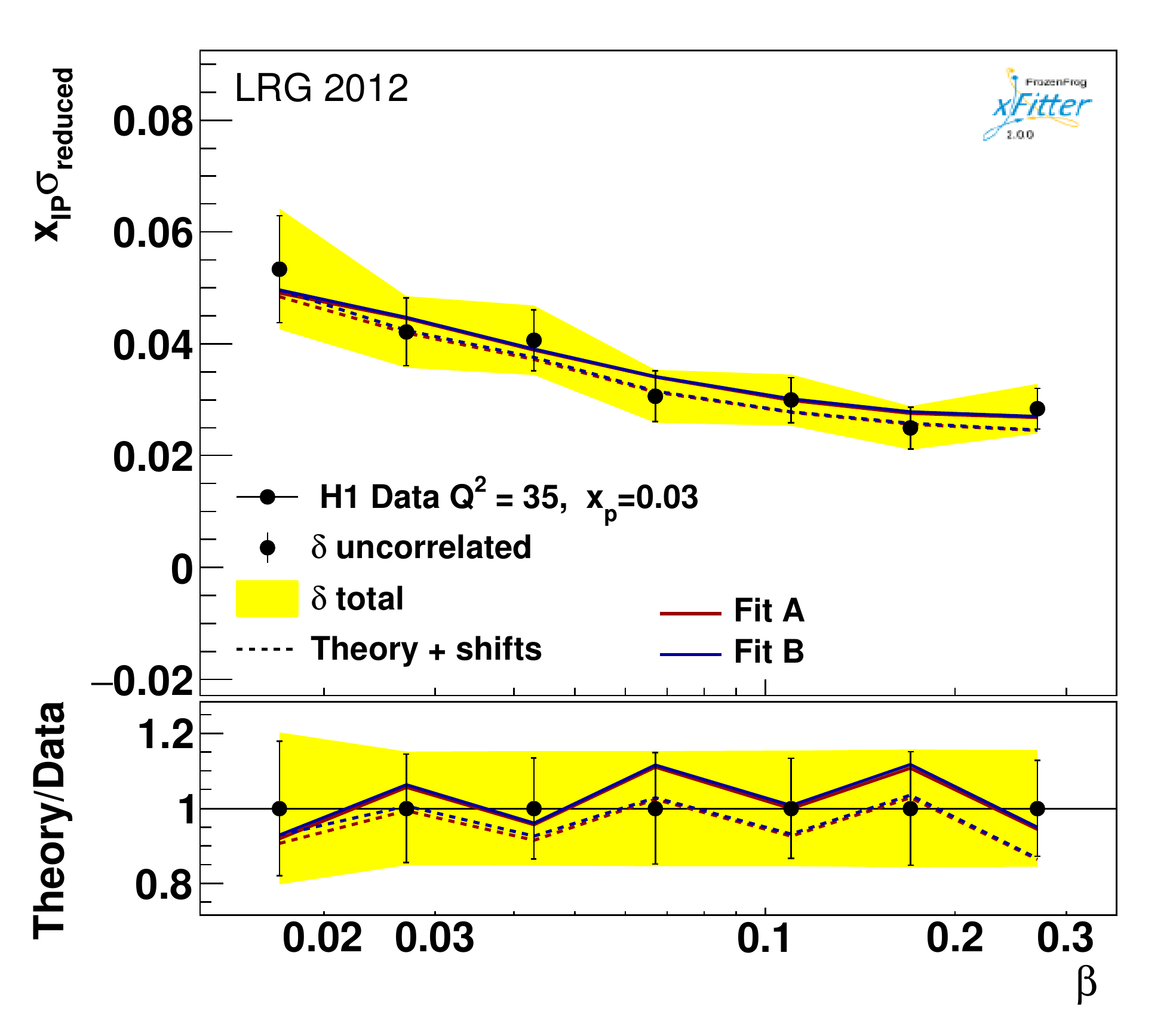}
\includegraphics[clip,width=0.30\textwidth]{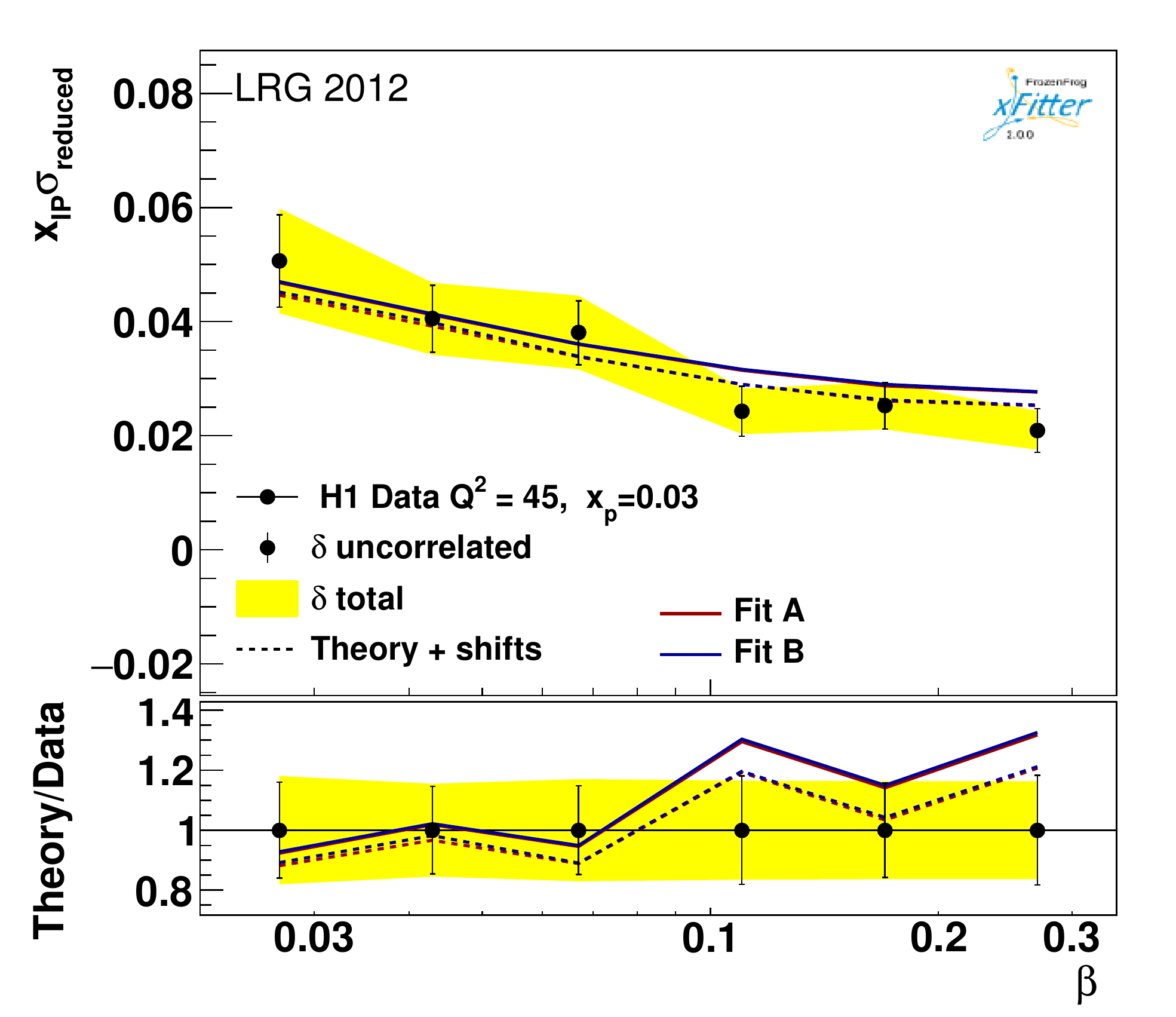}
\includegraphics[clip,width=0.30\textwidth]{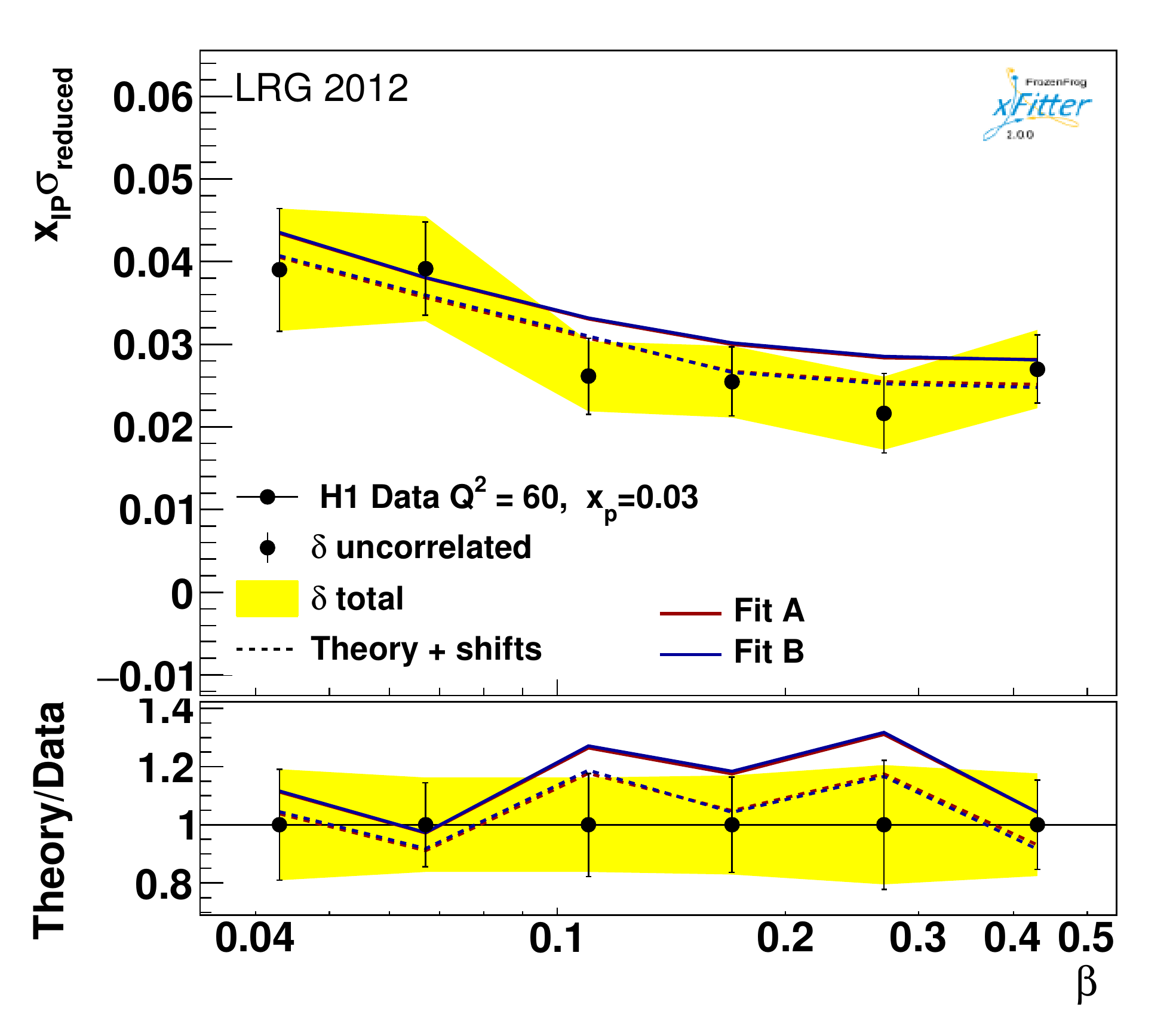}
\includegraphics[clip,width=0.30\textwidth]{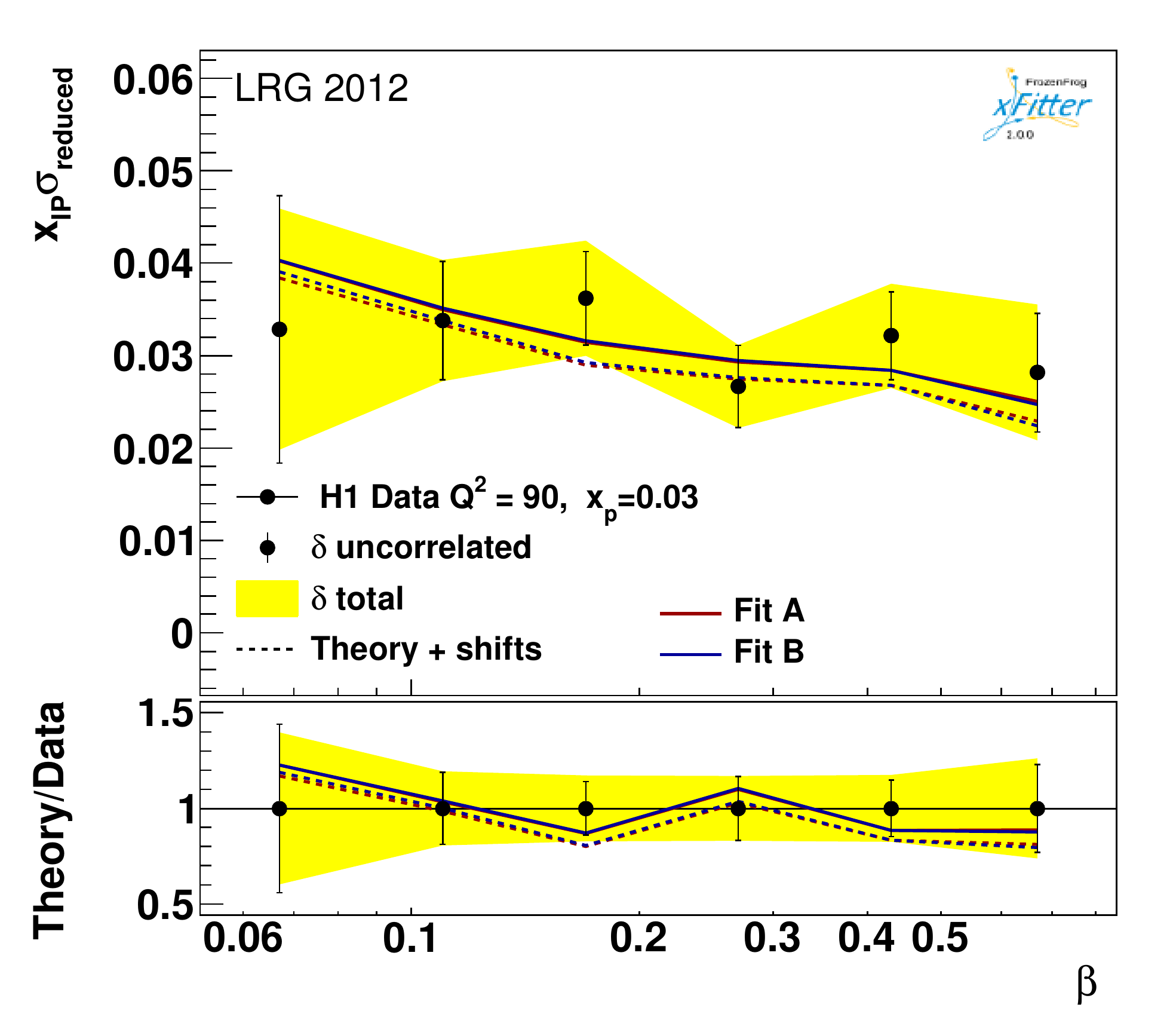}
\includegraphics[clip,width=0.30\textwidth]{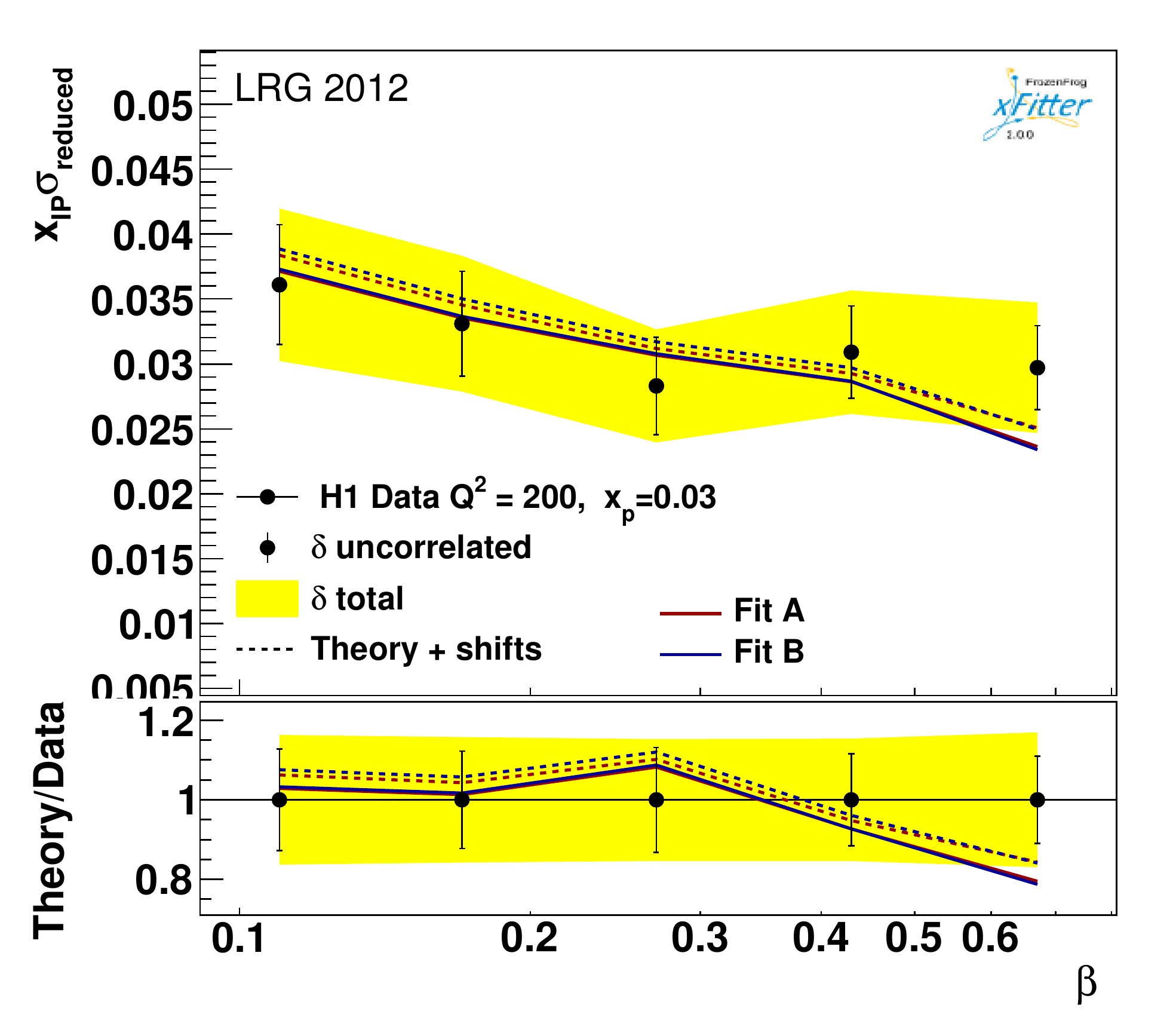}
\includegraphics[clip,width=0.30\textwidth]{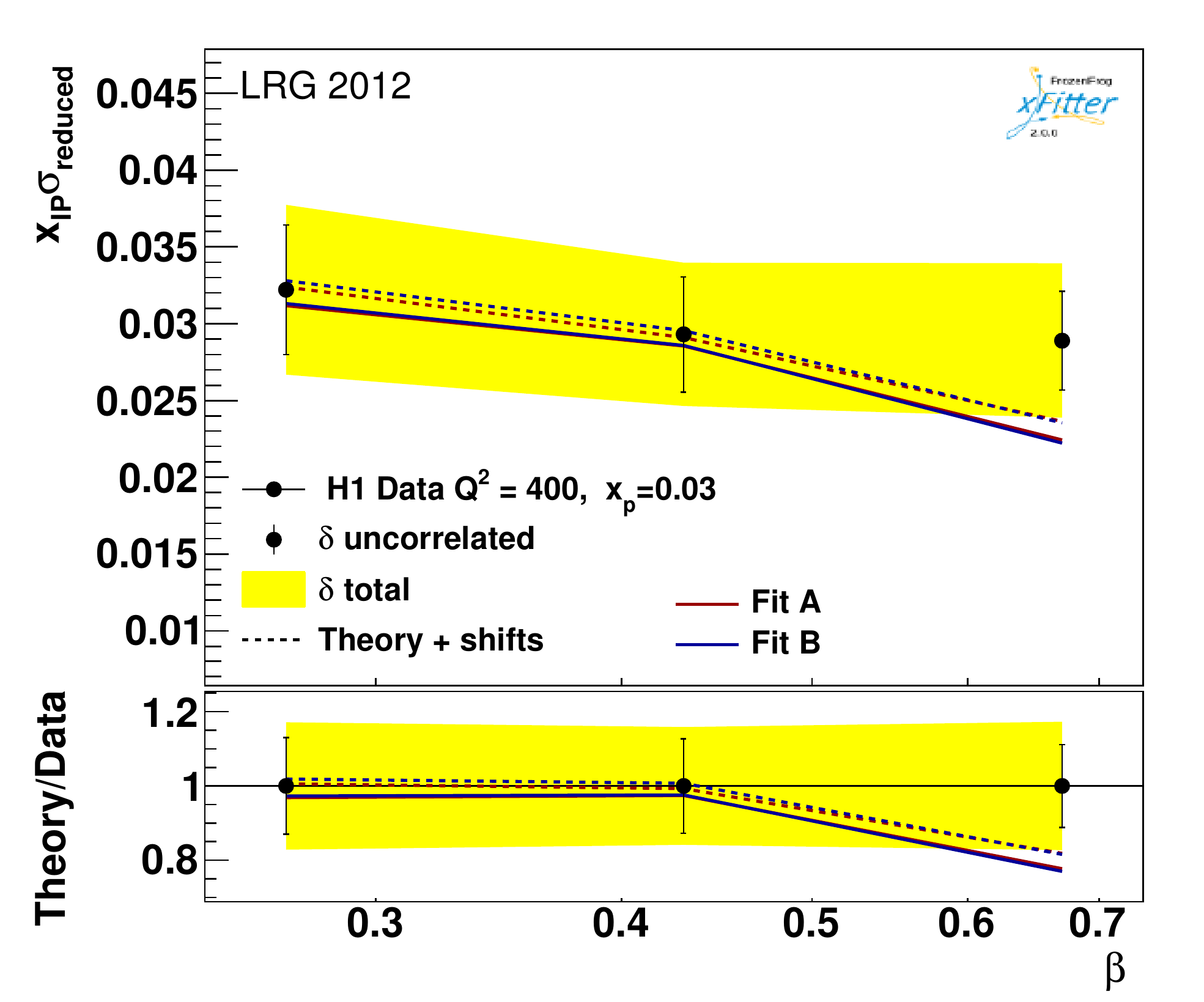}
\includegraphics[clip,width=0.30\textwidth]{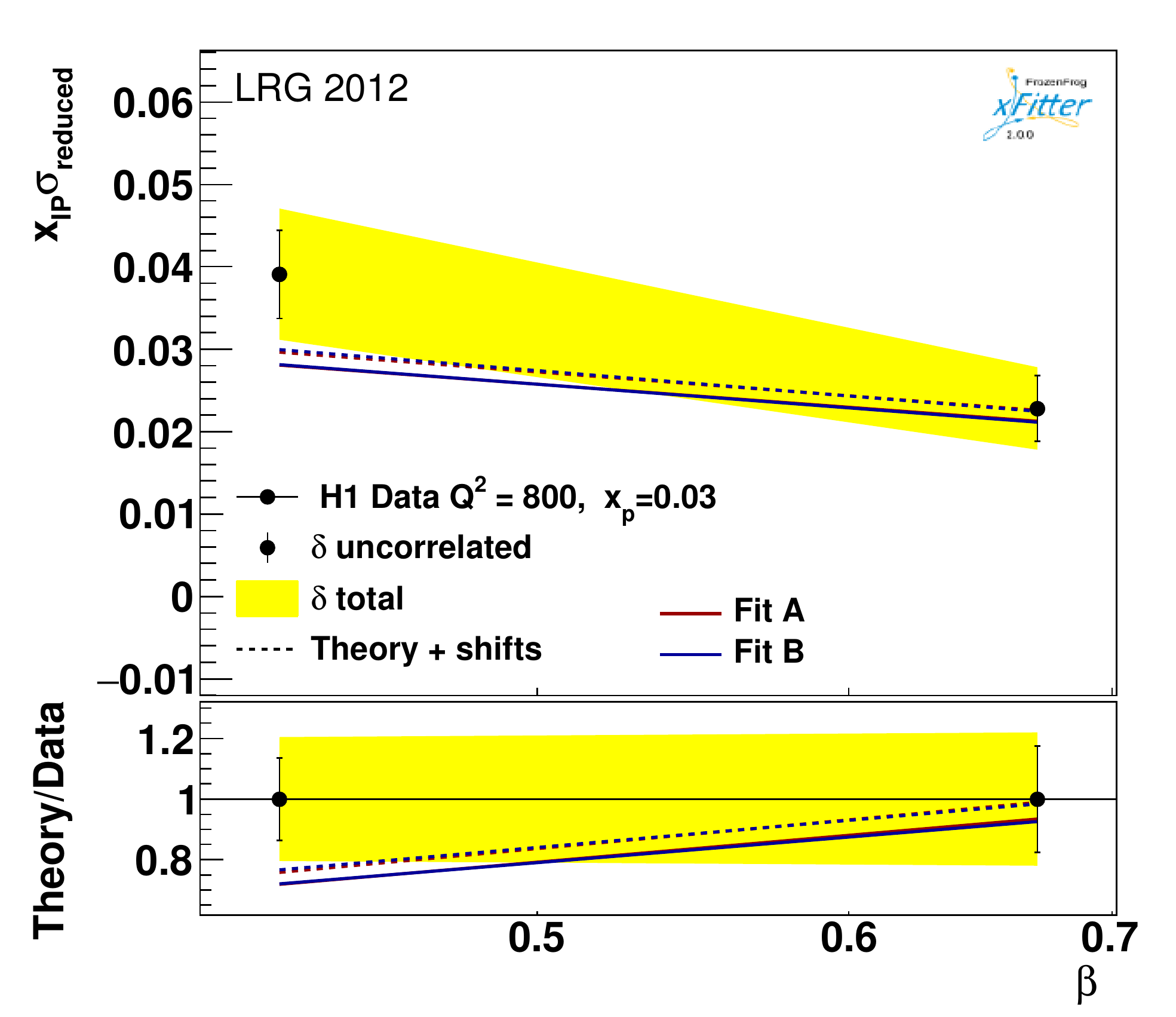}
\includegraphics[clip,width=0.30\textwidth]{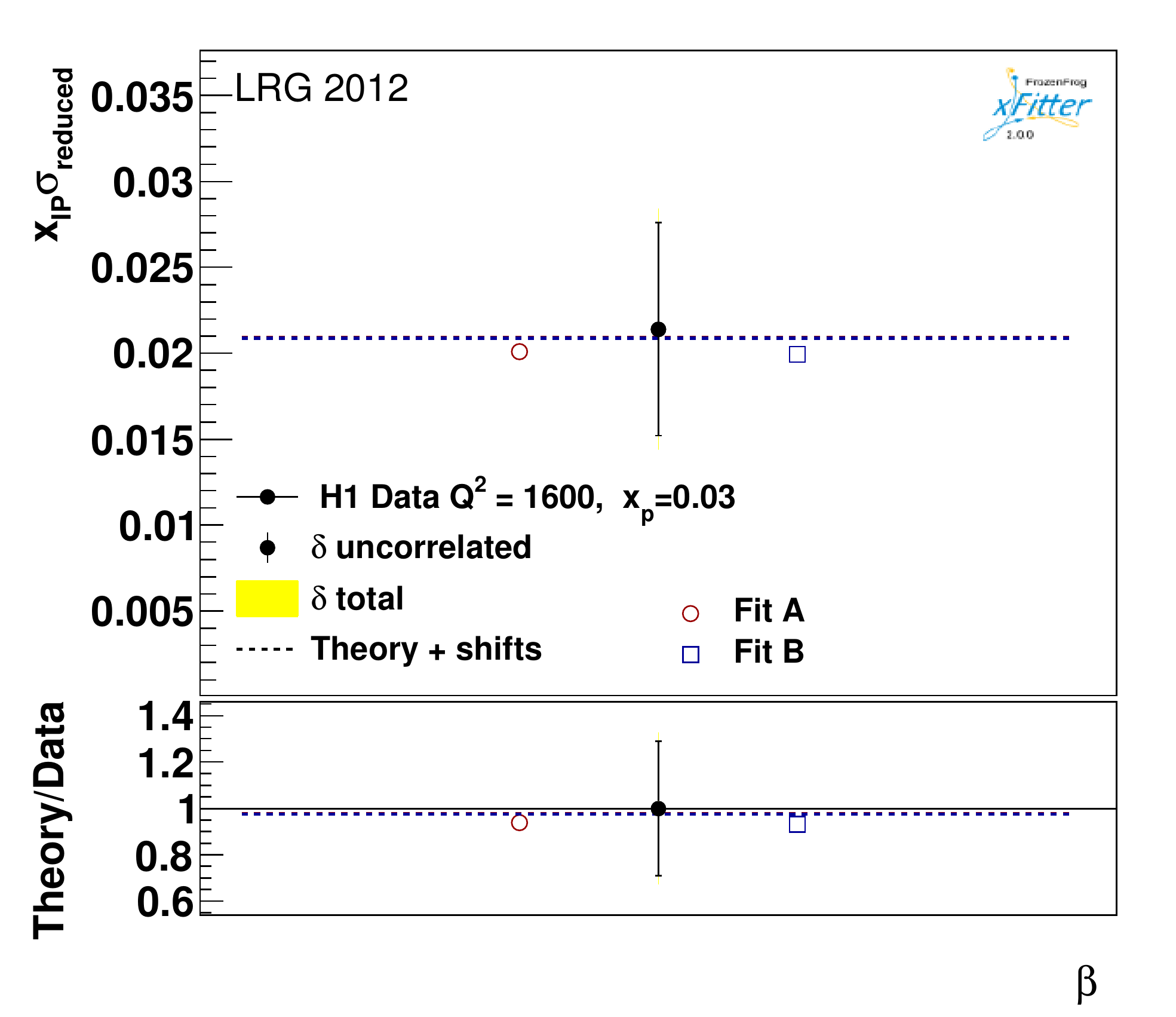}
\begin{center}
\caption{\small The results of our NLO pQCD fit based on {\tt Fit B} for the reduced diffractive cross section $x_{\pom} \sigma_r^{D(3)}$ as a function of $\beta$ for $x_{\pom} = 0.03$ in comparison with H1-LRG-2012 data~\cite{Aaron:2012ad}. See the caption of Fig.~\ref{LRG-2012-xp001} for further details.  } \label{LRG-2012-xp03}
\end{center}
\end{figure*}

In the following, using the results of  {\tt Fit A} and {\tt Fit B}, we compare the reduced diffractive cross section $x_{\pom} \sigma_r^{D(3)}$ with the H1-LRG-2012 and H1-LRG-2011 data sets. The plots have been shown as a function of $\beta$ for different values of $Q^2$ and $x_{\pom}$. The error bars on the data points represent the uncorrelated uncertainties and the yellow bands represent the total uncorrelated and correlated uncertainties.  
For comparison with H1-LRG-2012 diffractive DIS data and for a more detailed study of the $x_{\pom}$ dependence, we have presented our NLO pQCD results for the reduced diffractive cross section $x_{\pom} \sigma_r^{D(3)}$ in Figs.~\ref{LRG-2012-xp001}, \ref{LRG-2012-xp003}, \ref{LRG-2012-xp01} and \ref{LRG-2012-xp03} for
$x_{\pom} = 0.001, 0.003, 0.01$ and $0.03$, respectively. These data have been compared with the results of our analyses {\tt Fit A} and  {\tt Fit B} presented in section.~\ref{sec:results}. As can be seen, the results of our pQCD fits are in good agreement with the experimental data at all values of $x_{\pom}$.

\begin{figure*}[htb]
\vspace{1.0cm}
\includegraphics[clip,width=0.80\textwidth]{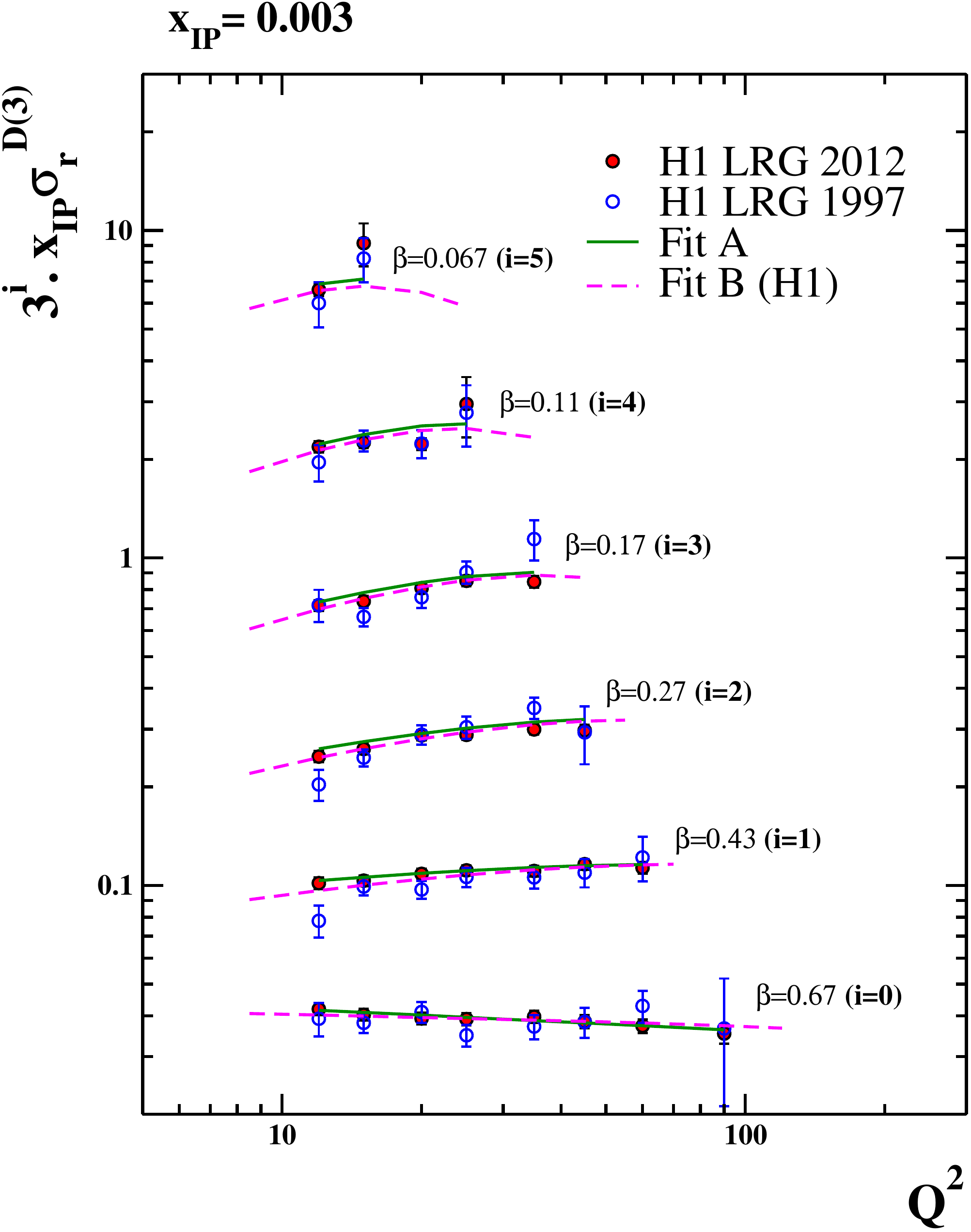}
\begin{center}
\caption{\small The results of our NLO pQCD fit based on {\tt Fit A} for the reduced diffractive cross section $x_{\pom} \sigma_r^{D(3)}$ as a function of $Q^2$ for different values of $\beta$ and $x_{\pom} = 0.003$. The data are correspond to the H1-LRG-2012~\cite{Aaron:2012ad} and H1-LRG-1997~\cite{Aktas:2006hy} measurements. The data are multiplied by a further factor of $3^i$ for visibility, with $i$ as indicated in parentheses. } \label{LRG-2012-xp003-Scaled}
\end{center}
\end{figure*}

\begin{figure*}[htb]
\vspace{1.0cm}
\includegraphics[clip,width=0.80\textwidth]{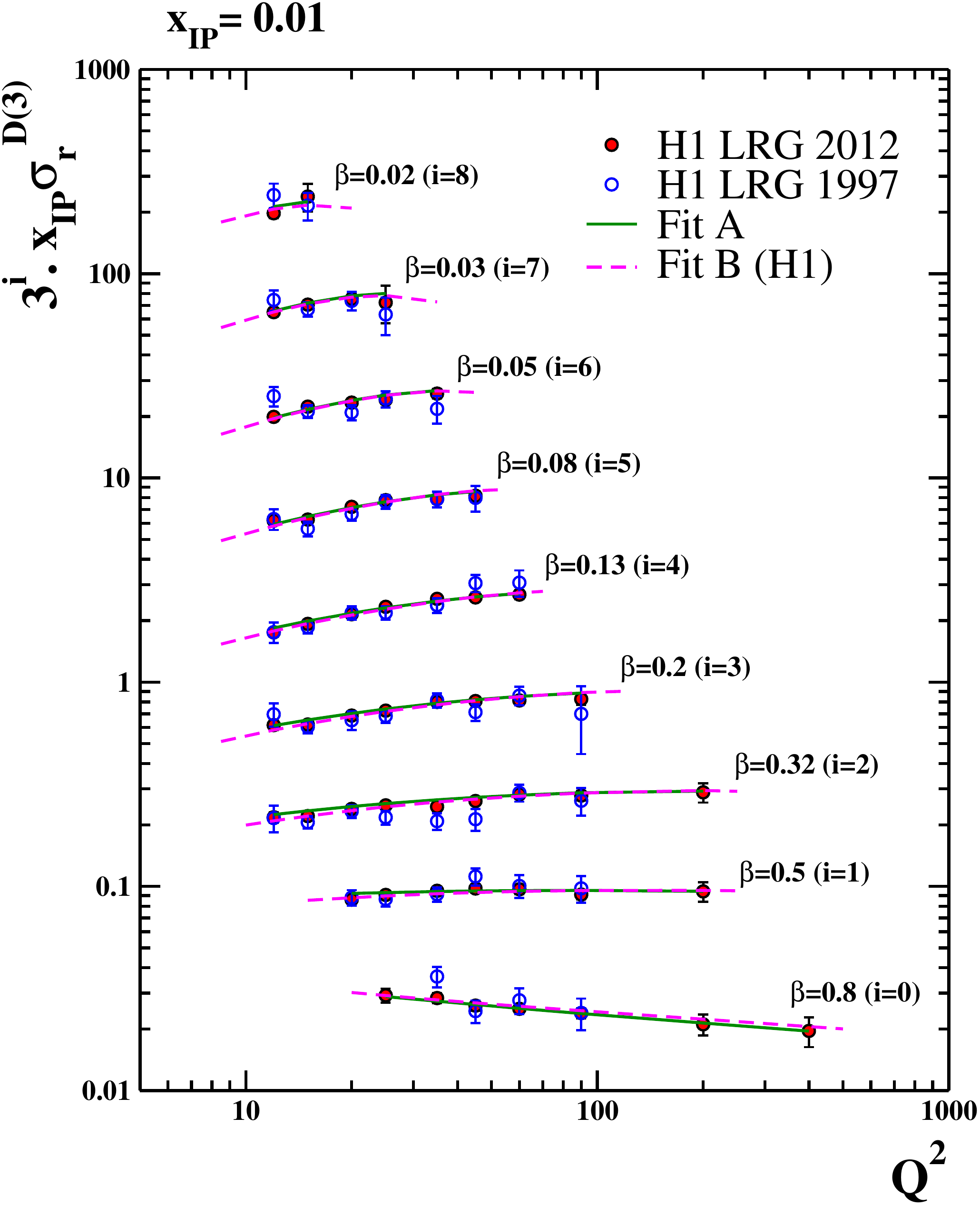}
\begin{center}
\caption{\small The results of our NLO pQCD fit based on {\tt Fit A} for the reduced diffractive cross section $x_{\pom} \sigma_r^{D(3)}$ as a function of $Q^2$ for different values of $\beta$ and $x_{\pom} = 0.01$. The data are correspond to the H1 LRG 2012~\cite{Aaron:2012ad} and H1-LRG-1997~\cite{Aktas:2006hy} measurements. See the caption of Fig.~\ref{LRG-2012-xp003-Scaled} for further details.  } \label{LRG-2012-xp01-Scaled}
\end{center}
\end{figure*}

In Figs.~\ref{LRG-2012-xp003-Scaled} and \ref{LRG-2012-xp01-Scaled}, we present our theory predictions
based on the results of {\tt Fit A} for the reduced diffractive cross section $x_{\pom} \sigma_r^{D(3)}$ as a function of $Q^2$ for different values of $\beta$ and at $x_{\pom} = 0.003$ and $0.01$, respectively. For the comparison, we have also shown in these figures the theory predictions based on the {\tt H1-Fit A} analysis and the old H1-LRG-1997~\cite{Aktas:2006hy} measurements.
The results clearly demonstrate a good agreement between the results of our pQCD fit and the H1-LRG-2012 data used in the analysis as well as the predictions based on the {\tt H1-2006 Fit B}. For the case of $x_{\pom} = 0.003$, as the values of $\beta$ and $Q^2$ are increased, a better agreement between our results and the {\tt H1-2006 Fit B} is observed. 

\begin{figure*}[htb]
\begin{center}
\vspace{0.5cm}
\resizebox{0.480\textwidth}{!}{\includegraphics{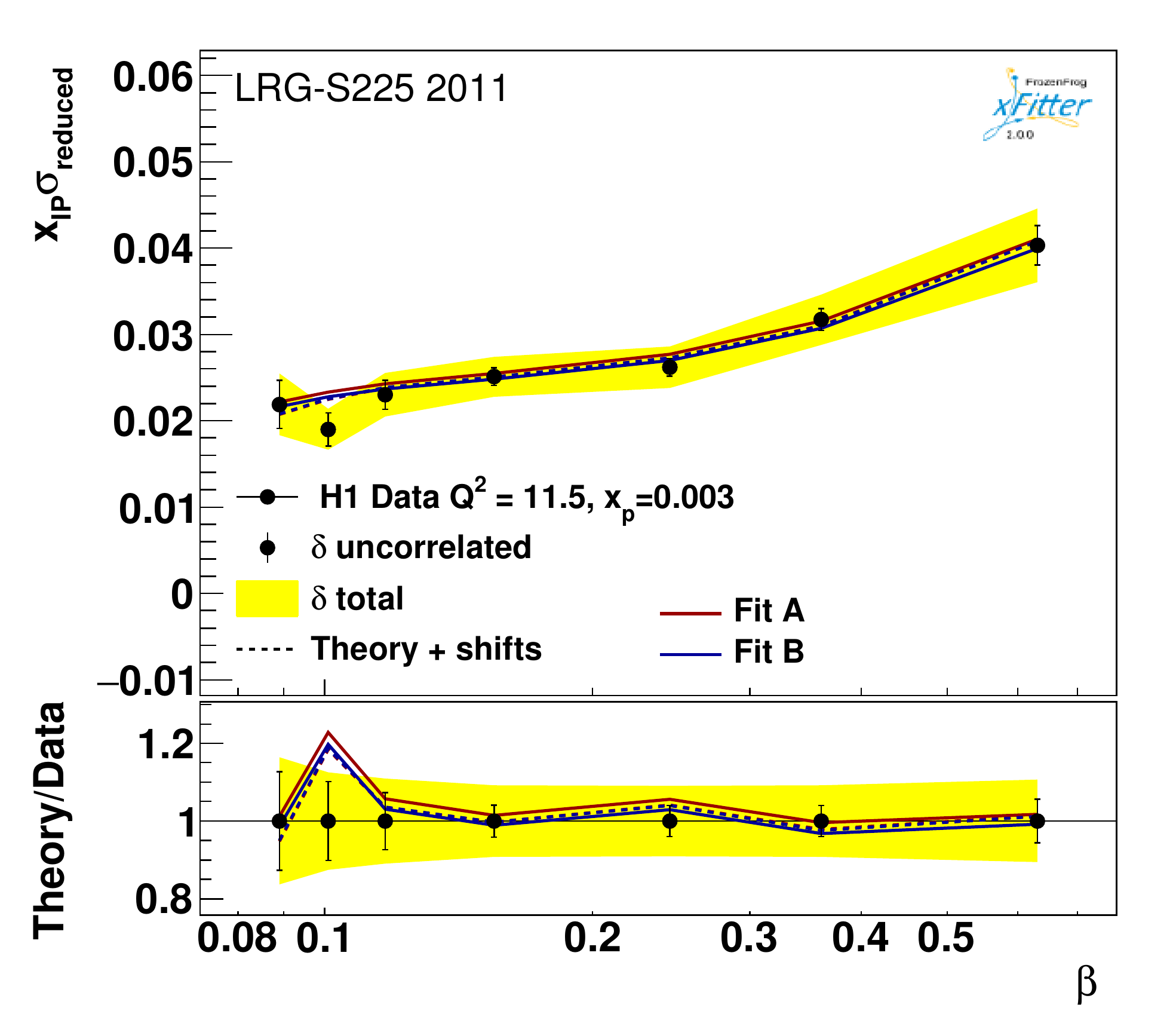}}   
\resizebox{0.480\textwidth}{!}{\includegraphics{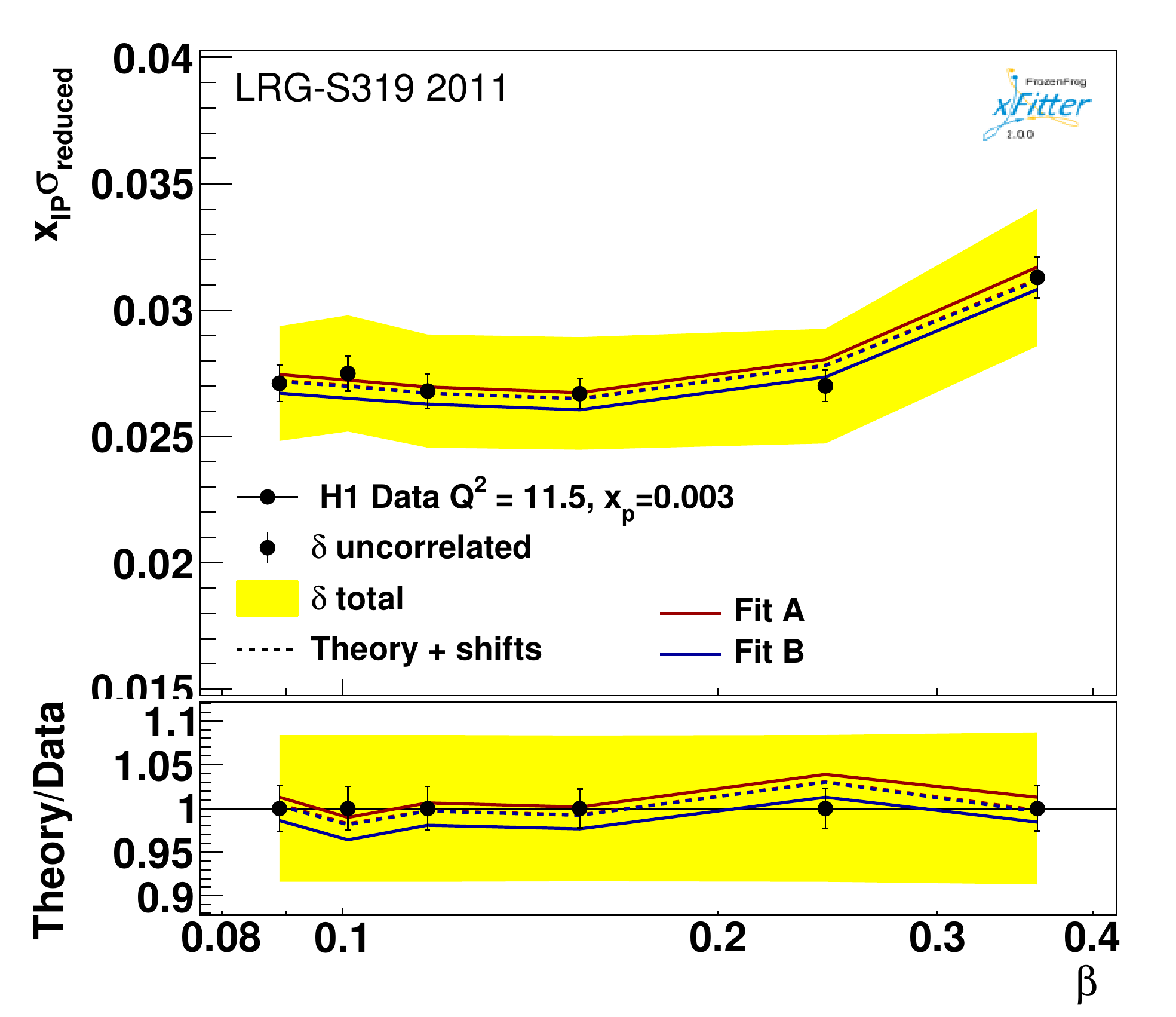}}   
\caption{\small The results of our NLO pQCD fit based on {\tt Fit A}  and  {\tt Fit B} for the reduced diffractive cross section $x_{\pom} \sigma_r^{D(3)}$ as a function of $\beta$ for $x_{\pom} = 0.003$ and $ Q^2=11.5 $ GeV$ ^2 $ in comparison with H1-LRG-2011 data~\cite{Aaron:2012zz} at $ \sqrt{s}=225 $ (left) and $ 319 $ (right). The error bars on the data points represent the uncorrelated uncertainties and the yellow bands represent the total uncorrelated and correlated uncertainties. }\label{fig:LRG-2011}
\end{center}
\end{figure*}

In the case of H1-LRG-2011 data~\cite{Aaron:2012zz}, we present in Fig.~\ref{fig:LRG-2011} the NLO pQCD fits of both our {\tt Fit A} and {\tt Fit B}. This figure shows, for instance, the NLO theory predictions for the reduced diffractive cross section $x_{\pom} \sigma_r^{D(3)}$ as a function of $\beta$ for $x_{\pom} = 0.003$ and $Q^2=11.5 \, {\rm GeV}^2$ in comparison with H1-LRG-2011 data at $\sqrt{s}=225 \, {\rm GeV}$ (left) and $319 \, {\rm GeV}$ (right). The error bars on the data points and the yellow bands represent the uncorrelated uncertainties and the total uncorrelated and correlated uncertainties, respectively. As can
be seen, in the kinematics considered, the theory is again in good agreement with the experiment.

From the results presented in this section, one can conclude that our NLO QCD predictions based on the DGLAP approach and using diffractive PDFs extracted from our QCD analysis of inclusive diffraction DIS data describe all analyzed data well.

%
\section{Summary and conclusions} \label{sec:Discussion}
%

In this paper, we have presented {\tt GKG18-DPDFs}, the first global QCD analysis of diffractive PDFs that makes use of 
the H1/ZEUS combined and the most recent H1 data sets on the reduced cross section of inclusive diffractive DIS.
Previous determinations of non-perturbative diffractive PDFs in the parton model of QCD~\cite{Aktas:2006hy,Chekanov:2009aa,Monfared:2011xf} were based on the older diffractive inclusive DIS data from H1 and ZEUS collaboration.
The advent of precise data from the H1~\cite{Aaron:2012zz,Aaron:2012ad} and H1/ZEUS combined~\cite{Aaron:2012hua} data sets as well as the widely used {\tt xFitter} package offer us the opportunity to obtain a new set of diffractive PDFs.
The TR GM-VFNS provides a rigorous theoretical framework for considering the heavy-quarks contributions and is employed here to determine diffractive PDFs of heavy quarks. The {\tt GKG18-DPDFs} delivers for the first time the optimized Hessian error analysis.

We study the impact of the new inclusive diffractive DIS data sets by producing two diffractive PDFs using two different scenarios.
Firstly, by considering simultaneously the $Q^2_{\rm min} = 9$ GeV$^2$ cut on all analyzed diffractive DIS data sets, and secondly by removing H1/ZEUS combined data with $Q^2_{\rm min} < 16$ GeV$^2$ in order to investigate possible tension between these data sets at small values of $Q^2$. In order to validate the efficiency and emphasize the phenomenological impact of this selection, the differences between these two diffractive PDFs sets are presented and discussed. 
We find that both of our diffractive PDFs determinations are in very good agreement with the results in the literature for the total quark singlet densities.
 
We also find differences between our results and the H1-2006 DPDFs fit  for the gluon density. There is much better agreement between {\tt GKG18} and ZUES-2010 for the gluon density. For the charm and bottom quark densities, there are insignificant discrepancies between {\tt GKG18-DPDFs} results and ZEUS-2010 for the small values of $z$; $z<0.01$.
Our theory predictions based on the determined diffractive  PDFs  for the reduced diffractive cross section are also in satisfactory agreements with the data sets analyzed as well as with the previous set of H1 data sets. The most significant changes are seen for the heavy quark densities at small values of $z$  and in the increased precision in the determination of the gluon diffractive PDF due to the inclusion of new precise data. 
For the future, our main aim is to include the very recent diffractive dijet production data, which could provide an additional constraint on the determination of the
diffractive gluon density.

A {\tt FORTRAN} subroutine, which evaluates the  leading order (LO) and NLO  diffractive PDFs presented here for given values of $\beta$, $x_{\pom}$ and $Q^2$,
can be obtained from the authors upon request via electronic mail.

%
\section*{Acknowledgments}
%

We are grateful to Sergey Levonian from H1 collaboration, Matthew Wing and Wojtek Slominski from ZEUS collaboration for many helpful discussions and comments.
We also thank Torbjörn Sjöstrand, Christine O. Rasmussen and Federico Alberto Ceccopieri for detailed discussions on the {\tt GKG18} diffractive PDFs.
Authors thank School of Particles and Accelerators, Institute for Research in Fundamental Sciences (IPM) for financial support of this project. HK also acknowledges the University of Science and Technology of Mazandaran for financial support provided for this research.


%
%

\clearpage

%

\end{document}